\documentclass[preprint,showpacs,12pt,floatfix,aps]{revtex4}
 
\usepackage{graphicx}
\usepackage{latexsym}
\usepackage{amsmath}
\input psfig.sty

\newcommand{\bq}{\begin{equation}}
\newcommand{\eq}{\end{equation}}
\newcommand{\bqn}{\begin{eqnarray}}
\newcommand{\eqn}{\end{eqnarray}}
\newcommand{\nb}{\nonumber}
\newcommand{\lb}{\label}
\newcommand{\rr}{\bf r}

\begin{document}
\title{Stable Gravastars of Anisotropic Dark Energy}
\author{R. Chan $^{1}$}
\email{chan@on.br}
\author{M.F.A. da Silva $^{2}$}
\email{mfasnic@gmail.com}
\author{P. Rocha $^{2,3,4}$}
\email{pedrosennarocha@gmail.com}
\author{Anzhong Wang $^{5}$}
\email{anzhong_wang@baylor.edu}
\affiliation{\small $^{1}$ Coordena\c{c}\~ao de Astronomia e Astrof\'{\i}sica, 
Observat\'orio Nacional, Rua General Jos\'e Cristino, 77, S\~ao Crist\'ov\~ao  
20921-400, Rio de Janeiro, RJ, Brazil\\
$^{2}$ Departamento de F\'{\i}sica Te\'orica, Instituto de F\'{\i}sica, 
Universidade do Estado do Rio de Janeiro, Rua S\~ao Francisco Xavier 524, 
Maracan\~a 20550-900, Rio de Janeiro - RJ, Brasil\\
$^{3}$ Universidade Est\'acio de S\'a, Rio de Janeiro, RJ, Brazil\\
$^{4}$ ICET/ITIC, Universidade Santa \'Ursula, Rua Fernando Ferrari, 
75 Botafogo 22231-020 , Rio de Janeiro , RJ, Brazil \\
$^{5}$ GCAP-CASPER, Department of Physics, Baylor University, Waco, TX 76798, USA}
 
\date{\today}

\begin{abstract}
Dynamical models of prototype gravastars made of phantom energy are constructed, in 
which an infinitely thin spherical shell of a perfect fluid with the equation of 
state $p = (1-\gamma)\sigma$ divides the whole spacetime into two regions, the 
internal region filled with a dark energy (or phantom) fluid, and the external Schwarzschild region. 
It is found that in some cases the  models represent the ``bounded excursion"  
stable gravastars, where  the thin shell is oscillating between two finite radii, 
while in other cases they collapse until the formation of black holes or  normal 
stars. In the phase space, the region for the ``bounded excursion"  gravastars is 
very small in comparison to that of black holes, but not empty, as found in our 
previous papers. Therefore, although the existence of gravastars can not be 
completely excluded  from current analysis, the opposite is not possible either,
that is, even if gravastars exist, they do not exclude the existence of black holes.
  
\end{abstract}

\pacs{98.80.-k,04.20.Cv,04.70.Dy}

\maketitle

\section{Introduction}

As alternatives to black holes,  gravastars have received some attention
recently \cite{grava}, partially due to the tight connection between the 
cosmological constant and a currently accelerating universe \cite{DEs}, 
although very strict observational constraints on the existence of such 
stars may exist \cite{BN07}. 

The pioneer model of gravastar was proposed by Mazur and Mottola (MM) \cite{MM01},
consisting of five layers: an internal core
$0 < r < r_1$ , described by the de Sitter universe, an intermediate thin layer of stiff fluid
$r_1 < r < r_2$ , an external region $r > r_2$ , described by the Schwarzschild solution, and two 
infinitely thin shells, appearing, respectively, on the hypersurfaces $r = r_1$ and
$r = r_2$. The intermediate layer is constructed in such way that $r_1$ is inner than the de Sitter horizon, while $r_2$ is outer than the Schwarzschild horizon, eliminating the apparent horizon. Configurations with a de Sitter interior have long history which we can find, for example, in the work of Dymnikova and Galaktionov \cite{irina}.  
After this work, Visser and Wiltshire \cite{VW04} pointed out that there are 
two different types of stable gravastars which are stable gravastars and 
``bounded excursion" gravastars. In the spherically symmetric case, the motion 
of the surface of the gravastar can be written in the form \cite{VW04},
\bq
\lb{1.4}
\frac{1}{2}\dot{a}^{2} + V(a) = 0,
\eq
where $a$ denotes the radius of the star, and $\dot{a} \equiv da/d\tau$, with 
$\tau$ being the proper time of the surface. Depending on the properties of 
the potential $V(a)$, the two kinds of gravastars are defined as follows. 

{\bf Stable gravastars}: In this case,  there must exist a radius $a_{0}$ such that
\bq
\lb{1.5}
V\left(a_{0}\right) = 0, \;\;\; V'\left(a_{0}\right) = 0, \;\;\;
V''\left(a_{0}\right) > 0,
\eq
where a prime denotes the ordinary differentiation with respect to the indicated argument.
If and only if there exists such a radius $a_{0}$ for which the above conditions are satisfied,
the model is said to be stable. 
Among other things, VW found that there are many equations of state for which the gravastar
configurations are stable, while others are not \cite{VW04}. Carter studied  the same
problem and found new equations of state for which the gravastar is stable \cite{Carter05}, 
while De Benedictis {\em et al} \cite{DeB06} and Chirenti and Rezzolla \cite{CR07} 
investigated the stability of the original model
of  Mazur and  Mottola against axial-perturbations, and found that gravastars are stable to
these perturbations, too. Chirenti and Rezzolla also showed that their quasi-normal modes 
differ from those of black holes with the same mass, and thus can be used to discern a gravastar 
from a black hole. 

{\bf "Bounded excursion" gravastars}: As VW noticed, there is a less stringent notion of 
stability, the so-called ``bounded excursion" models, in which there exist two radii $a_{1}$ 
and $a_{2}$ such that
\bq
\lb{1.6}
V\left(a_{1}\right) = 0, \;\;\; V'\left(a_{1}\right) \le 0, \;\;\;
V\left(a_{2}\right) = 0, \;\;\; V'\left(a_{2}\right) \ge 0,
\eq
with $V(a) < 0$ for $a \in \left(a_{1}, a_{2}\right)$, where $a_{2} > a_{1}$. 

Lately, we studied both types of gravastars \cite{JCAP,JCAP1}, and found that, 
such configurations can indeed be constructed, although   the region for the formation 
of them is very small in comparison to that of black holes.

Based on the discussions about the gravastar picture some authors have proposed
alternative models \cite{Chan}. Among them, we can find a Chaplygin dark star \cite{Paramos},
a gravastar supported by non-linear electrodynamics \cite{Lobo07},
a gravastar with continuous anisotropic pressure \cite{CattoenVisser05}
and recently, Dzhunushaliev et al. worked on spherically symmetric configurations
of a phantom scalar field and they found something like a gravastar but it was unstable\cite{singleton}.
In addition, Lobo \cite{Lobo} studied two models for a dark energy fluid. One of them
describes a homogeneous energy density and the other describes  an
ad-hoc monotonically decreasing energy density, although both of them are with anisotropic
pressure.  In order to match an exterior Schwarzschild spacetime he  
introduced a thin shell between the interior and the exterior spacetimes.

In this paper, we generalize our previous works \cite{JCAP,JCAP1} to the case where the 
equation of state of the infinitely thin shell is given by $p= (1-\gamma)  \sigma$ with 
$\gamma$ being a constant, the interior consists of a phantom energy fluid \cite{Lobo},
while the exterior is still the Schwarzschild space. 
We shall first construct three-layer dynamical models,  and then show both types of
gravastars and black holes exist for various situations. 
The rest of 
the paper is  organized as follows: In Sec. II we present the metrics of the 
interior and exterior spacetimes, and write down the motion of the thin shell
in the form of Eq.(\ref{1.4}).  In Sec. III we show the definitions of dark and 
phantom energy, for the
interior fluid and for the shell.  In Sec. IV we discuss the formation of
black holes from standard or phantom energy. In Sec. V we analyze the formation of
gravastar or normal star from standard or phantom energy. In Sec. VI we
study special cases where we can not have the "bounded excursion".
Finally, in Sec. VII we present our conclusions.

\section{ Dynamical Three-layer Prototype Gravastars}

The interior fluid is made of an anisotropic dark energy fluid with a metric
given by \cite{Lobo}
\bq
ds^2_{-}=-f_1 dt^2 + f_2 dr^2 + r^2 d\Omega^2,
\lb{ds2-}
\eq
where  $d\Omega^2 \equiv d\theta^2 + \sin^2(\theta)d\phi^2$, and 
\bqn
f_1 &=& (1+b r^2)^{\frac{1-\omega}{2}}(1+2 b r^2)^{\omega},\nb\\
f_2 &=& \frac{1+2 b r^2}{1 + b r^2},
\eqn  
where $\omega$ is a constant, and its physical meaning can be seen from the
following equation (\ref{prpt}). Since the mass is given by 
$\bar m(r)=b r^3/[2(1+2br^2)]$ then we have that $b > 0$.
The corresponding energy density $\rho$, radial  and tangential  pressures $p_r$ and 
$p_t$ are given, respectively, by
\begin{eqnarray}
p_r&=&\omega \rho =\left(\frac{\omega
b}{8\pi}\right)\left(\frac{3+2b r^2}{(1+2b r^2)^2}\right), \nb \\
p_t&=& -\left(\frac{b}{8\pi}\right)\left(\frac{\omega(3+2b
r^2)}{(1+2b r^2)^2}\right)
+ \frac{b^2r^2}{32\pi\left[(1+2b r^2)^3(1+b r^2)\right]}\times \nb \\
&&\Big\{(1+\omega)(3+2b r^2)\left[(1+3\omega)+2br^2(1+\omega)\right] \nb \\
&&-8\omega(5+2br^2)(1+br^2)\Big\}.
\lb{prpt}
\end{eqnarray}
The exterior spacetime is given by the Schwarzschild metric
\bq
ds^2_{+}= - f dv^2 + f^{-1} d{\rr}^2 + {\rr}^2 d\Omega^2,
\lb{ds2+}
\eq
where $f=1 - {2m}/{\rr}$.
The metric of the hypersurface  on the shell is given by
\bq
ds^2_{\Sigma}= -d\tau^2 + R^2(\tau) d\Omega^2.
\lb{ds2Sigma}
\eq
Since $ds^2_{-} = ds^2_{+} = ds^2_{\Sigma}$, we find that  $r_{\Sigma}={\rr}_{\Sigma}=R$,
and  
\bqn
\lb{dott2}
f_1\dot t^2 - f_2 \dot R^2 &=&  1,\\
\lb{dotv2}
f\dot v^2 - \frac{\dot R^2}{f} &=& 1,
\eqn
where the dot denotes the ordinary differentiation with respect to the proper time.
On the other hand,  the interior and exterior normal vectors to the thin shell are given by
\bqn
\lb{nalpha-}
n^{-}_{\alpha} &=& (-\dot R, \dot t, 0 , 0 ),\nb\\
n^{+}_{\alpha} &=& (-\dot R, \dot v, 0 , 0 ).
\eqn
Then, the interior and exterior extrinsic curvature are given by
\bqn
K^{-}_{\tau\tau}&=&\frac{1}{2} (1+b R^2)^{-\omega/2} \dot t \left\{ \left[ 4 (1+b R^2
)^{\omega/2} b R^2 \dot R^2+2 (1+b R^2)^{\omega/2} \dot R^2- \right. \right. \nb \\
& &\left. \left. (1+2 b R^2)^\omega \sqrt{1+b R^2} b R^2 \dot t^2-(1+2 b R^2)^ 
\omega \sqrt{1+b R^2} \dot t^2\right] (2 b R^2 \omega+2 b R^2+3 \omega+1)- \right. \nb\\
& &\left. 2 (1+b R^ 2)^{\omega/2} (1+2 b R^2) \dot R^2 \right\} (1+2 b R^2
)^{-2} (1+b R^2)^{-1} b R+ \dot R \ddot t- \ddot R \dot t,\\
\lb{Ktautau-}
K^{-}_{\theta\theta} &=& \frac{\dot t(1+b R^2) R}{1 + 2 b R^2},\\
\lb{Kthetatheta-}
K^{-}_{\phi\phi} &=& K^{-}_{\theta\theta}\sin^2(\theta),\\
\lb{Kphiphi-}
K^{+}_{\tau\tau} &=& \dot v(4 m^2 \dot v^2-4 m R \dot v^2-3 R^2 \dot R^2+
R^2 \dot v^2) (2 m-R)^{-1} m R^{-3}+\dot R \ddot v- \ddot R \dot v,\\
\lb{Ktautau+}
K^{+}_{\theta\theta} &=& -\dot v (2 m-R),\\
\lb{Kthetatheta+}
K^{+}_{\phi\phi} &=& K^{+}_{\theta\theta}\sin^2(\theta).
\lb{Kphiphi+}
\eqn
Since \cite{Lake}
\bq
[K_{\theta\theta}]= K^{+}_{\theta\theta}-K^{-}_{\theta\theta} = - M,
\lb{M}
\eq
where $M$ is the mass of the shell, we find that
\bq
M=\dot v (2 m-R)+\frac{\dot t(1+b R^2) R}{1 + 2 b R^2}.
\lb{M1}
\eq
Then, substituting equations (\ref{dott2}) and (\ref{dotv2}) into (\ref{M1}) 
we get
\bq
M=-R\left(1-\frac{2m}{R} + \dot R^2 \right)^{1/2} + 
R\frac{ \left[ 1 + b R^2 + \dot R^2 (1 + 2 b R^2) \right]^{1/2}}
{(1+b R^2)^{-(\omega+1)/4}(1+2b R^2)^{(\omega+2)/2}}.
\lb{M2}
\eq
In order to keep the ideas of MM as much as possible, we consider the thin 
shell as consisting
of a fluid with the equation of state, $p=(1-\gamma)\sigma$, where $\sigma$ and $p$ denote, 
respectively, the surface energy density and pressure of the shell and $\gamma$ is a constant. 
Then, the equation of motion of the shell is given by \cite{Lake}
\bq
\dot M + 8\pi R \dot R p = 4 \pi R^2 [T_{\alpha\beta}u^{\alpha}n^{\beta}]=
\pi R^2 \left(T^+_{\alpha\beta}u_+^{\alpha}n_+^{\beta}-T^-_{\alpha\beta}u_-^{\alpha}n_-^{\beta} \right),
\lb{dotM}
\eq
where $u^{\alpha}$ is the four-velocity.  Since the interior fluid is made
of an anisotropic fluid and the exterior is vacuum, we get
\bq
\dot M + 8\pi R \dot R (1-\gamma)\sigma = 0.
\lb{dotM1}
\eq
Recall that $\sigma = M/(4\pi R^2)$, we find that Eq.(\ref{dotM1}) has the solution
\bq
M=k R^{2(\gamma-1)},
\lb{Mk}
\eq
where $k$ is an integration constant. Substituting Eq.(\ref{Mk}) into Eq.(\ref{M2}),
and rescaling $m, \; b$ and $R$ as,
\bqn
m &\rightarrow& mk^{-\frac{1}{2\gamma-3}},\nb\\
b &\rightarrow& b k^{\frac{2}{2\gamma-3}},\nb\\
R &\rightarrow& Rk^{-\frac{1}{2\gamma-3}},
\eqn
we find that it can be written in the form of Eq.(\ref{1.4}) with $a$ replaced by $R$, and
\bqn
V(R,m,\omega,b,\gamma)&=& -\frac{1}{2R^2 b_2 \left[b_2^{(\omega+1)}-b_1^{(\omega+1)/2}\right]^2} 
\left\{b_2^{(\omega+2)} R^{4(\gamma-1)} b_1^{(\omega+1)/2}\right.\\
& & -2 b_2^{(3\omega+4)/2} R^{2(\gamma-1)} b_1^{(\omega+1)/4} 
\left[b_2^{(-\omega)} b_1^{(\omega+1)/2} R^2 -b_2^{-(\omega+1)} b_1^{(\omega+3)/2} R^2\right.\nb\\
& & \left. - 2 b_2^{(-\omega)} b_1^{(\omega+1)/2} m R+b_1 R^2+b_2 R^2 
+ 2 b_2 m R+b_2 R^{4(\gamma-1)}\right]^{1/2}   \nb \\
& & + b_2^{(\omega+2)} R^2 b_1^{(\omega+1)/2}
  - b_2^{(2\omega+3)} R^2-2 b_2^{(\omega+2)} m R b_1^{(\omega+1)/2}  \nb \\
& &\left. +2 b_2^{(2\omega+3)} m R+b_2^{(2\omega+3)} R^{4(\gamma-1)}-b_1^{(\omega+2)} R^2
 +b_2^{(\omega+1)} b_1^{(\omega+3)/2} R^2 \right\}.
\lb{VR}
\eqn
where
\bq
\lb{b1}
b_1 \equiv 1+b R^2,\;\;\; 
b_2 \equiv 1+2 b R^2.
\eq
Clearly, for any given constants $m$, $\omega$, $b$ and $\gamma$, equation (\ref{VR}) uniquely 
determines the collapse of the prototype  gravastar. Depending on the initial value $R_{0}$,  
the collapse can form either a black hole,  a gravastar,   a Minkowski, or a spacetime filled with
phantom fluid. In the last case, the thin shell
first collapses to a finite non-zero minimal radius and then expands to infinity.  To  guarantee
that initially the spacetime does not have any kind of horizons,  cosmological or event,
we must restrict $R_{0}$ to the range,
\bq
\lb{2.2b}
R_{0} > 2 m,
\eq
where $R_0$ is the initial collapse radius. When $m = 0= b$, the thin shell disappears,
and the whole spacetime is Minkowski. So, in the following we shall not consider this case.

Since the potential  (\ref{VR}) is so complicated, it is too difficult to study it
analytically. Instead, in the following we shall study it numerically. Before doing so, we
shall  

\section{Classifications of Matter, Dark Energy, and Phantom Energy  for Anisotropic Fluids}

Recently \cite{Chan08}, the classification of matter, dark and phantom energy 
for an anisotropic fluid was given  in terms of the energy conditions. Such a classification
is necessary for systems where anisotropy is important, and  the pressure components 
may play very important roles and  can have quite different contributions.
In this paper, we will use this classification to study the collapse of the
dynamical prototype gravastars, constructed in the last section. 
In particular, we define dark
energy  as a fluid which violates the strong energy
condition
(SEC).  From the Raychaudhuri equation, we can see that such defined
dark energy always exerts  divergent forces on  time-like or null geodesics.
On the other hand,  we define phantom energy as a fluid  that violates at least
one of the null energy conditions (NEC's). We shall further distinguish phantom
energy that satisfies the SEC
from that which does not satisfy the SEC. We call
the former attractive 
phantom energy, and the latter
 repulsive phantom energy.
Such a classification is summarized in Table I.

For the sake of completeness, in Table II we apply it to the matter field
located on the thin shell, while in Table III we combine all the results of Tables I 
and II, and present all the possibilities.   

\begin{table}
\caption{\label{tab:table1} This table summarizes the  classification of
the interior matter field, based on the energy conditions \cite{HE73}, where 
we assume that $\rho \ge 0$.}
\begin{ruledtabular}
\begin{tabular}{cccc}
Matter & Condition 1 & Condition 2  & Condition 3 \\
\hline
Normal Matter           & $\rho+p_r+2p_t\ge 0$ & $\rho+p_r\ge 0$ & $\rho+p_t\ge 0$ \\
Dark Energy               & $\rho+p_r+2p_t <  0$ & $\rho+p_r\ge 0$ & $\rho+p_t\ge 0$ \\
Repulsive Phantom Energy  & $\rho+p_r+2p_t <  0$ & $\rho+p_r <  0$ & $\rho+p_t\ge 0$ \\
Repulsive Phantom Energy  & $\rho+p_r+2p_t <  0$ & $\rho+p_r\ge 0$ & $\rho+p_t <  0$ \\
Repulsive Phantom Energy  & $\rho+p_r+2p_t <  0$ & $\rho+p_r <  0$ & $\rho+p_t <  0$ \\
Attractive Phantom Energy & $\rho+p_r+2p_t\ge 0$ & $\rho+p_r <  0$ & $\rho+p_t\ge 0$ \\
Attractive Phantom Energy & $\rho+p_r+2p_t\ge 0$ & $\rho+p_r\ge 0$ & $\rho+p_t <  0$ \\
Attractive Phantom Energy & $\rho+p_r+2p_t\ge 0$ & $\rho+p_r <  0$ & $\rho+p_t <  0$ \\
\end{tabular}
\end{ruledtabular}
\end{table}

\begin{table}
\caption{\label{tab:table2} This table summarizes the  classification of matter
on the thin shell, based on the energy conditions \cite{HE73}. The last column indicates
the particular values of the parameter $\gamma$, where we assume that $\rho \ge 0$.}
\begin{ruledtabular}
\begin{tabular}{cccc}
Matter & Condition 1 & Condition 2  & $\gamma$ \\
\hline
Normal Matter           & $\sigma+2p\ge 0$ & $\sigma+p\ge 0$ & -1 or 0  \\
Dark Energy               & $\sigma+2p <  0$ & $\sigma+p\ge 0$ &  7/4 \\
Repulsive Phantom Energy  & $\sigma+2p <  0$ & $\sigma+p <  0$ &   3  \\
Attractive Phantom Energy & $\sigma+2p\ge 0$ & $\sigma+p <  0$ &  Not possible   \\    
\end{tabular}
\end{ruledtabular}
\end{table}

\begin{figure}
\includegraphics[width=15cm]{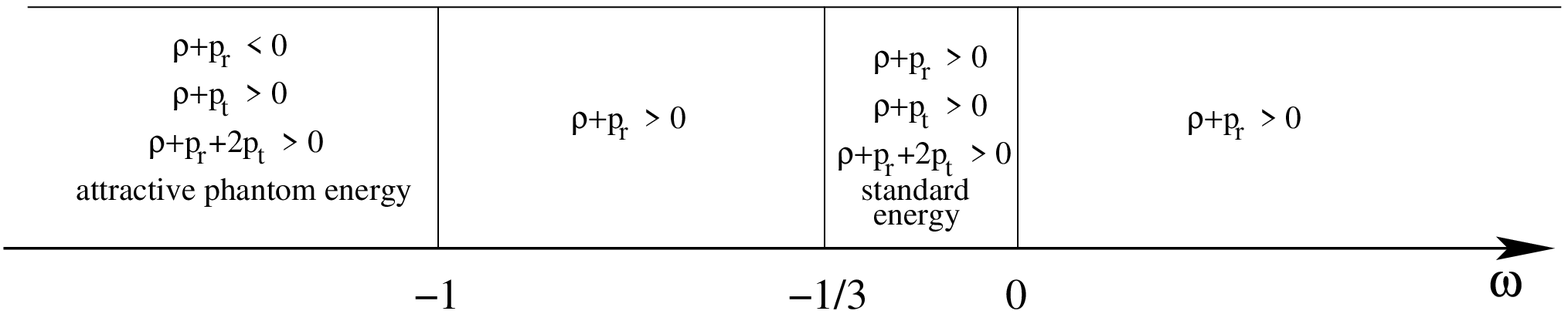}
\caption{In this figure we show the intervals of $\omega$ for which the weak and strong 
energy conditions are independent of the coordinate $R$ and the parameter $b$. The
condition $\rho+p_r>0$ is violated for $\omega<-1$ and fulfilled for $\omega>-1$, 
for any values of $R$ and $b$.
The conditions $\rho+p_t>0$ and $\rho+p_r+2p_t>0$ are satisfied for $\omega<-1$
and $-1/3<\omega<0$, for any values of $R$ and $b$.  For the others intervals 
of $\omega$ the analysis of the
energy conditions depends on a complex relation of $R$ and $b$.} 
\label{energyconditions}
\end{figure}

In order to consider the equations (\ref{ds2-}) and (\ref{prpt}) for describing dark energy
stars we must analyze carefully the ranges of the parameter $\omega$ that in
fact furnish the expected fluids.  It can be shown that the
condition $\rho+p_r>0$ is violated for $\omega<-1$ and fulfilled for $\omega>-1$, 
for any values of $R$ and $b$.
The conditions $\rho+p_t>0$ and $\rho+p_r+2p_t>0$ are satisfied for $\omega<-1$
and $-1/3<\omega<0$, for any values of $R$ and $b$.  
For the other intervals of
$\omega$  the
 energy conditions depend
on very complicated relations of $R$ and $b$.  See figure
\ref{energyconditions}. 
This provides an explicit example, in which the definition of dark energy must be
dealed with great care.  Another case was provided in a previous work \cite{Chan08}.  
Note that in the paper \cite{Lobo} where the solution
is used for the first time to model dark energy star, the author presented a
particular case for $\omega=-0.5$.  However, it is easy to see graphically that
for this value of
$\omega$ the weak and strong energy conditions are satisfied for any positive
value of the parameter $b$ and the coordinate $R$.  Thus, the corresponding solution
does not represent a dark energy star as it is claimed by the author.
Taking several values of $\omega$ in the intervals $-1<\omega<-1/3$ and 
$\omega>0$, we could not found any case where the interior dark energy exist.

In order to fulfill the energy condition $\sigma+2p\ge0$ of the shell
and assuming that
$p=(1-\gamma)\sigma$ we must have $\gamma \le 3/2$. On the other hand, in order
to satisfy the condition $\sigma+p\ge 0$, we obtain $\gamma \le 2$.
Hereinafter, we will use only some particular values of the parameter
$\gamma$ which are analyzed in this work. See Table II.

In the next sections we will discuss three physical possibilities for the
type of system that can be formed from the study of the potential
$V(R,m,\omega,b,\gamma)$: (a) Black hole or dispersion of the matter,
(b) Gravastar or normal star and (c) Black hole or phantom gravastar.

\section{Black Hole or Dispersion of the Matter}

For $m>m_{c}$ the potential $V(R)$ is strictly negative as shown in figures 
4, 5, 8, 9, 12, 13, 16, 17, 20, 21, 24, 25, 28, 29, 32, 33, 36, 37, 38 and 39.
Then, the collapse always forms black holes.
For $m=m_{c}$, there are two different possibilities, depending on the choice of the initial
radius $R_{0}$. In particular, if the star begins to collapse with $R_{0}>R_{c}$, it 
will approach to the minimal radius $R_{c}$. Once it reaches this point, the
shell will stop collapsing.
However, this point is unstable and any small perturbations will lead the star
either to expand for ever and leave behind a flat spacetime, or to collapse until $R=0$,
whereby a Schwarzschild black hole is finally formed. On the other hand, if the star begins
to collapse with $2m_{c}<R_{0}<R_{c}$ as shown in these figures, the star will collapse until a
black hole is formed. For $m<m_{c}$,
the potentials $V(R)$ for each case have a positive maximal, and the equation $V(R,m<m_{c})=0$
has two positive roots $R_{1,2}$ with $R_{2}>R_{1}>0$. There are also two possibilities here,
depending on the choice of the initial radius $R_{0}$. If $R_{0}>R_{2}$, the star will first
contract to its minimal radius $R=R_{2}$ and then expand to infinity, whereby a Minkowski
spacetime is finally formed. If $2m<R_{0}<R_{1}$, the star will collapse continuously until
$R=0$, and a black hole will be finally formed.

\section{Gravastar or Normal Star}

In this case the potential takes the shape given by figures 2, 3, 6, 7, 10, 11, 14, 15, 18, 
19, 22, 23, 26, 27, 30, 31, 34 and 35, from which we can see that
$V(R) = 0$ now can have  one, two  or three real roots, depending on the mass of the shell. 
For $m>m_c$ we have, say, $R_{i}$, where $R_{i+1} > R_{i}$.
If we choose $R_{0} > R_{3}$ (for $m=m_c$ we have $R_2=R_3$), then the star 
will not be allowed in this region
because the potential is greater than the zero.  However,
if we choose $R_{1} < R_{0} < R_{2}$, the collapse will bounce back and forth 
between $R = R_{1}$
and $R = R_{2}$. Such a  possibility is shown in these figures. This is exactly
the  so-called "bounded excursion" model mentioned in \cite{VW04}, and studied in some details
in \cite{JCAP,JCAP1}.  Of course, in a realistic situation, the star will emit both gravitational waves and
particles, and the potential shall be self-adjusted to produce a minimum at $R = R_{static}$ where
$V\left(R=R_{static}\right) = 0 = V'\left(R=R_{static}\right)$
whereby a gravastar or a normal star is finally formed \cite{VW04,JCAP,JCAP1},
although in \cite{JCAP,JCAP1}  the potential tends to $-\infty$ when $R$ tends to $\infty$.
Here it is completely different since the potential now tends to $+\infty$ when
$R$ tends to $\infty$. 
Thus, in the cases studied here we do not have situations where the star expands
leaving behind a flat spacetime, as in  \cite{JCAP,JCAP1}.

\section{Black Hole or Phantom Gravastar}

In this case the potential takes the shape given by figures 40, 41, 42 and 43, 
from which we can see that
$V(R) = 0$ now has four real roots, say, $R_{i}$, where $R_{i+1} > R_{i}$.
If we choose $R_{0} > R_{4}$, then again the star will not be allowed in this 
region because the potential is greater than   zero.
However, if we choose $R_{3} < R_{0} < R_{4}$, the collapse will bounce back and 
forth between $R = R_{3}$
and $R = R_{4}$, as in the previous case. But, if we choose 
$R_{2} < R_{0} < R_{3}$, we can note that this region is forbidden because either
the potential is imaginary or   greater than zero. 
However,
if we choose $R_{1} < R_{0} < R_{2}$, the collapse will bounce back and forth
between $R = R_{1}$
and $R = R_{2}$. If $R_0 < R_1$ the system will collapse until $R=0$,
whereby a Schwarzschild black hole is finally formed.

\begin{table}
\caption{\label{tab:table3}This table summarizes all possible kind of energy
of the interior fluid and of the shell. The boldface figure numbers represent
stable structures.}
\begin{ruledtabular}
\begin{tabular}{ccccc}
Case & Interior Energy & Shell Energy & Figures & Structures \\
\hline
A & Standard          & Standard           & {\bf 2},{\bf 3},{\bf 6},{\bf 7},{\bf 10},
{\bf 11},{\bf 14},{\bf 15},{\bf 18},{\bf 19} & Normal Star \\
B & Standard           & Dark              & 4,8,12,16,20 & Black Hole/Dispersion\\
C & Standard          & Repulsive Phantom  & 5,9,13,17,21 & Black Hole/Dispersion\\
D & Dark              & Standard           &  & Interior not found\\
E & Dark              & Dark               &  & Interior not found\\
F & Dark              & Repulsive Phantom  &  & Interior not found\\
G & Repulsive Phantom & Standard           & {\bf 22},{\bf 23},{\bf 26},{\bf 27},
{\bf 30},{\bf 31},{\bf 34},{\bf 35} & Gravastar \\
H & Repulsive Phantom & Dark               & 24,28,32,36 & Black Hole/Dispersion\\
I & Repulsive Phantom & Repulsive Phantom  & 25,29,33,37 & Black Hole/Dispersion\\
J & Attractive Phantom & Standard          & {\bf 41}, {\bf 43}, 40, 42 & Gravastar or Black Hole \\
K & Attractive Phantom & Dark              & 38 & Black Hole/Dispersion\\
L & Attractive Phantom & Repulsive Phantom & 39 & Black Hole/Dispersion\\
\end{tabular}
\end{ruledtabular}
\end{table}

\section{Conclusions}

In this paper, we have studied the problem of the stability of gravastars by
constructing dynamical three-layer models  of VW \cite{VW04},
which consists of an internal phantom fluid, a dynamical infinitely thin  shell of
perfect fluid with the equation of state $p = (1-\gamma)\sigma$, and an external Schwarzschild space.

It must be noted that, although phantoms have been used widely to explain the
late cosmic acceleration of the universe, microscopic models of them, either
for an anisotropic or a perfect fluid, has not been constructed, yet, if there
is any.

We have shown explicitly that the final output can be a black
hole, a "bounded excursion" stable gravastar, a Minkowski, or a phantom 
spacetime, depending on the total mass $m$ of the system, the parameter 
$\omega$, 
the constant $b$, the parameter   $\gamma$ and
the initial position $R_{0}$ of the dynamical shell. All these possibilities
have non-zero measurements in the phase space of $m$, $b$, $\omega$, $\gamma$ 
and $R_{0}$, although the region  of gravastars is very small in comparison with
that of black holes. All the results can be summarized in Table III. 
An interesting result that we can deduce from Table III is that we can have
black hole formation even with an interior phantom energy for any given $\gamma$.
The results obtained in this paper further confirm our previous conclusion:
{\em even though the existence of gravastars cannot be completely 
excluded in these dynamical models, our results do indicate that, even if
gravastars indeed exist, they do not exclude the existence of black holes}.

\begin{figure}
\vspace{.2in}
\centerline{\psfig{figure=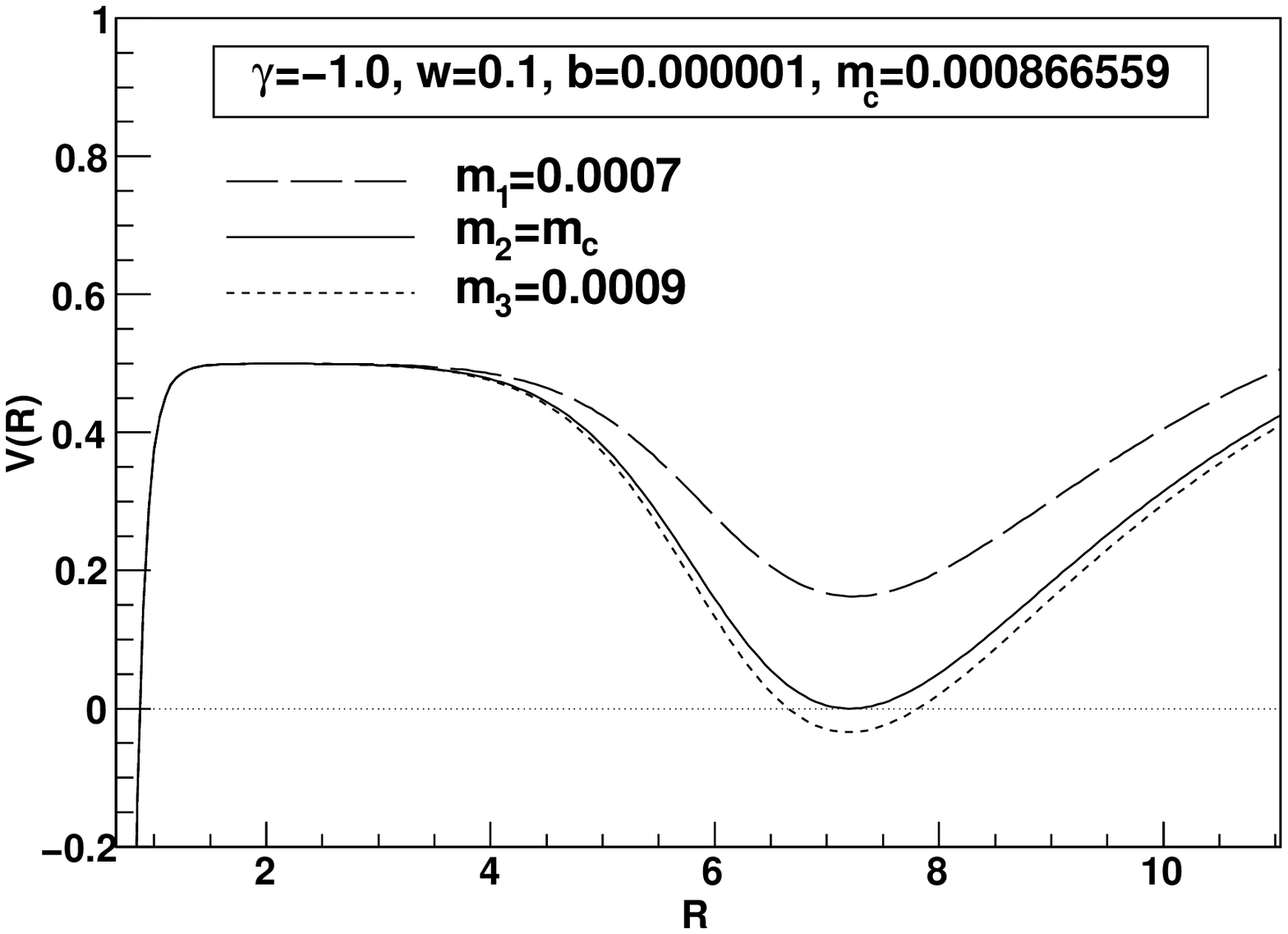,width=3.3truein,height=3.0truein}\hskip
.25in \psfig{figure=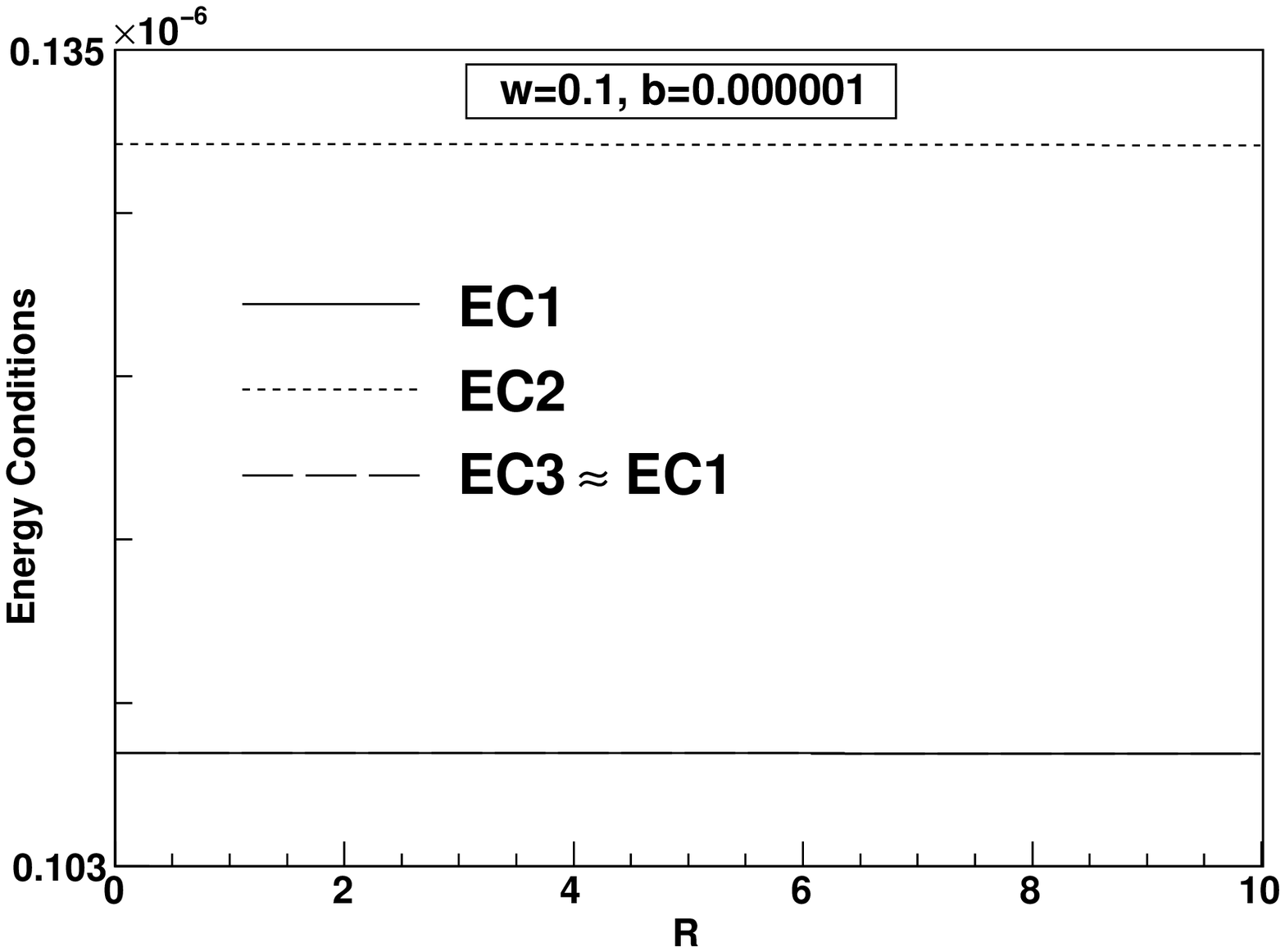,width=3.3truein,height=3.0truein}
\hskip .5in} \caption{The potential $V(R)$ and the energy conditions EC1$\equiv \rho+p_r+2p_t$, 
EC2$\equiv \rho+p_r$ and EC3$\equiv \rho+p_t$, for $\gamma=-1$,
$\omega=0.1$, $b=0.000001$ and $m_c=0.000866559$. {\bf Case A}}
\label{fig2}
\end{figure}

\begin{figure}
\vspace{.2in}
\centerline{\psfig{figure=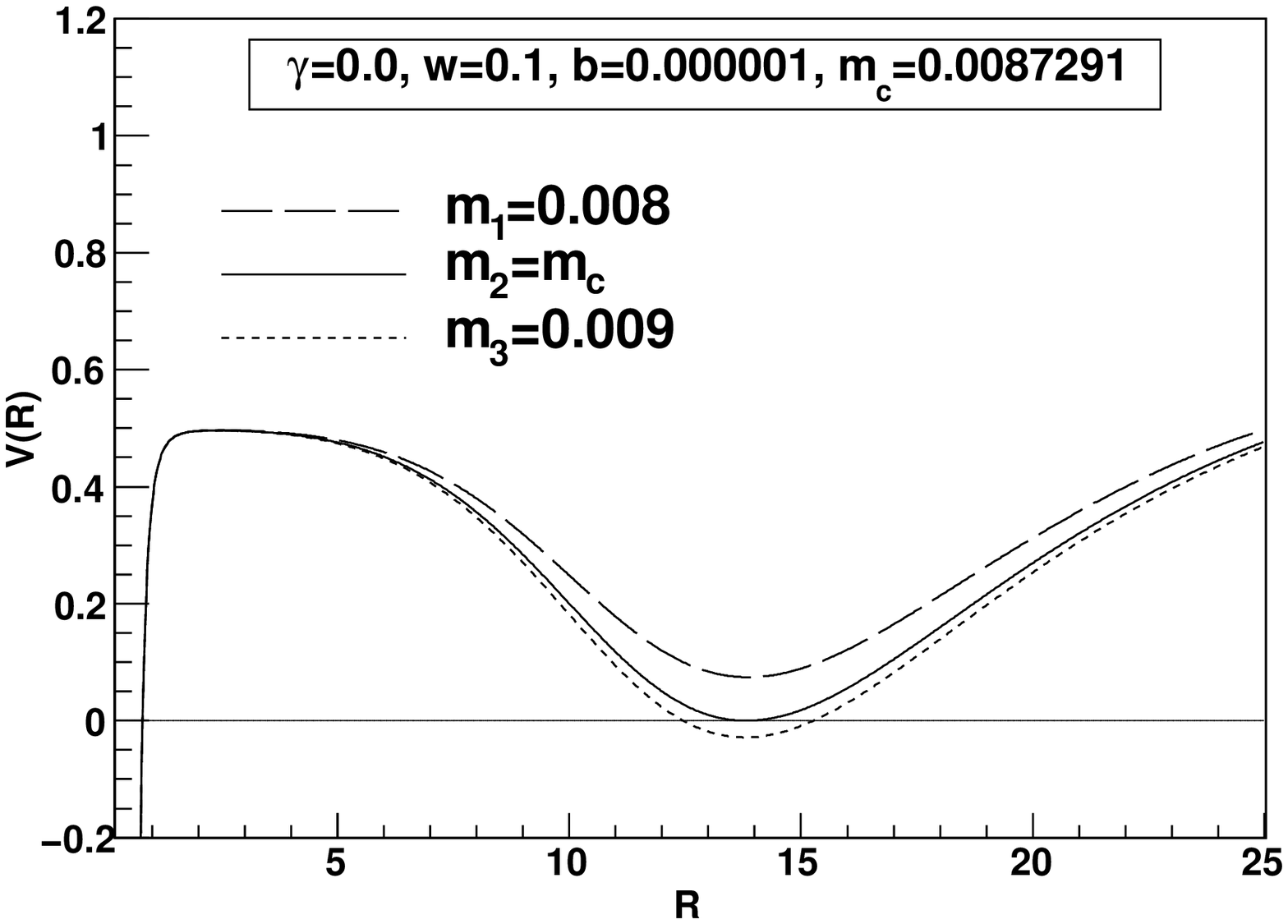,width=3.3truein,height=3.0truein}\hskip
.25in \psfig{figure=ECw0v1b0v000001.eps,width=3.3truein,height=3.0truein}
\hskip .5in} \caption{The potential $V(R)$ and the energy conditions EC1$\equiv \rho+p_r+2p_t$, 
EC2$\equiv \rho+p_r$ and EC3$\equiv \rho+p_t$, for $\gamma=0$,
$\omega=0.1$, $b=0.000001$ and $m_c=0.0087291$. {\bf Case A}}
\label{fig3}
\end{figure}

\begin{figure}
\vspace{.2in}
\centerline{\psfig{figure=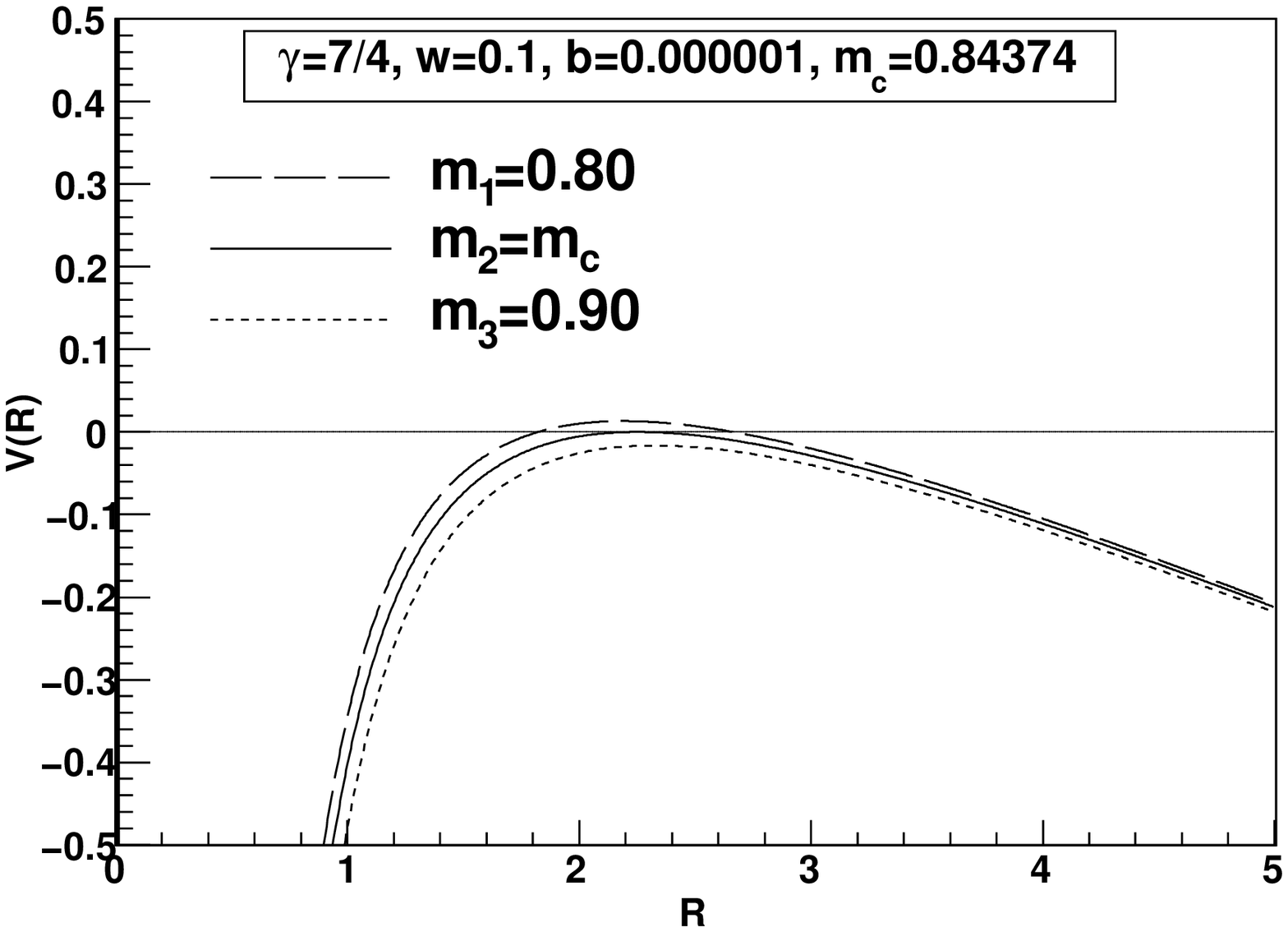,width=3.3truein,height=3.0truein}\hskip
.25in \psfig{figure=ECw0v1b0v000001.eps,width=3.3truein,height=3.0truein}
\hskip .5in} \caption{The potential $V(R)$ and the energy conditions EC1$\equiv \rho+p_r+2p_t$, 
EC2$\equiv \rho+p_r$ and EC3$\equiv \rho+p_t$, for $\gamma=7/4$,
$\omega=0.1$, $b=0.000001$ and $m_c=0.84374$. {\bf Case B}}
\label{fig4}
\end{figure}

\begin{figure}
\vspace{.2in}
\centerline{\psfig{figure=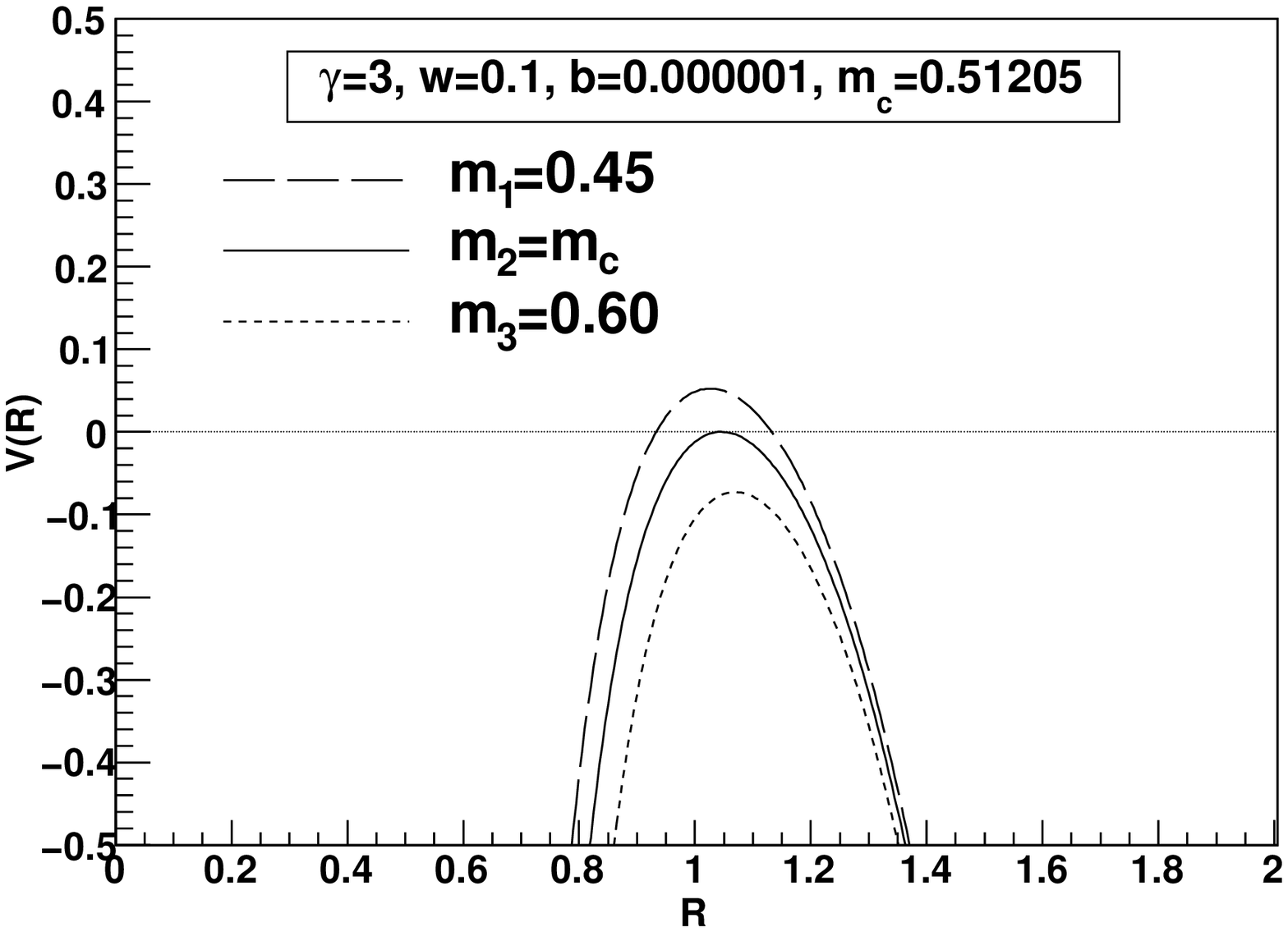,width=3.3truein,height=3.0truein}\hskip
.25in \psfig{figure=ECw0v1b0v000001.eps,width=3.3truein,height=3.0truein}
\hskip .5in} \caption{The potential $V(R)$ and the energy conditions EC1$\equiv \rho+p_r+2p_t$, 
EC2$\equiv \rho+p_r$ and EC3$\equiv \rho+p_t$, for $\gamma=3$,
$\omega=0.1$, $b=0.000001$ and $m_c=0.51205$. {\bf Case C}}
\label{fig5}
\end{figure}

\begin{figure}
\vspace{.2in}
\centerline{\psfig{figure=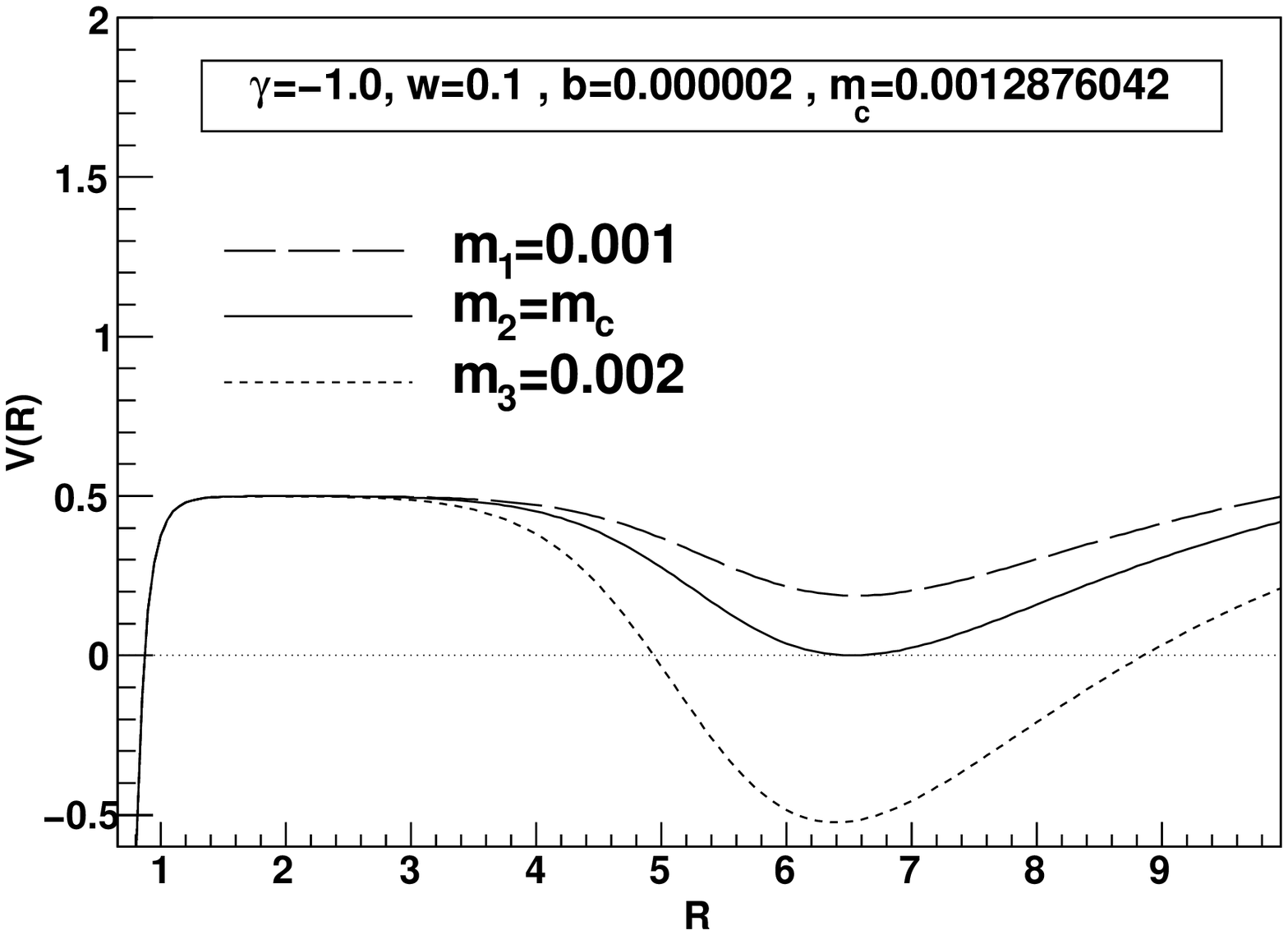,width=3.3truein,height=3.0truein}\hskip
.25in \psfig{figure=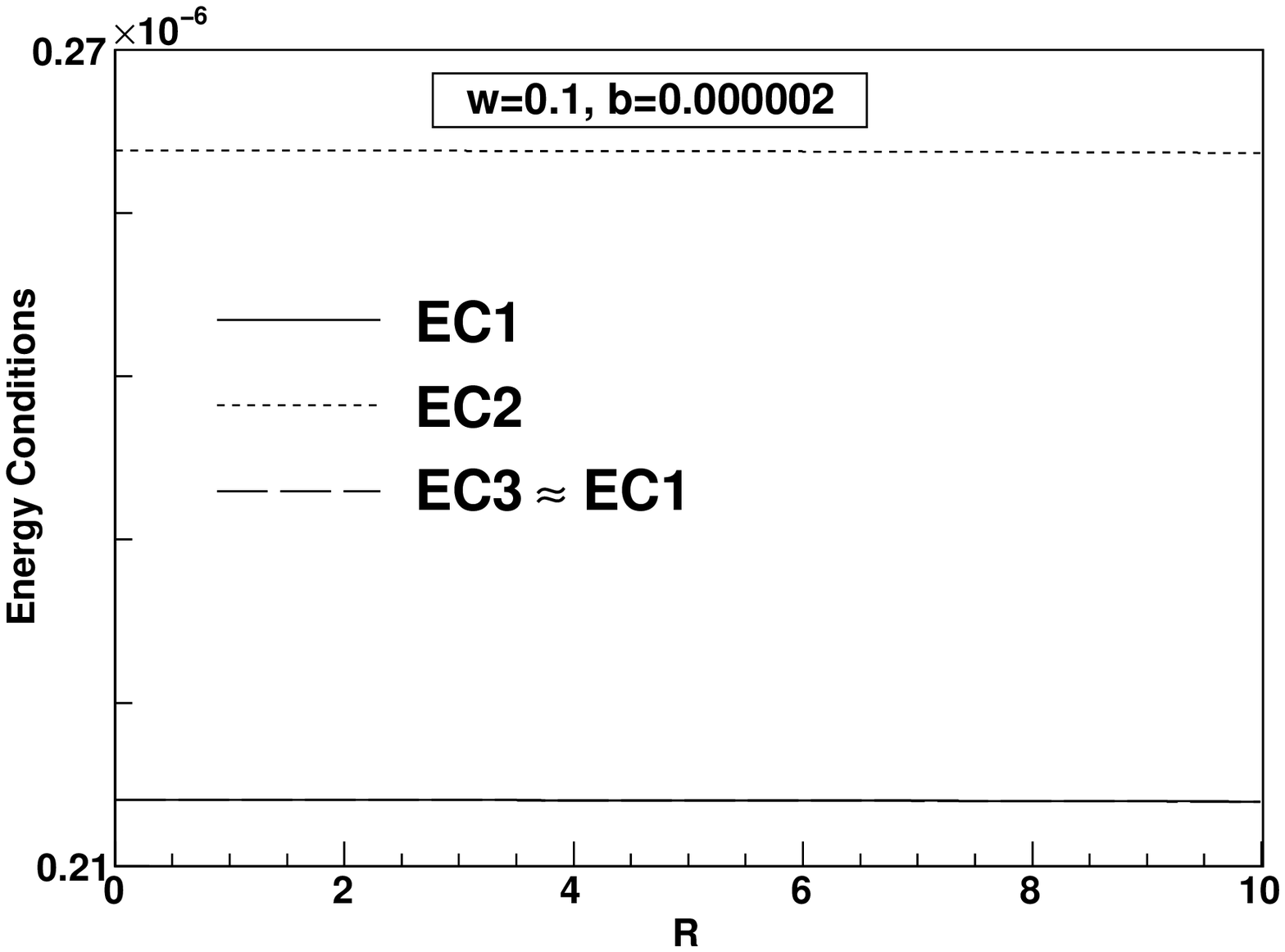,width=3.3truein,height=3.0truein}
\hskip .5in} \caption{The potential $V(R)$ and the energy conditions EC1$\equiv \rho+p_r+2p_t$, 
EC2$\equiv \rho+p_r$ and EC3$\equiv \rho+p_t$, for $\gamma=-1$,
$\omega=0.1$, $b=0.000002$ and $m_c=0.0012876042$. {\bf Case A}}
\label{fig6}
\end{figure}

\begin{figure}
\vspace{.2in}
\centerline{\psfig{figure=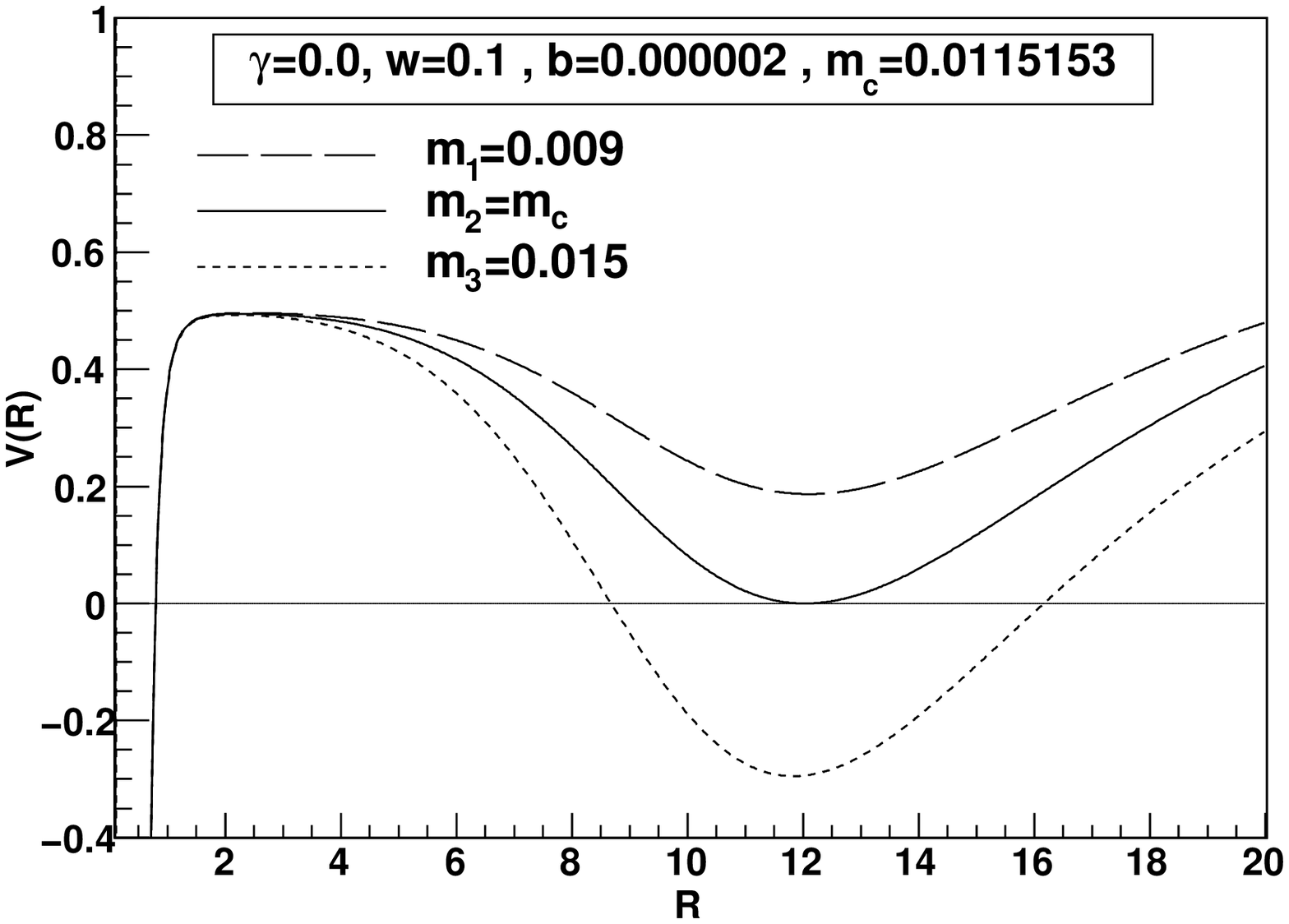,width=3.3truein,height=3.0truein}\hskip
.25in \psfig{figure=ECw0v1b0v000002.eps,width=3.3truein,height=3.0truein}
\hskip .5in} \caption{The potential $V(R)$ and the energy conditions EC1$\equiv \rho+p_r+2p_t$, 
EC2$\equiv \rho+p_r$ and EC3$\equiv \rho+p_t$, for $\gamma=0$,
$\omega=0.1$, $b=0.000002$ and $m_c=0.0115153$. {\bf Case A}}
\label{fig7}
\end{figure}

\begin{figure}
\vspace{.2in}
\centerline{\psfig{figure=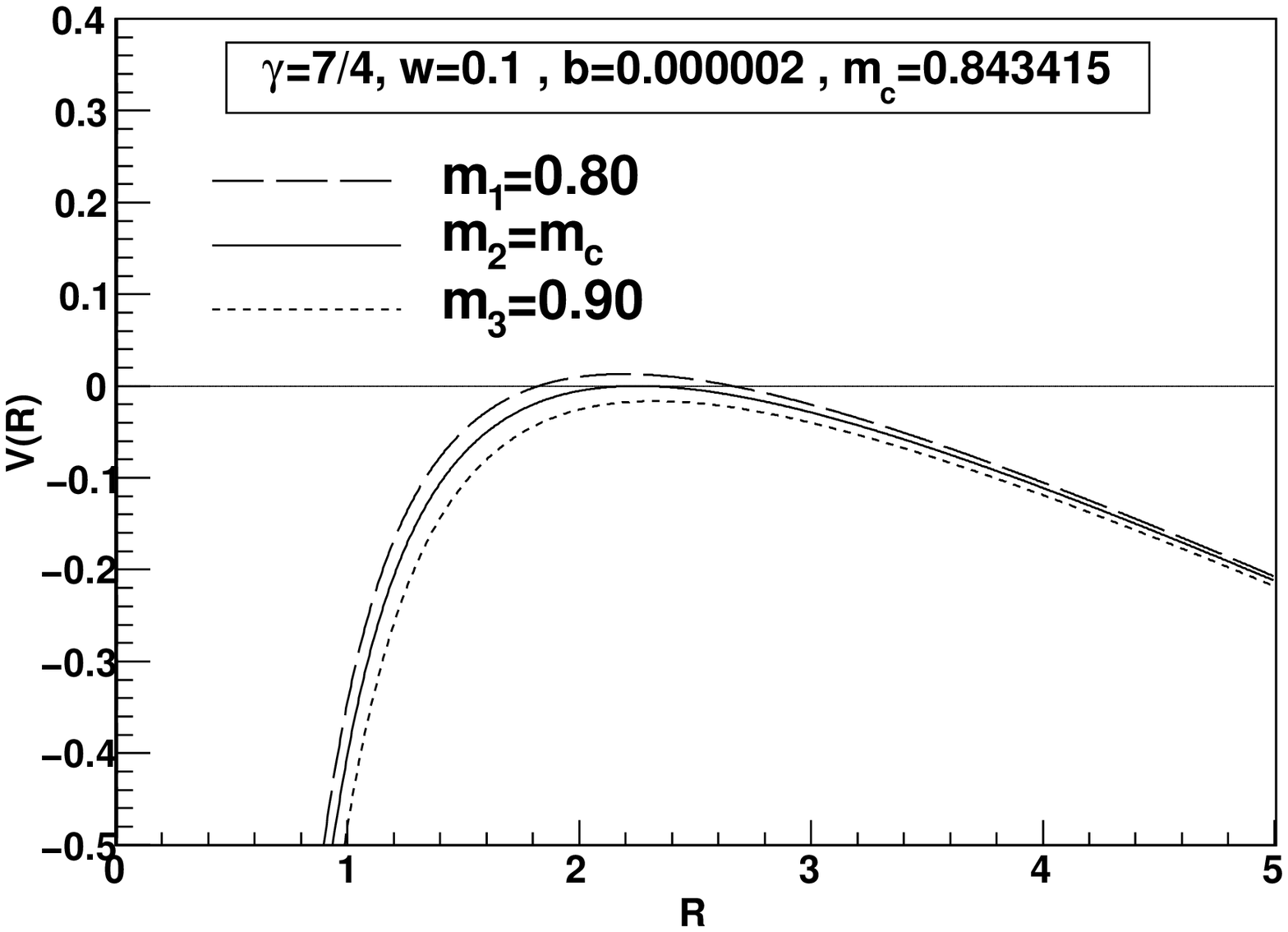,width=3.3truein,height=3.0truein}\hskip
.25in \psfig{figure=ECw0v1b0v000002.eps,width=3.3truein,height=3.0truein}
\hskip .5in} \caption{The potential $V(R)$ and the energy conditions EC1$\equiv \rho+p_r+2p_t$, 
EC2$\equiv \rho+p_r$ and EC3$\equiv \rho+p_t$, for $\gamma=7/4$,
$\omega=0.1$, $b=0.000002$ and $m_c=0.843415$. {\bf Case B}}
\label{fig8}
\end{figure}

\begin{figure}
\vspace{.2in}
\centerline{\psfig{figure=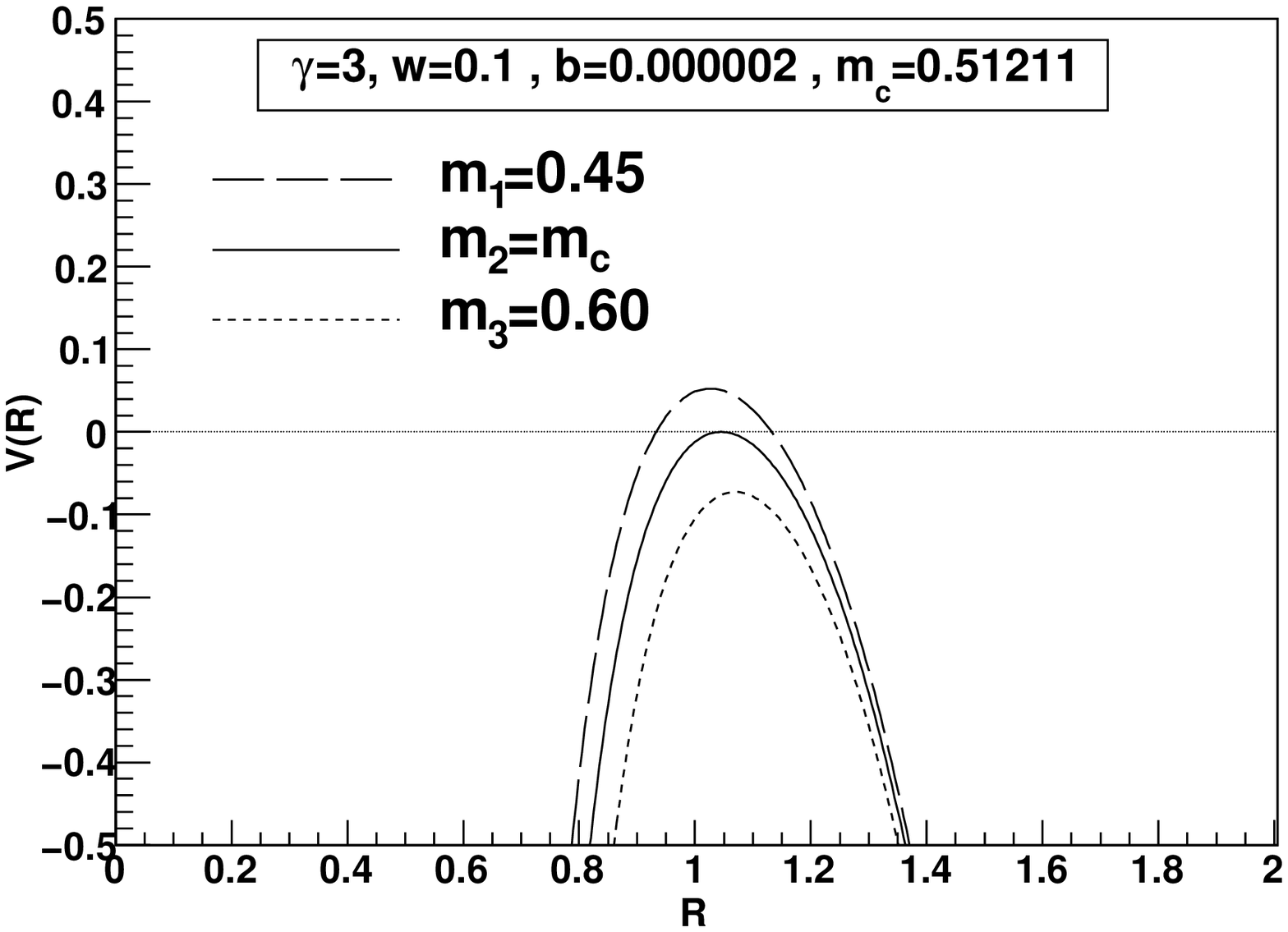,width=3.3truein,height=3.0truein}\hskip
.25in \psfig{figure=ECw0v1b0v000002.eps,width=3.3truein,height=3.0truein}
\hskip .5in} \caption{The potential $V(R)$ and the energy conditions EC1$\equiv \rho+p_r+2p_t$, 
EC2$\equiv \rho+p_r$ and EC3$\equiv \rho+p_t$, for $\gamma=3$,
$\omega=0.1$, $b=0.000002$ and $m_c=0.51211$. {\bf Case C}}
\label{fig9}
\end{figure}

\begin{figure}
\vspace{.2in}
\centerline{\psfig{figure=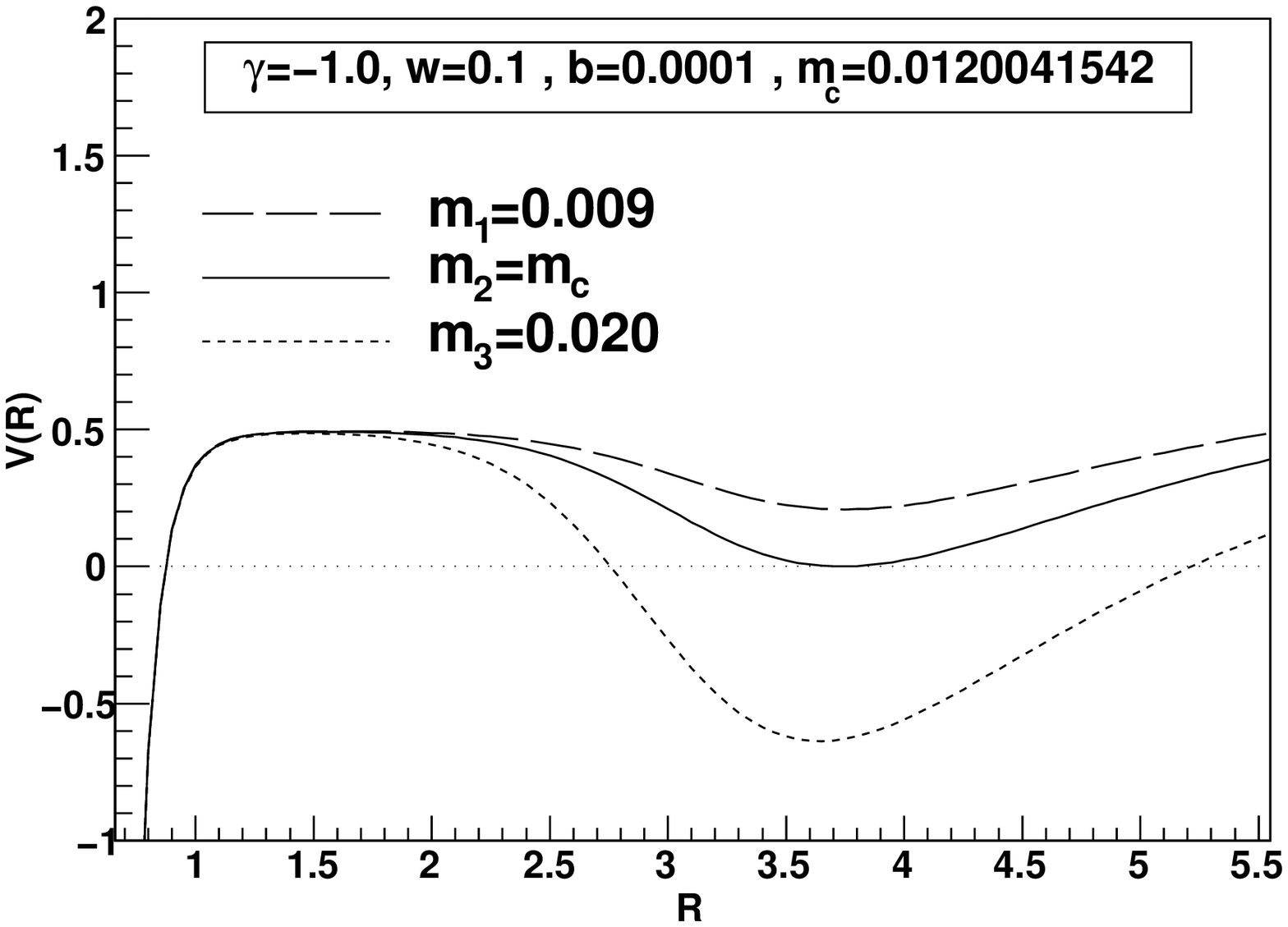,width=3.3truein,height=3.0truein}\hskip
.25in \psfig{figure=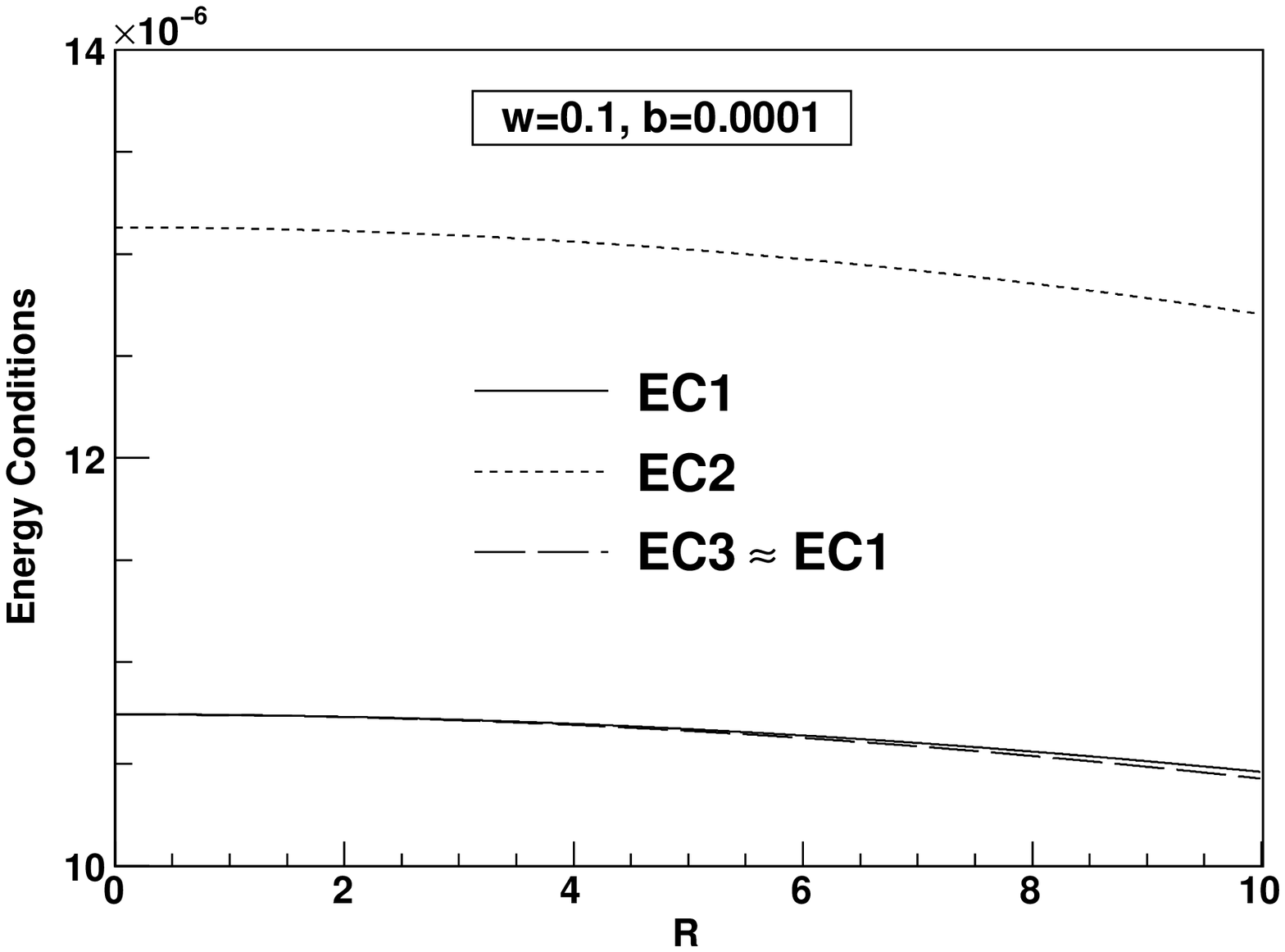,width=3.3truein,height=3.0truein}
\hskip .5in} \caption{The potential $V(R)$ and the energy conditions EC1$\equiv \rho+p_r+2p_t$, 
EC2$\equiv \rho+p_r$ and EC3$\equiv \rho+p_t$, for $\gamma=-1$,
$\omega=0.1$, $b=0.0001$ and $m_c=0.0120041542$. {\bf Case A}}
\label{fig10}
\end{figure}

\begin{figure}
\vspace{.2in}
\centerline{\psfig{figure=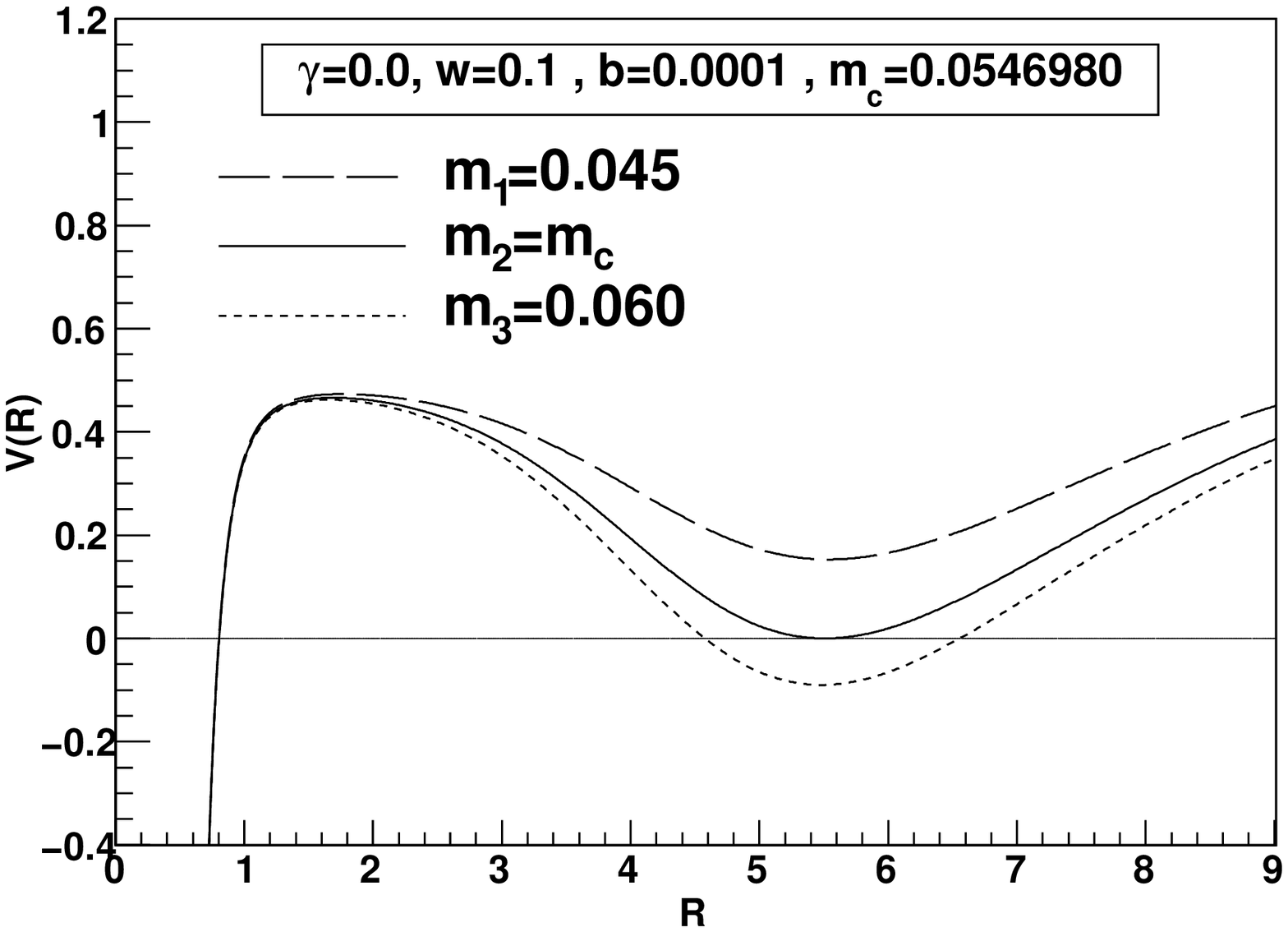,width=3.3truein,height=3.0truein}\hskip
.25in \psfig{figure=ECw0v1b0v0001.eps,width=3.3truein,height=3.0truein}
\hskip .5in} \caption{The potential $V(R)$ and the energy conditions EC1$\equiv \rho+p_r+2p_t$, 
EC2$\equiv \rho+p_r$ and EC3$\equiv \rho+p_t$, for $\gamma=0$,
$\omega=0.1$, $b=0.0001$ and $m_c=0.0546980$. {\bf Case A}}
\label{fig11}
\end{figure}

\begin{figure}
\vspace{.2in}
\centerline{\psfig{figure=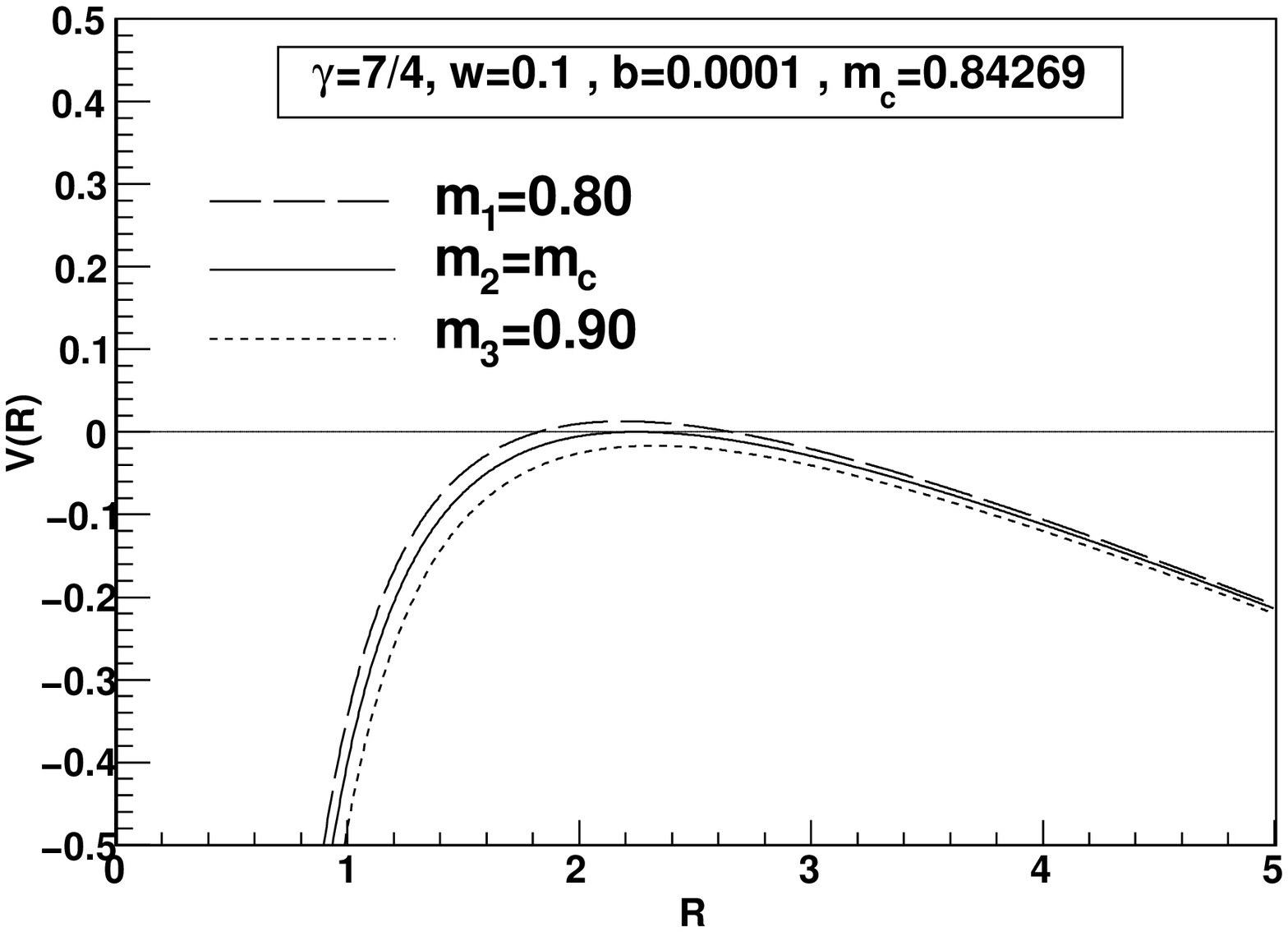,width=3.3truein,height=3.0truein}\hskip
.25in \psfig{figure=ECw0v1b0v0001.eps,width=3.3truein,height=3.0truein}
\hskip .5in} \caption{The potential $V(R)$ and the energy conditions EC1$\equiv \rho+p_r+2p_t$, 
EC2$\equiv \rho+p_r$ and EC3$\equiv \rho+p_t$, for $\gamma=7/4$,
$\omega=0.1$, $b=0.0001$ and $m_c=0.84269$. {\bf Case B}}
\label{fig12}
\end{figure}

\begin{figure}
\vspace{.2in}
\centerline{\psfig{figure=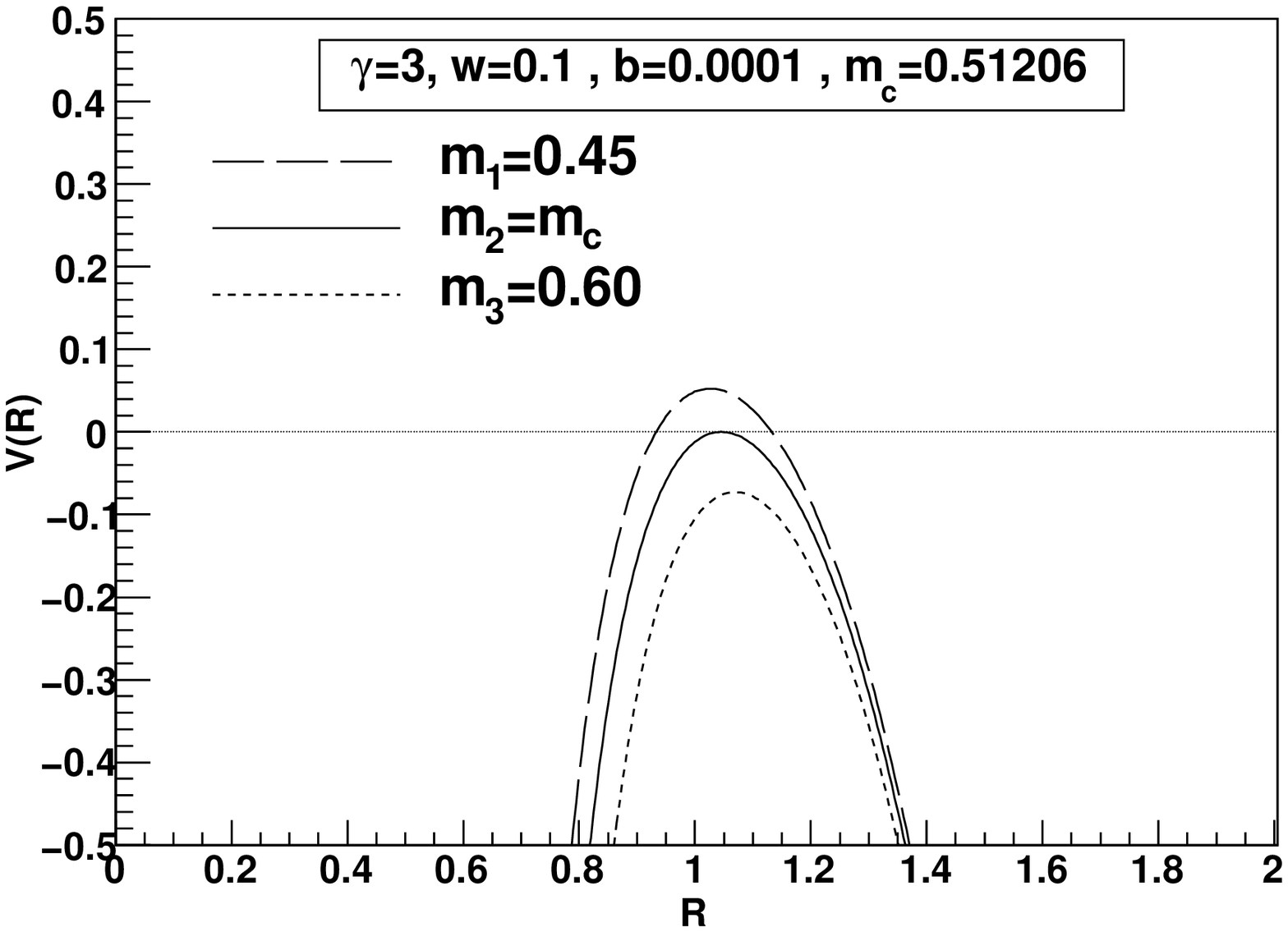,width=3.3truein,height=3.0truein}\hskip
.25in \psfig{figure=ECw0v1b0v0001.eps,width=3.3truein,height=3.0truein}
\hskip .5in} \caption{The potential $V(R)$ and the energy conditions EC1$\equiv \rho+p_r+2p_t$, 
EC2$\equiv \rho+p_r$ and EC3$\equiv \rho+p_t$, for $\gamma=3$,
$\omega=0.1$, $b=0.0001$ and $m_c=0.51206$. {\bf Case C}}
\label{fig13}
\end{figure}

\begin{figure}
\vspace{.2in}
\centerline{\psfig{figure=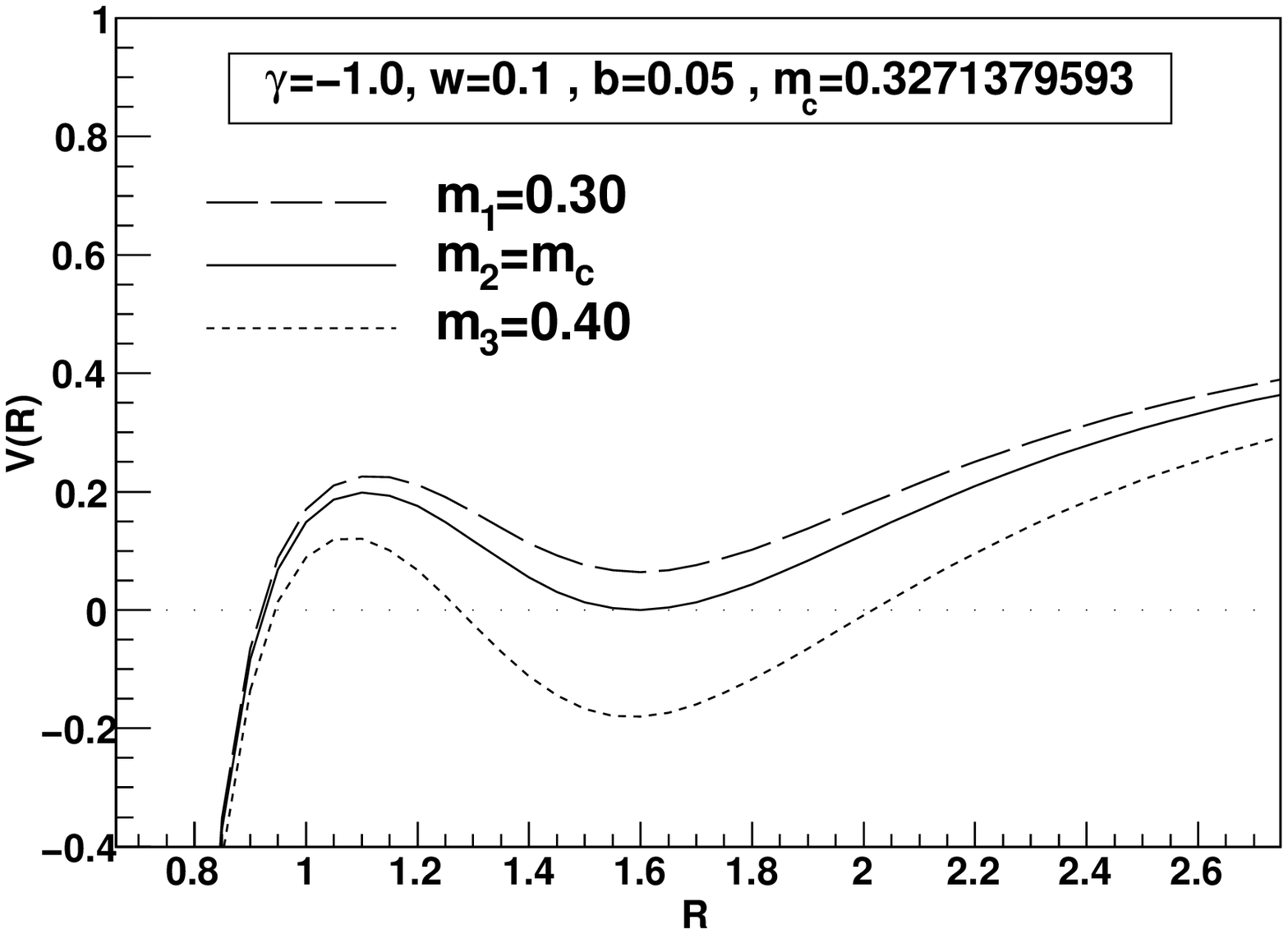,width=3.3truein,height=3.0truein}\hskip
.25in \psfig{figure=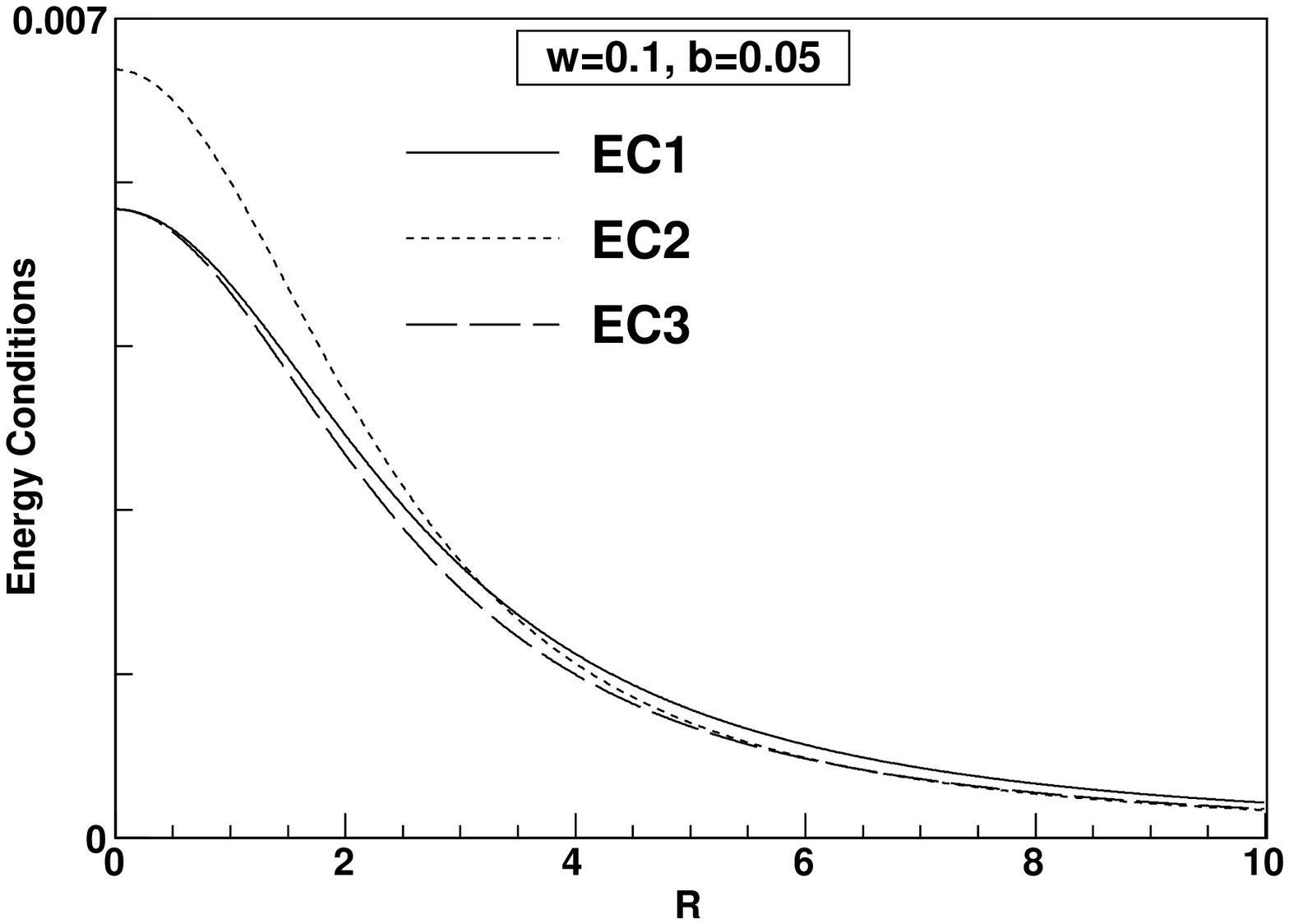,width=3.3truein,height=3.0truein}
\hskip .5in} \caption{The potential $V(R)$ and the energy conditions EC1$\equiv \rho+p_r+2p_t$, 
EC2$\equiv \rho+p_r$ and EC3$\equiv \rho+p_t$, for $\gamma=-1$,
$\omega=0.1$, $b=0.05$ and $m_c=0.3271379593$. {\bf Case A}}
\label{fig14}
\end{figure}

\begin{figure}
\vspace{.2in}
\centerline{\psfig{figure=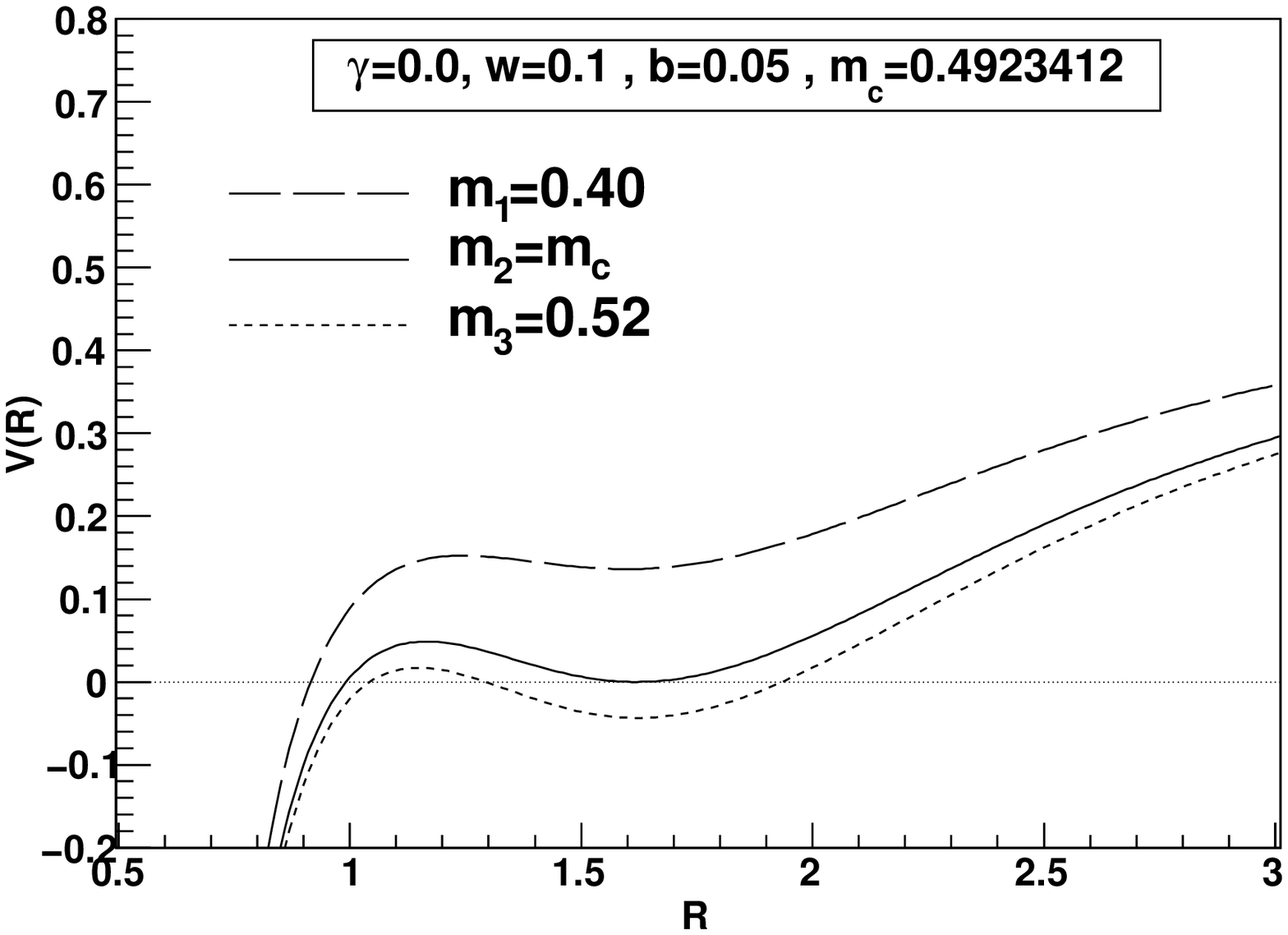,width=3.3truein,height=3.0truein}\hskip
.25in \psfig{figure=ECw0v1b0v05.eps,width=3.3truein,height=3.0truein}
\hskip .5in} \caption{The potential $V(R)$ and the energy conditions EC1$\equiv \rho+p_r+2p_t$, 
EC2$\equiv \rho+p_r$ and EC3$\equiv \rho+p_t$, for $\gamma=0$,
$\omega=0.1$, $b=0.05$ and $m_c=0.4923412$. {\bf Case A}}
\label{fig15}
\end{figure}

\begin{figure}
\vspace{.2in}
\centerline{\psfig{figure=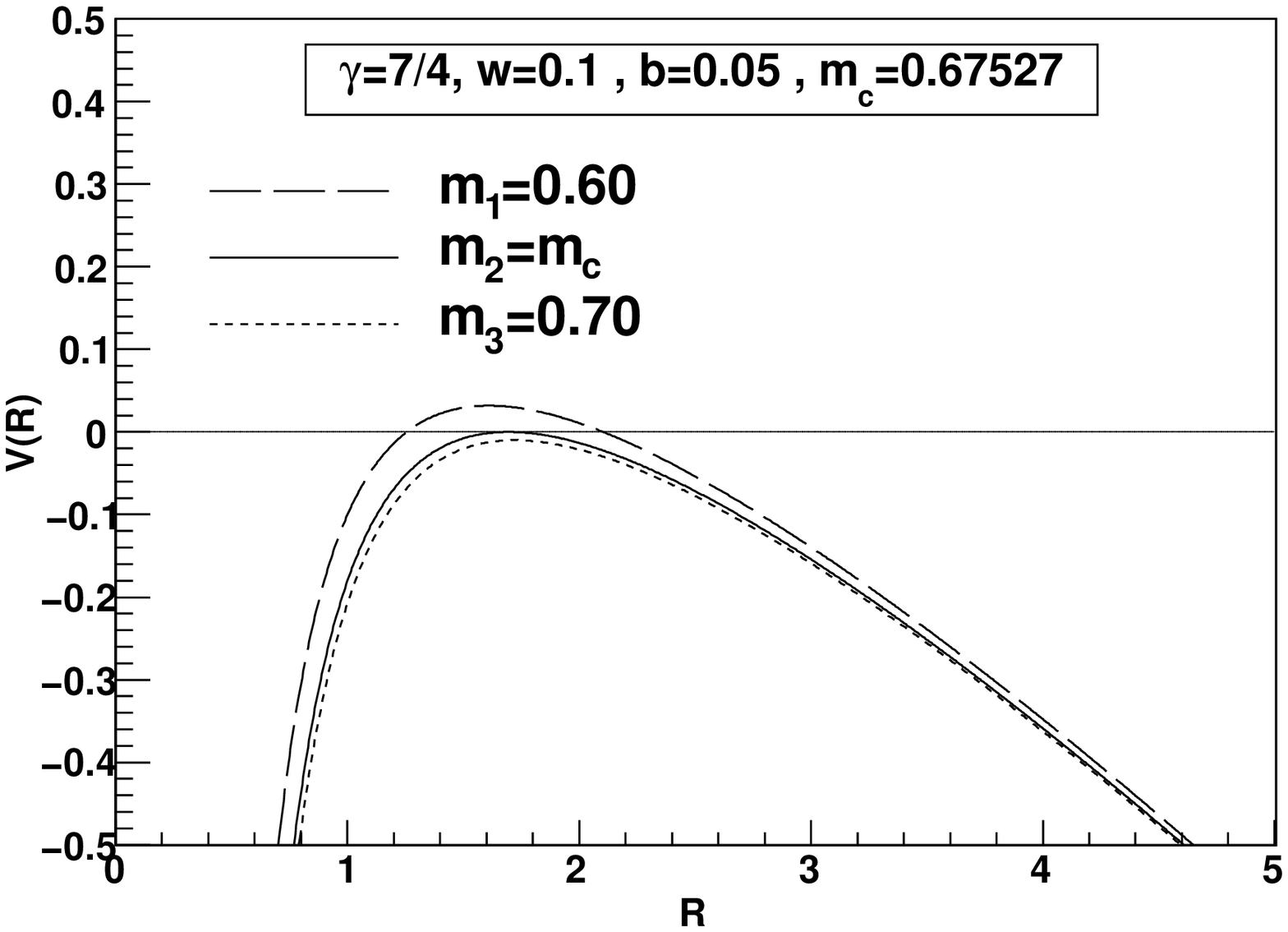,width=3.3truein,height=3.0truein}\hskip
.25in \psfig{figure=ECw0v1b0v05.eps,width=3.3truein,height=3.0truein}
\hskip .5in} \caption{The potential $V(R)$ and the energy conditions EC1$\equiv \rho+p_r+2p_t$, 
EC2$\equiv \rho+p_r$ and EC3$\equiv \rho+p_t$, for $\gamma=7/4$,
$\omega=0.1$, $b=0.05$ and $m_c=0.67527$. {\bf Case B}}
\label{fig16}
\end{figure}

\begin{figure}
\vspace{.2in}
\centerline{\psfig{figure=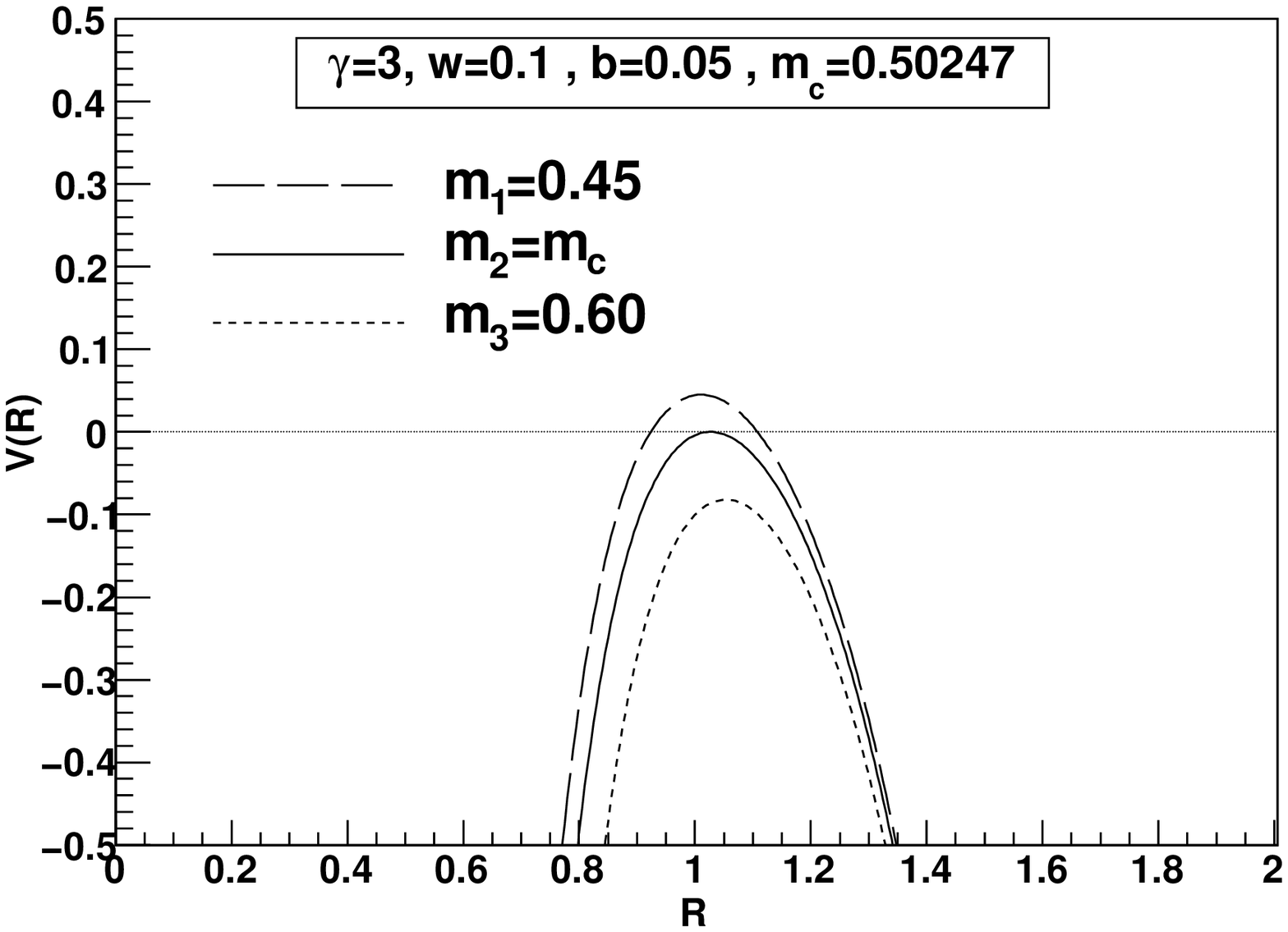,width=3.3truein,height=3.0truein}\hskip
.25in \psfig{figure=ECw0v1b0v05.eps,width=3.3truein,height=3.0truein}
\hskip .5in} \caption{The potential $V(R)$ and the energy conditions EC1$\equiv \rho+p_r+2p_t$, 
EC2$\equiv \rho+p_r$ and EC3$\equiv \rho+p_t$, for $\gamma=3$,
$\omega=0.1$, $b=0.05$ and $m_c=0.50247$. {\bf Case C}}
\label{fig17}
\end{figure}

\begin{figure}
\vspace{.2in}
\centerline{\psfig{figure=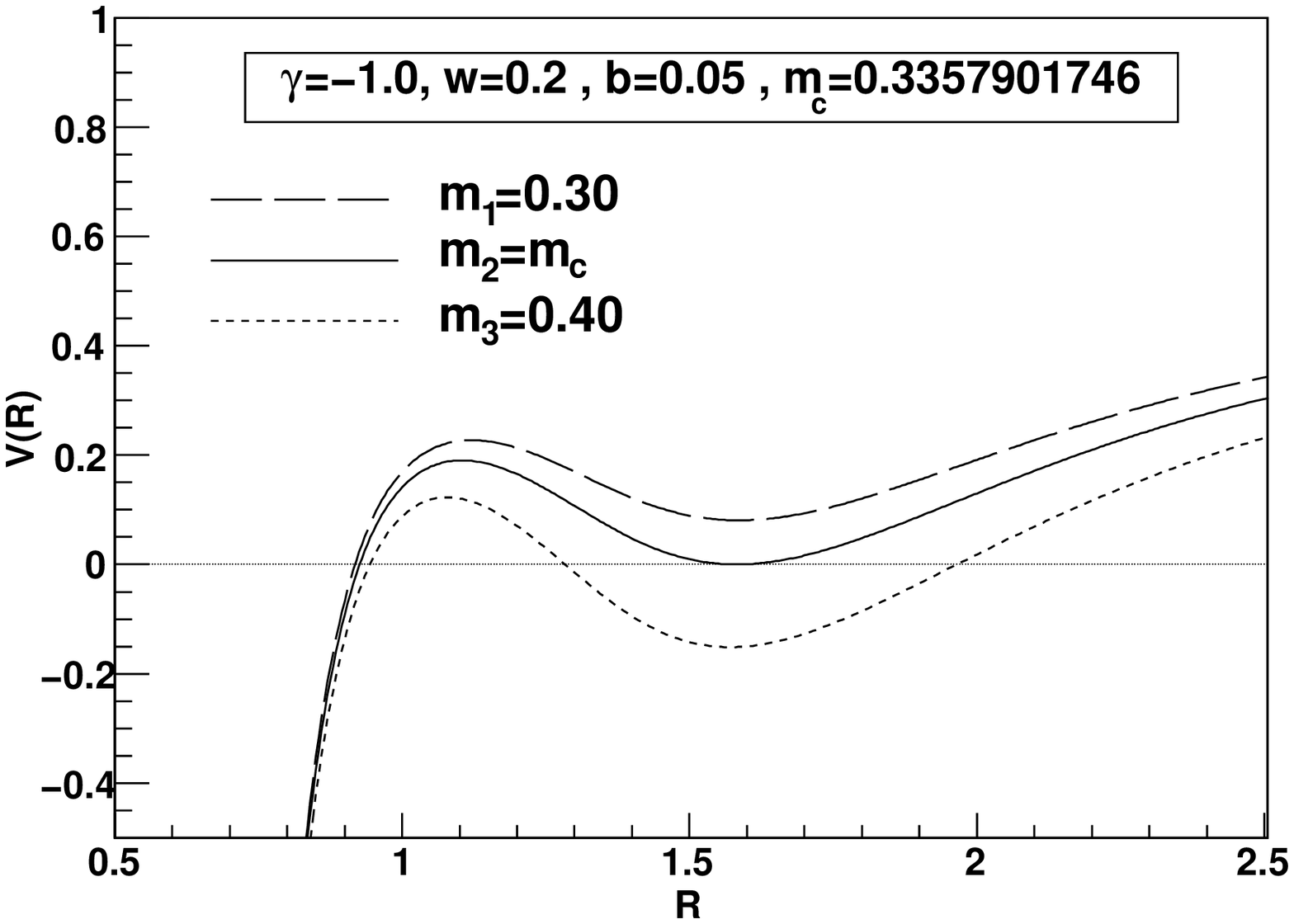,width=3.3truein,height=3.0truein}\hskip
.25in \psfig{figure=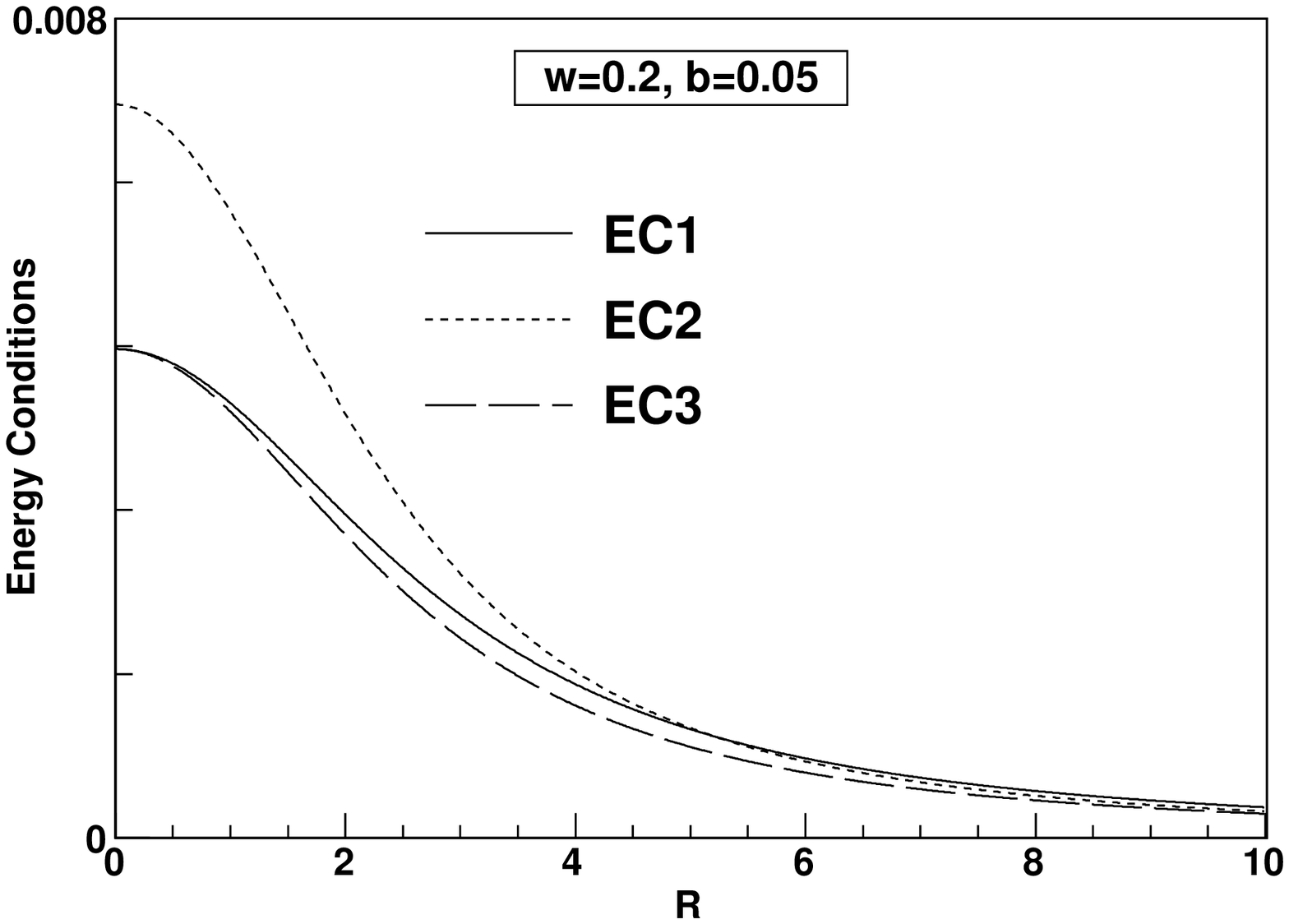,width=3.3truein,height=3.0truein}
\hskip .5in} \caption{The potential $V(R)$ and the energy conditions EC1$\equiv \rho+p_r+2p_t$, 
EC2$\equiv \rho+p_r$ and EC3$\equiv \rho+p_t$, for $\gamma=-1$,
$\omega=0.2$, $b=0.05$ and $m_c=0.3357901746$. {\bf Case A}}
\label{fig18}
\end{figure}

\begin{figure}
\vspace{.2in}
\centerline{\psfig{figure=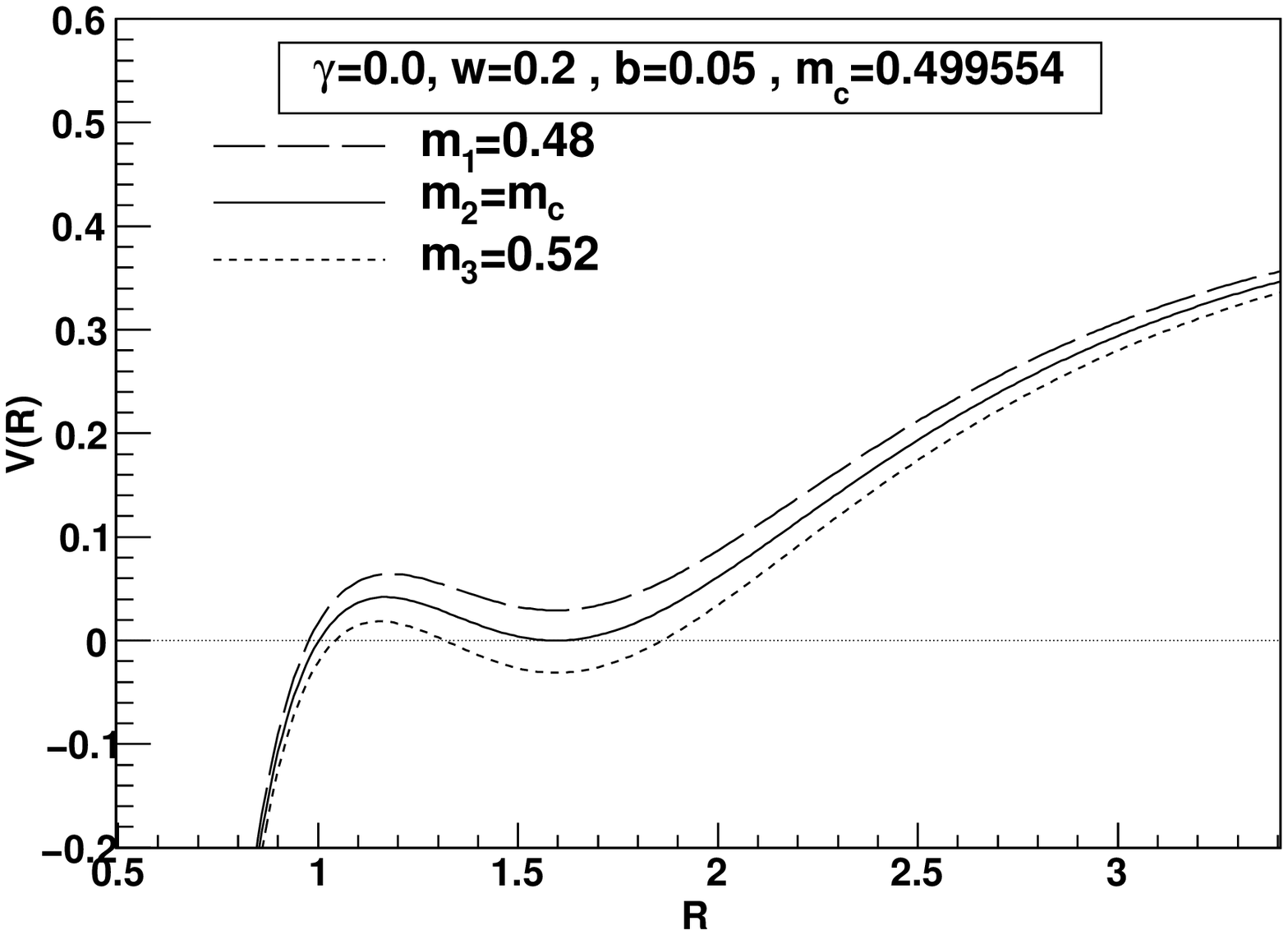,width=3.3truein,height=3.0truein}\hskip
.25in \psfig{figure=ECw0v2b0v05.eps,width=3.3truein,height=3.0truein}
\hskip .5in} \caption{The potential $V(R)$ and the energy conditions EC1$\equiv \rho+p_r+2p_t$, 
EC2$\equiv \rho+p_r$ and EC3$\equiv \rho+p_t$, for $\gamma=0$,
$\omega=0.2$, $b=0.05$ and $m_c=0.499554$. {\bf Case A}}
\label{fig19}
\end{figure}

\begin{figure}
\vspace{.2in}
\centerline{\psfig{figure=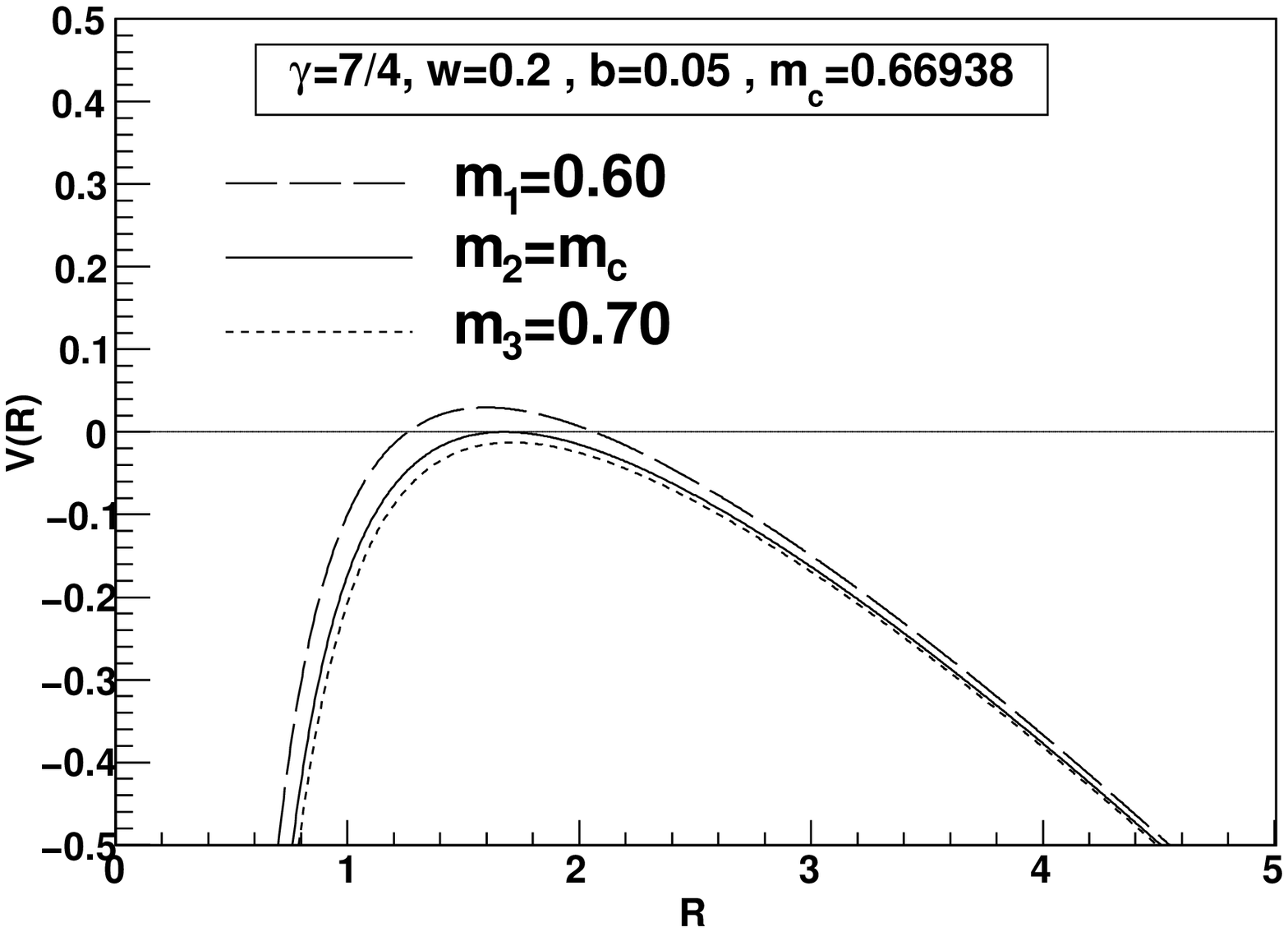,width=3.3truein,height=3.0truein}\hskip
.25in \psfig{figure=ECw0v2b0v05.eps,width=3.3truein,height=3.0truein}
\hskip .5in} \caption{The potential $V(R)$ and the energy conditions EC1$\equiv \rho+p_r+2p_t$, 
EC2$\equiv \rho+p_r$ and EC3$\equiv \rho+p_t$, for $\gamma=7/4$,
$\omega=0.2$, $b=0.05$ and $m_c=0.66938$. {\bf Case B}}
\label{fig20}
\end{figure}

\begin{figure}
\vspace{.2in}
\centerline{\psfig{figure=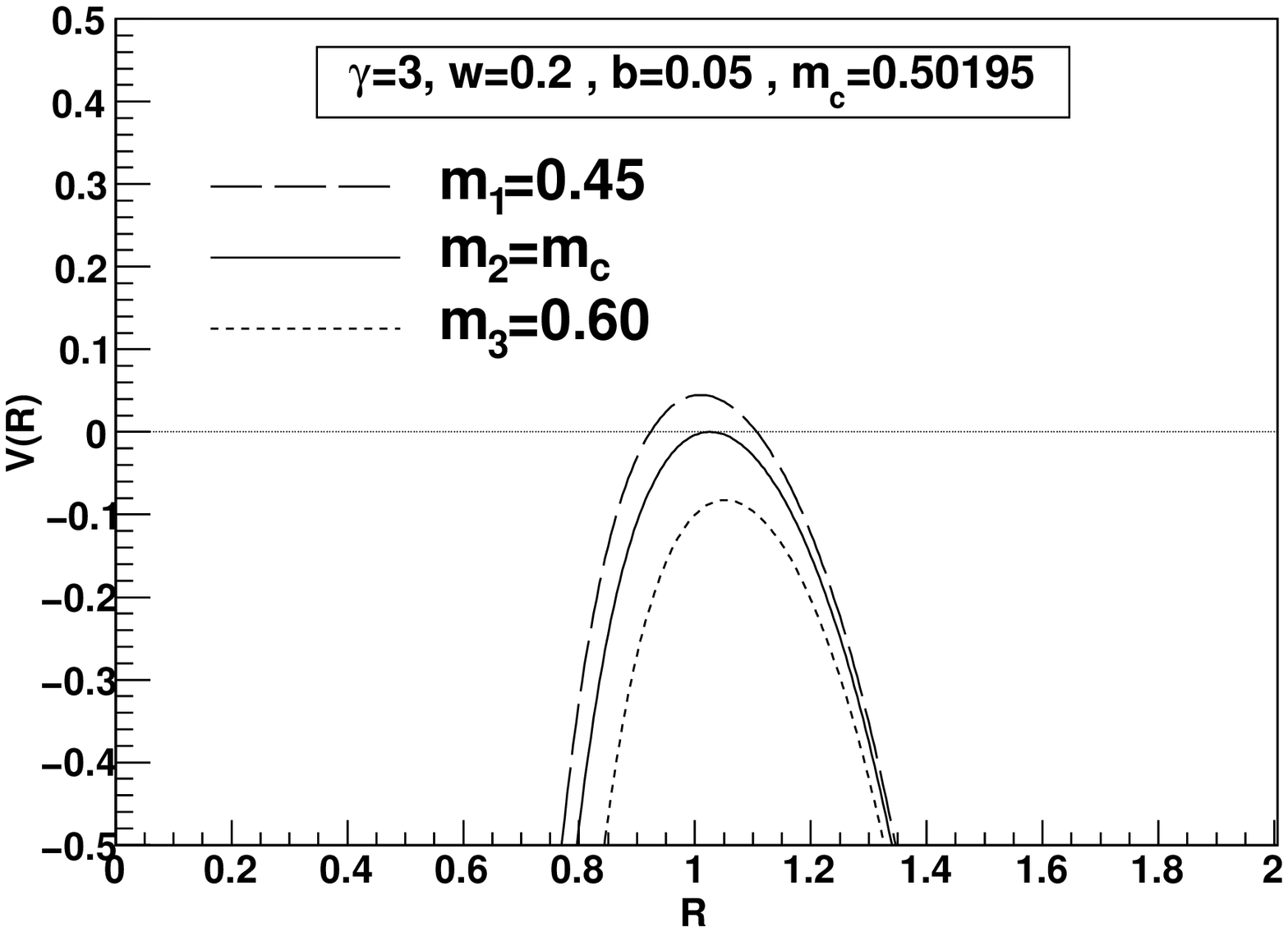,width=3.3truein,height=3.0truein}\hskip
.25in \psfig{figure=ECw0v2b0v05.eps,width=3.3truein,height=3.0truein}
\hskip .5in} \caption{The potential $V(R)$ and the energy conditions EC1$\equiv \rho+p_r+2p_t$, 
EC2$\equiv \rho+p_r$ and EC3$\equiv \rho+p_t$, for $\gamma=3$,
$\omega=0.2$, $b=0.05$ and $m_c=0.50195$. {\bf Case C}}
\label{fig21}
\end{figure}

\begin{figure}
\vspace{.2in}
\centerline{\psfig{figure=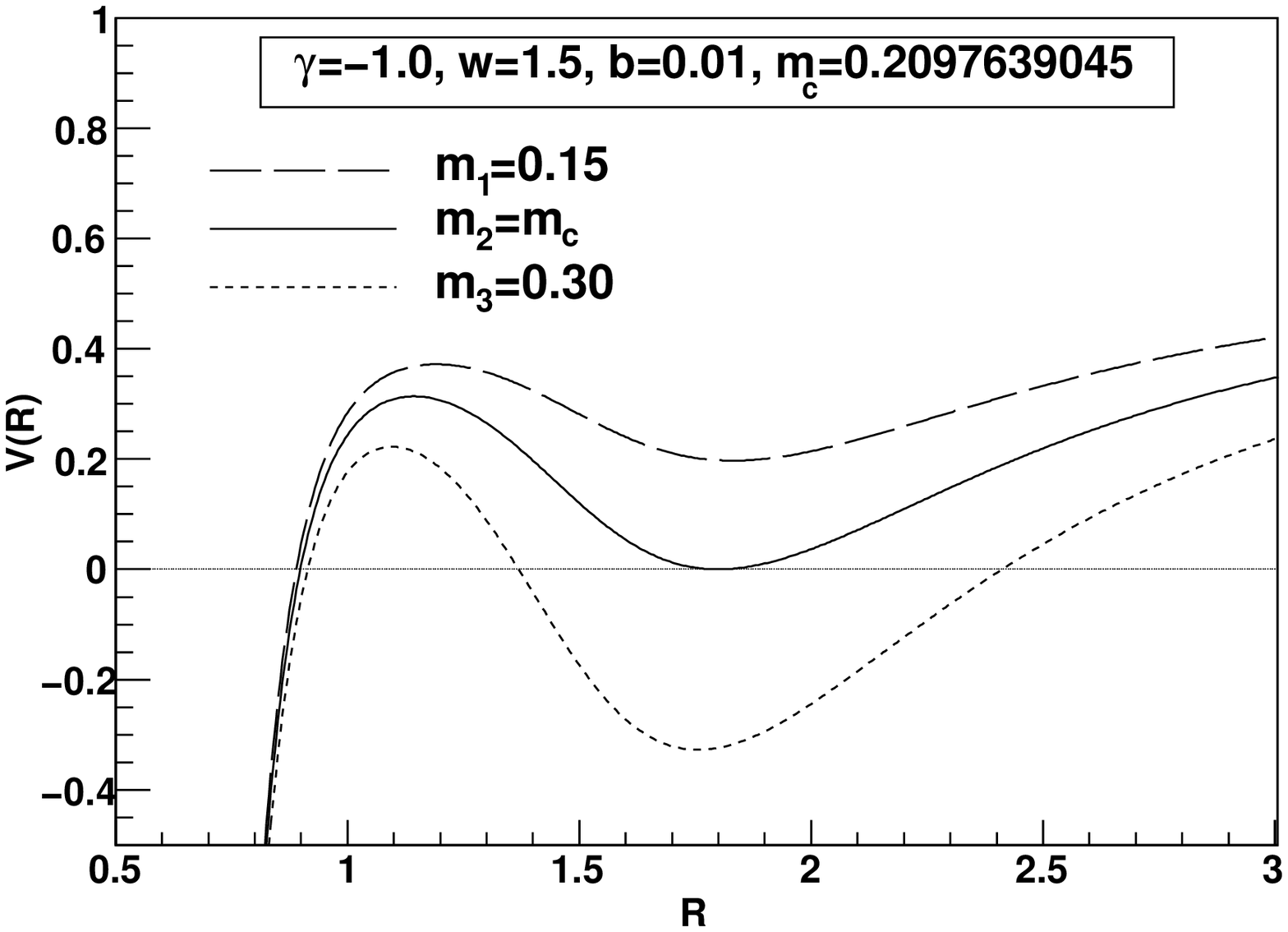,width=3.3truein,height=3.0truein}\hskip
.25in \psfig{figure=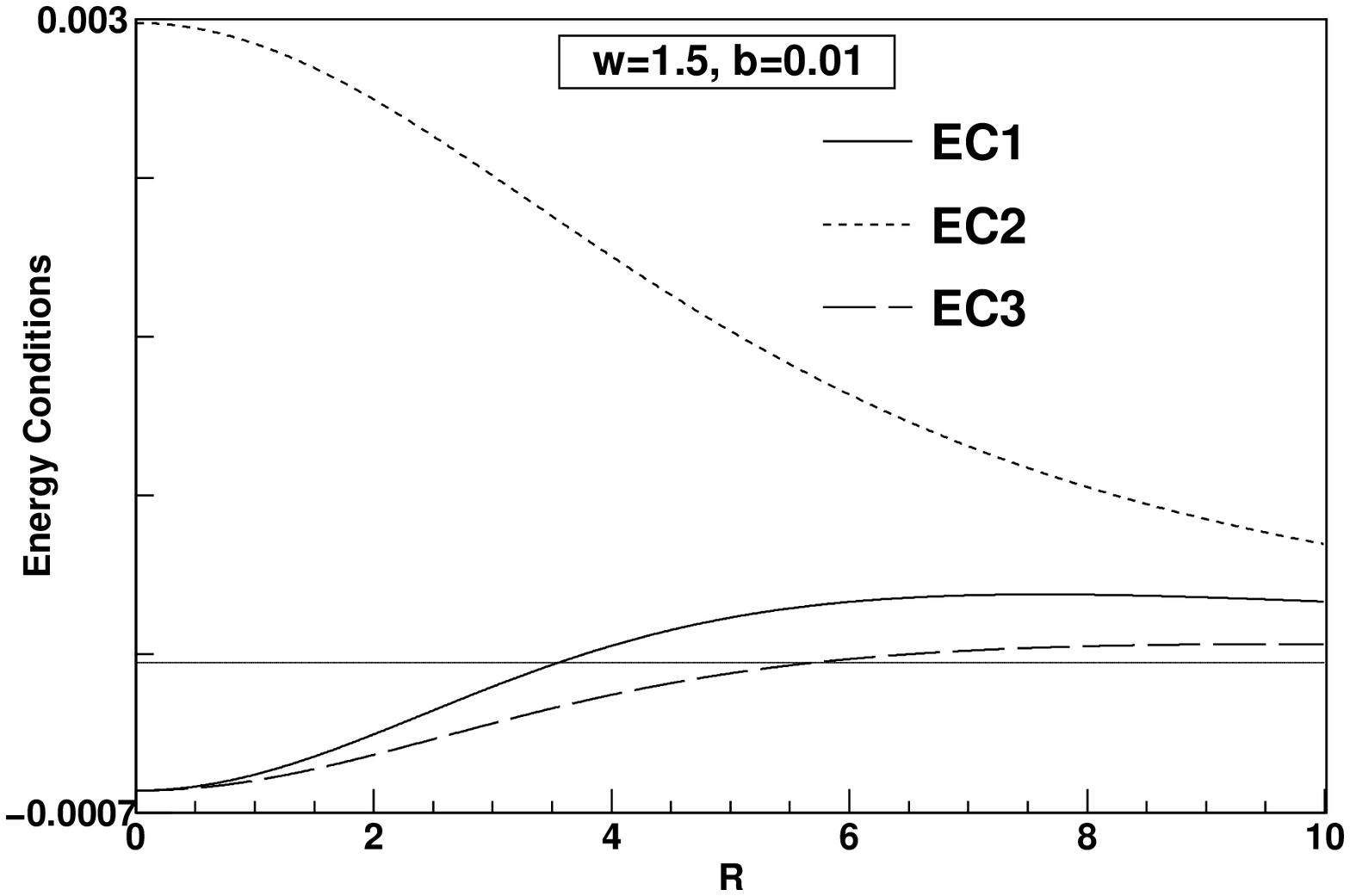,width=3.3truein,height=3.0truein}
\hskip .5in} \caption{The potential $V(R)$ and the energy conditions EC1$\equiv \rho+p_r+2p_t$, 
EC2$\equiv \rho+p_r$ and EC3$\equiv \rho+p_t$, for $\gamma=-1$,
$\omega=1.5$, $b=0.01$ and $m_c=0.2097639045$. {\bf Case G}}
\label{fig22}
\end{figure}

\begin{figure}
\vspace{.2in}
\centerline{\psfig{figure=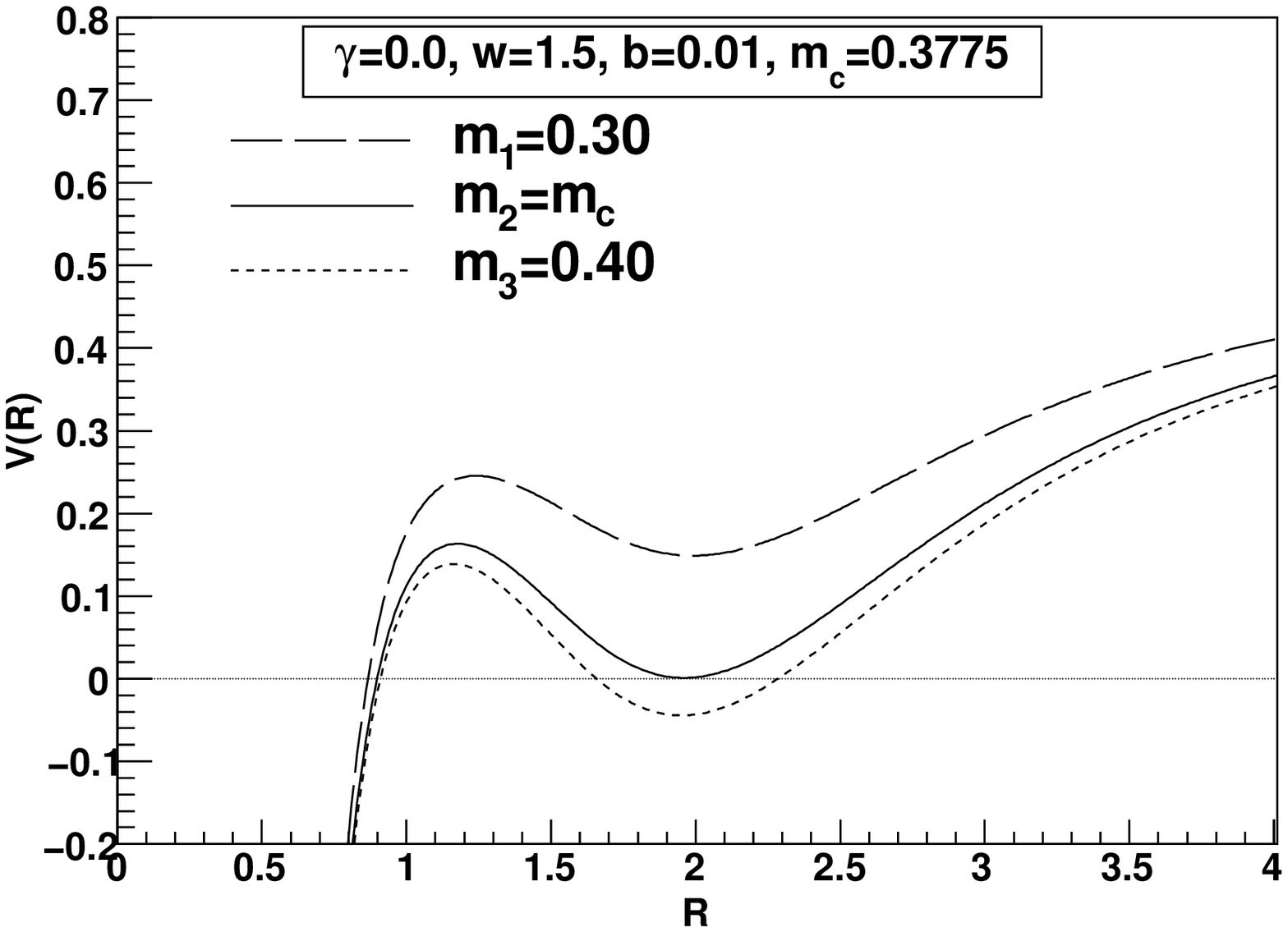,width=3.3truein,height=3.0truein}\hskip
.25in \psfig{figure=ECw1v5b0v01.eps,width=3.3truein,height=3.0truein}
\hskip .5in} \caption{The potential $V(R)$ and the energy conditions EC1$\equiv \rho+p_r+2p_t$, 
EC2$\equiv \rho+p_r$ and EC3$\equiv \rho+p_t$, for $\gamma=0$,
$\omega=1.5$, $b=0.01$ and $m_c=0.3775$. {\bf Case G}}
\label{fig23}
\end{figure}

\begin{figure}
\vspace{.2in}
\centerline{\psfig{figure=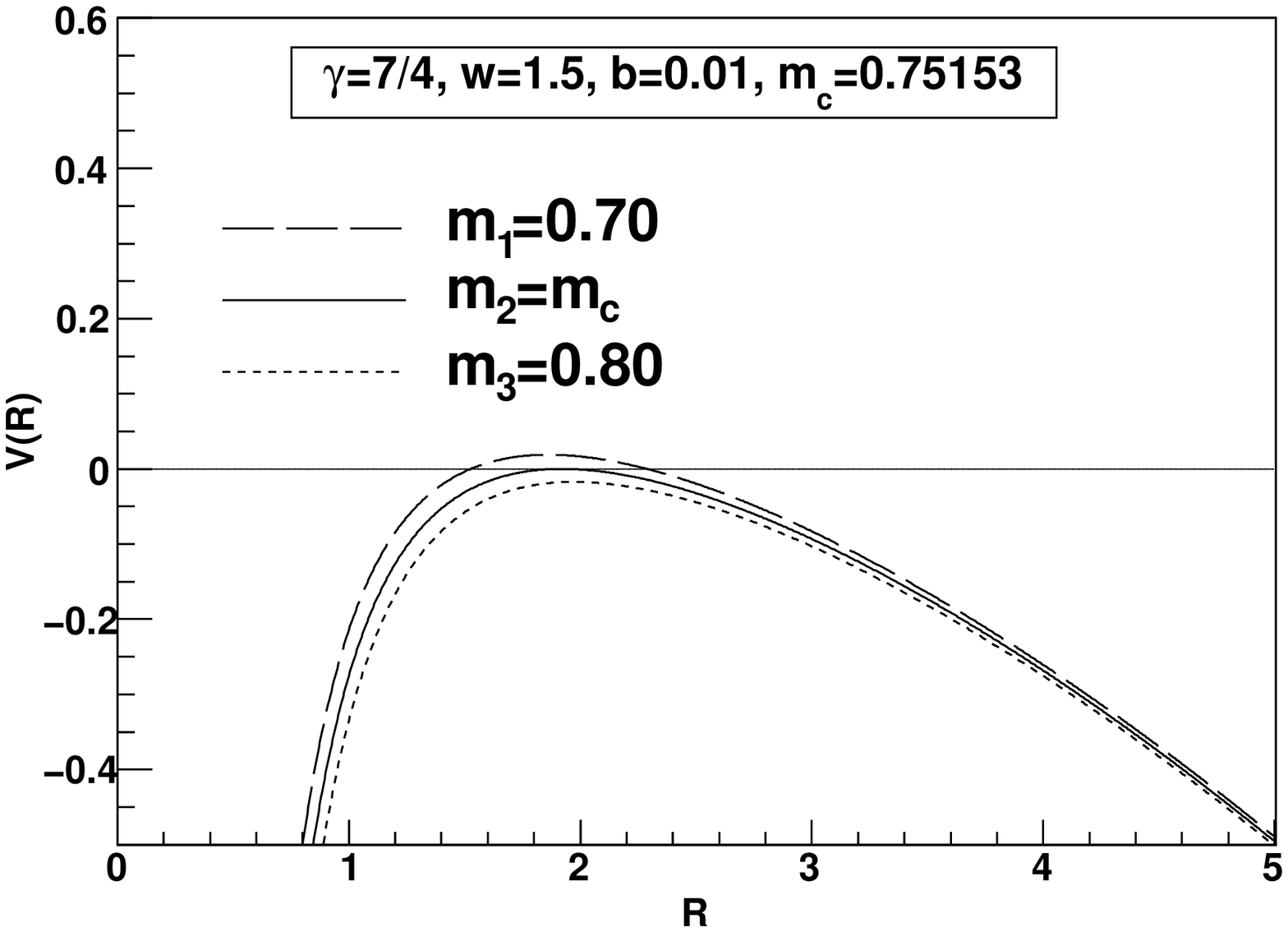,width=3.3truein,height=3.0truein}\hskip
.25in \psfig{figure=ECw1v5b0v01.eps,width=3.3truein,height=3.0truein}
\hskip .5in} \caption{The potential $V(R)$ and the energy conditions EC1$\equiv \rho+p_r+2p_t$, 
EC2$\equiv \rho+p_r$ and EC3$\equiv \rho+p_t$, for $\gamma=7/4$,
$\omega=1.5$, $b=0.01$ and $m_c=0.75153$. {\bf Case H}}
\label{fig24}
\end{figure}

\begin{figure}
\vspace{.2in}
\centerline{\psfig{figure=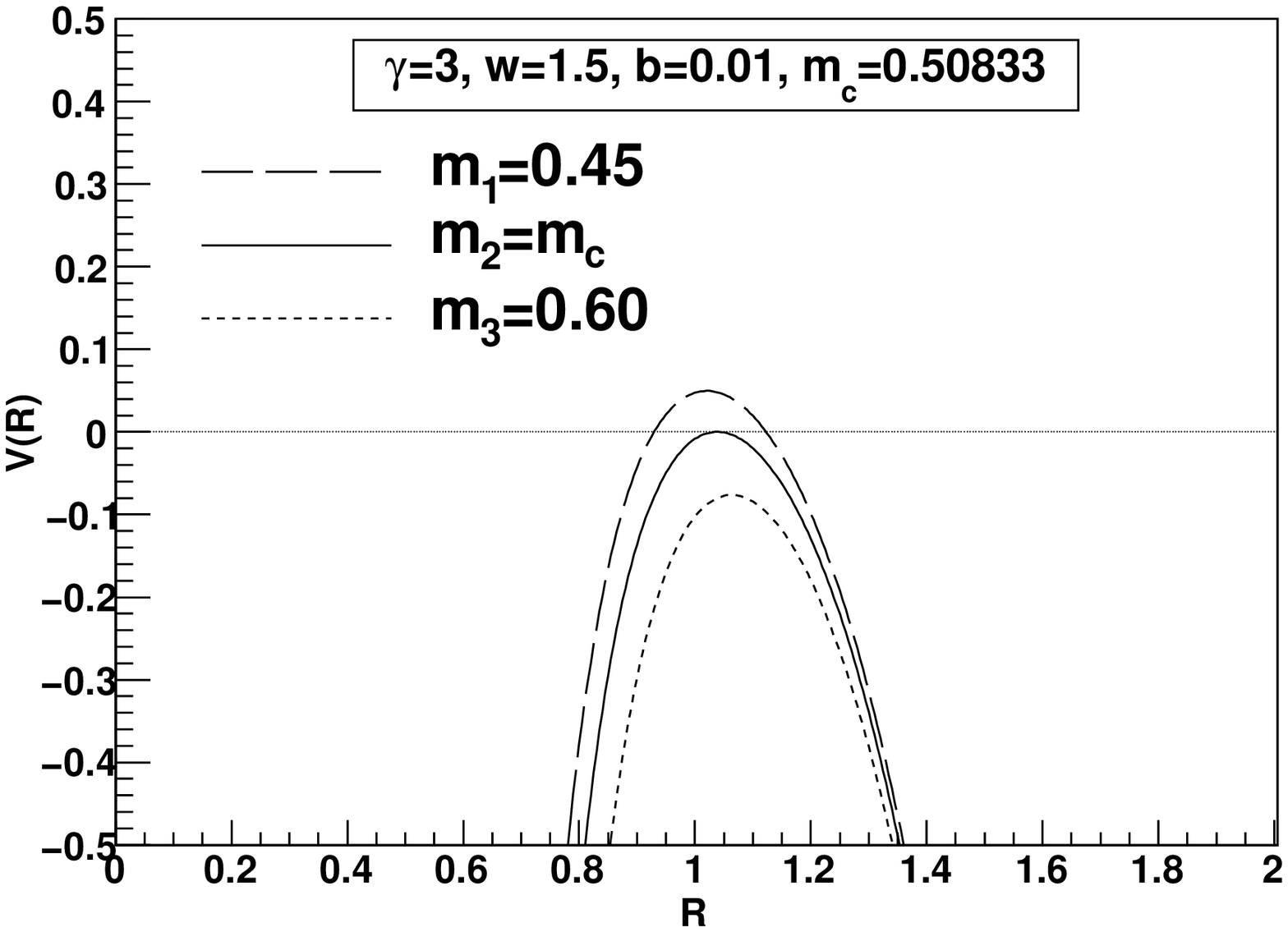,width=3.3truein,height=3.0truein}\hskip
.25in \psfig{figure=ECw1v5b0v01.eps,width=3.3truein,height=3.0truein}
\hskip .5in} \caption{The potential $V(R)$ and the energy conditions EC1$\equiv \rho+p_r+2p_t$, 
EC2$\equiv \rho+p_r$ and EC3$\equiv \rho+p_t$, for $\gamma=3$,
$\omega=1.5$, $b=0.01$ and $m_c=0.50833$. {\bf Case I}}
\label{fig25}
\end{figure}

\begin{figure}
\vspace{.2in}
\centerline{\psfig{figure=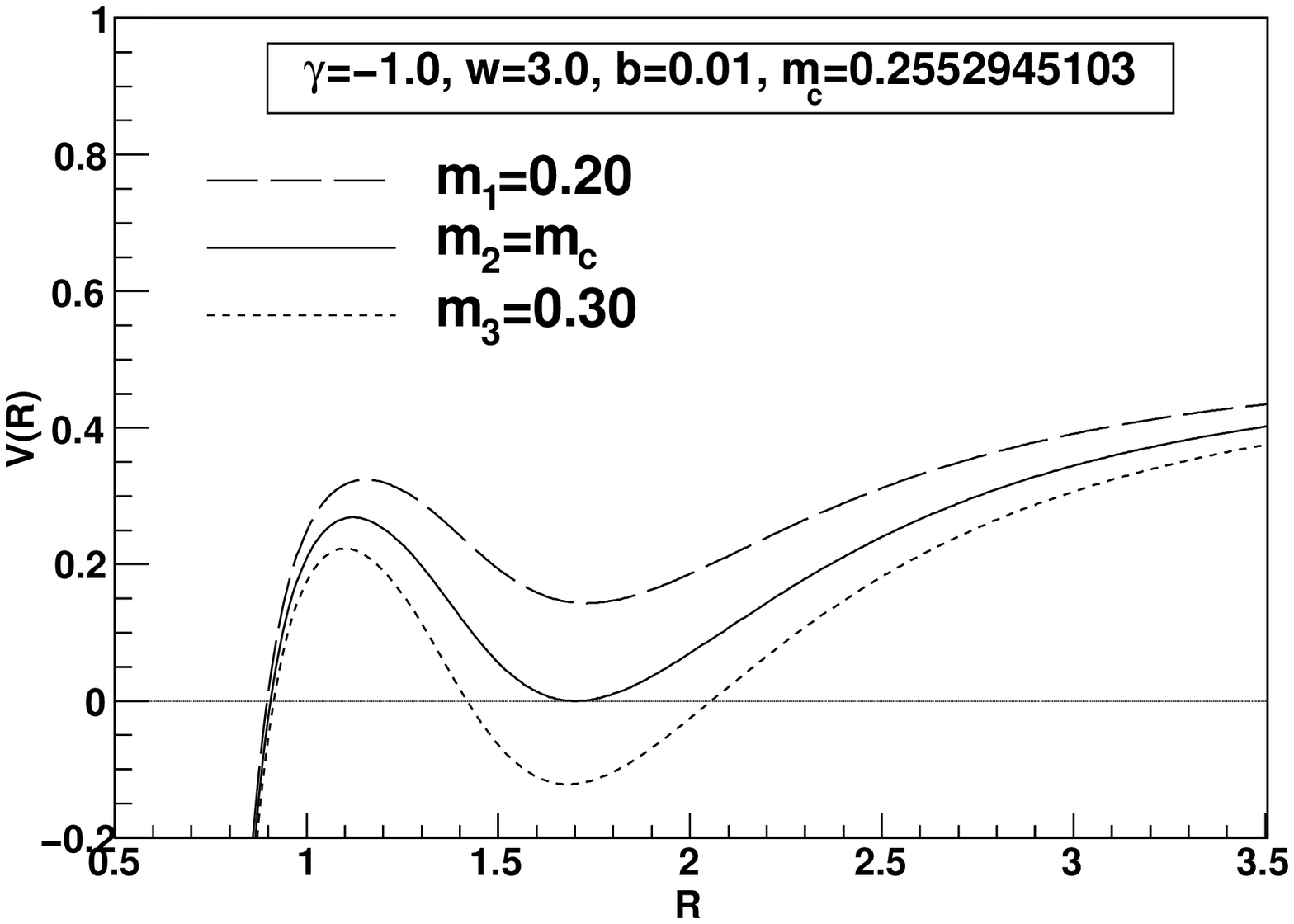,width=3.3truein,height=3.0truein}\hskip
.25in \psfig{figure=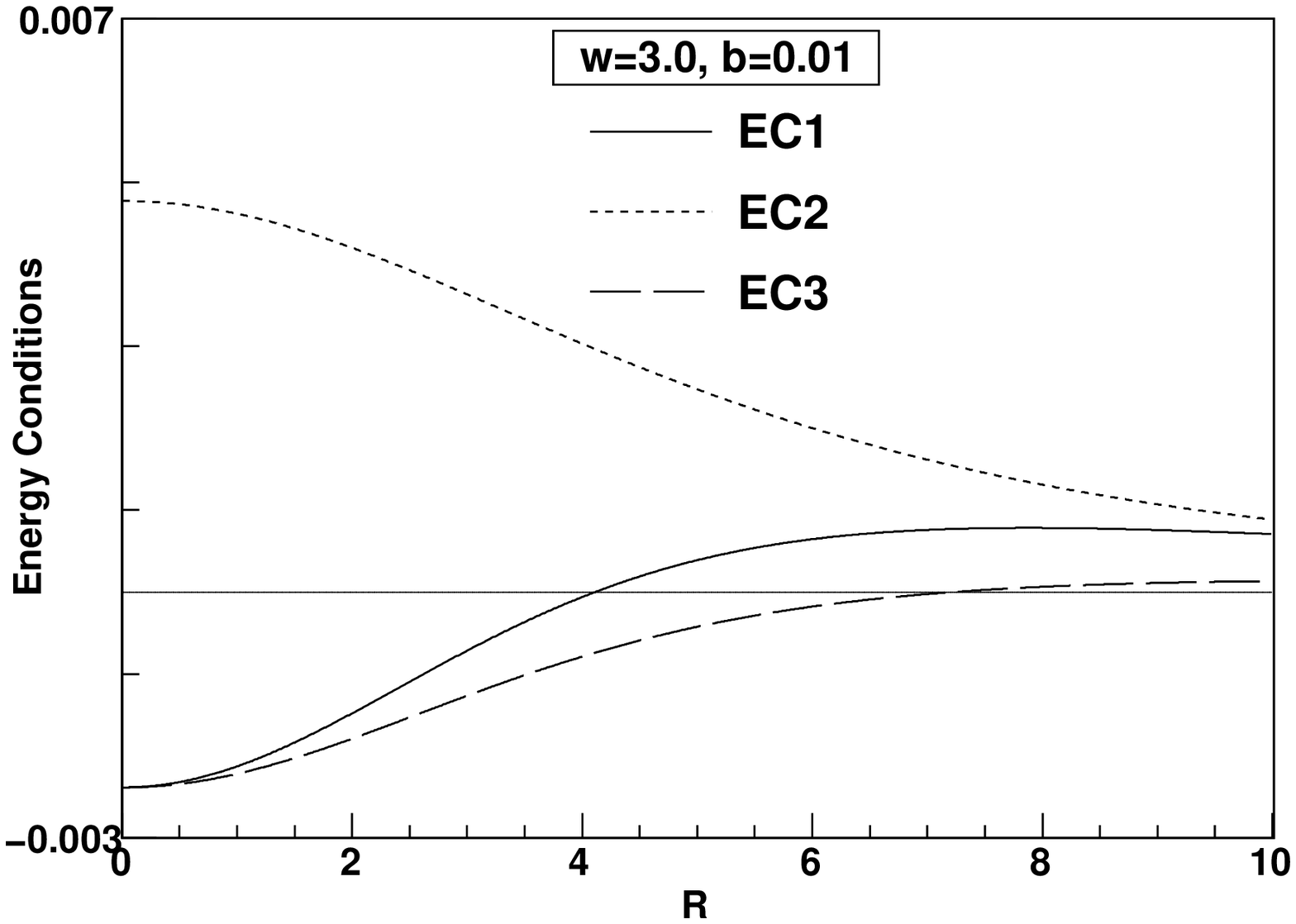,width=3.3truein,height=3.0truein}
\hskip .5in} \caption{The potential $V(R)$ and the energy conditions EC1$\equiv \rho+p_r+2p_t$, 
EC2$\equiv \rho+p_r$ and EC3$\equiv \rho+p_t$, for $\gamma=-1$,
$\omega=3$, $b=0.01$ and $m_c=0.2552945103$. {\bf Case G}}
\label{fig26}
\end{figure}

\begin{figure}
\vspace{.2in}
\centerline{\psfig{figure=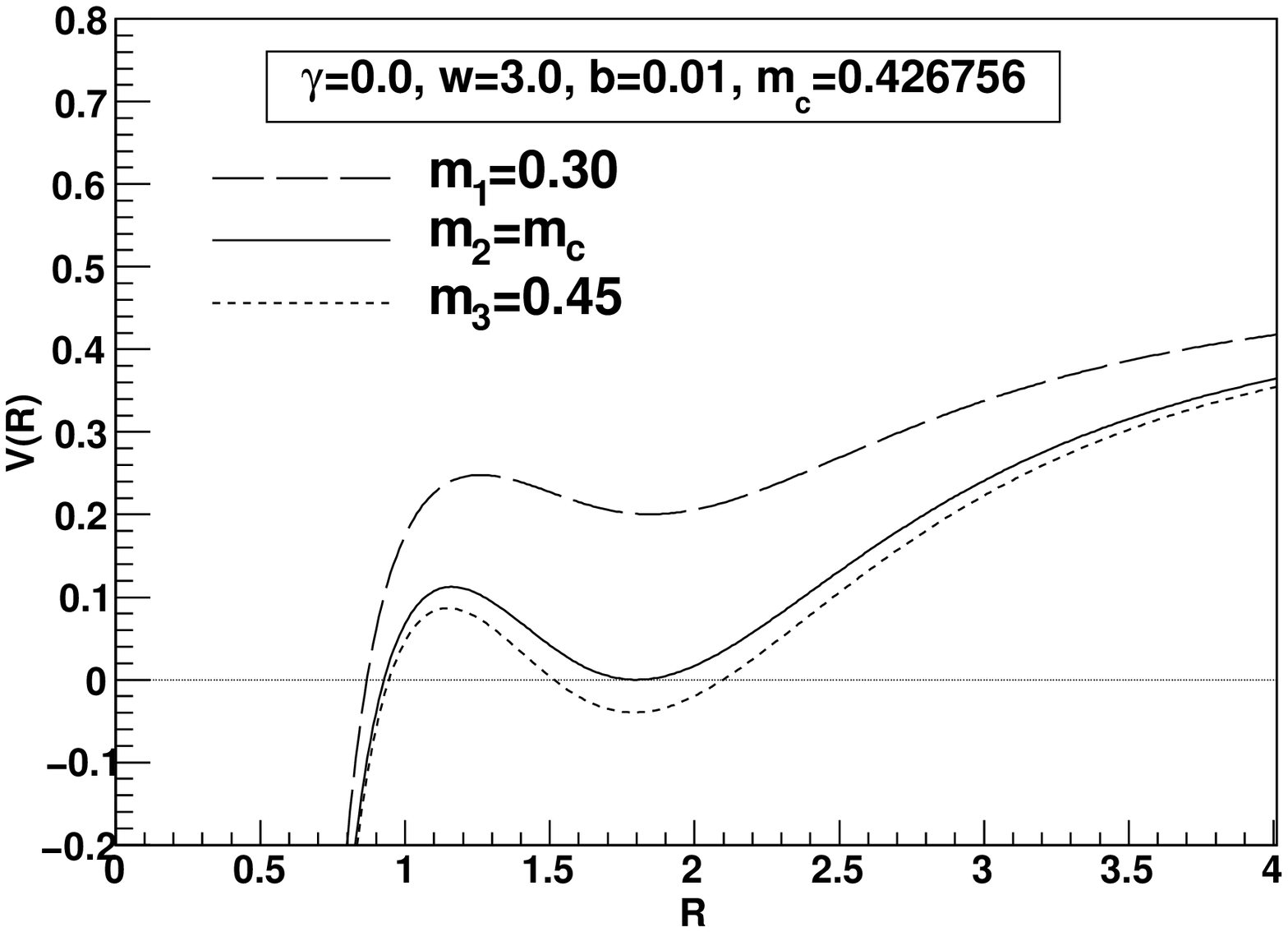,width=3.3truein,height=3.0truein}\hskip
.25in \psfig{figure=ECw3v0b0v01.eps,width=3.3truein,height=3.0truein}
\hskip .5in} \caption{The potential $V(R)$ and the energy conditions EC1$\equiv \rho+p_r+2p_t$, 
EC2$\equiv \rho+p_r$ and EC3$\equiv \rho+p_t$, for $\gamma=0$,
$\omega=3$, $b=0.01$ and $m_c=0.426756$. {\bf Case G}}
\label{fig27}
\end{figure}

\begin{figure}
\vspace{.2in}
\centerline{\psfig{figure=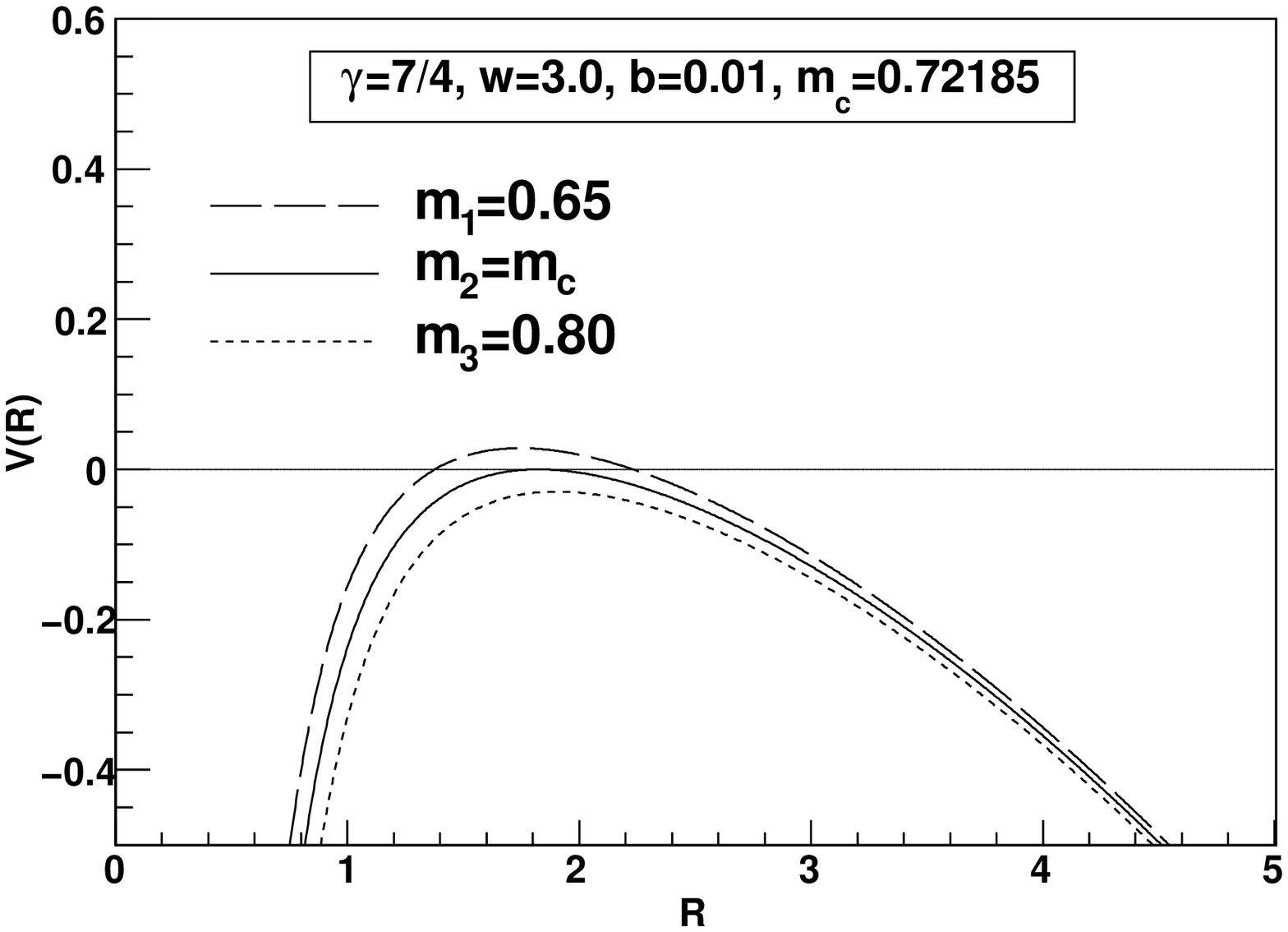,width=3.3truein,height=3.0truein}\hskip
.25in \psfig{figure=ECw3v0b0v01.eps,width=3.3truein,height=3.0truein}
\hskip .5in} \caption{The potential $V(R)$ and the energy conditions EC1$\equiv \rho+p_r+2p_t$, 
EC2$\equiv \rho+p_r$ and EC3$\equiv \rho+p_t$, for $\gamma=7/4$,
$\omega=3$, $b=0.01$ and $m_c=0.72185$. {\bf Case H}}
\label{fig28}
\end{figure}

\begin{figure}
\vspace{.2in}
\centerline{\psfig{figure=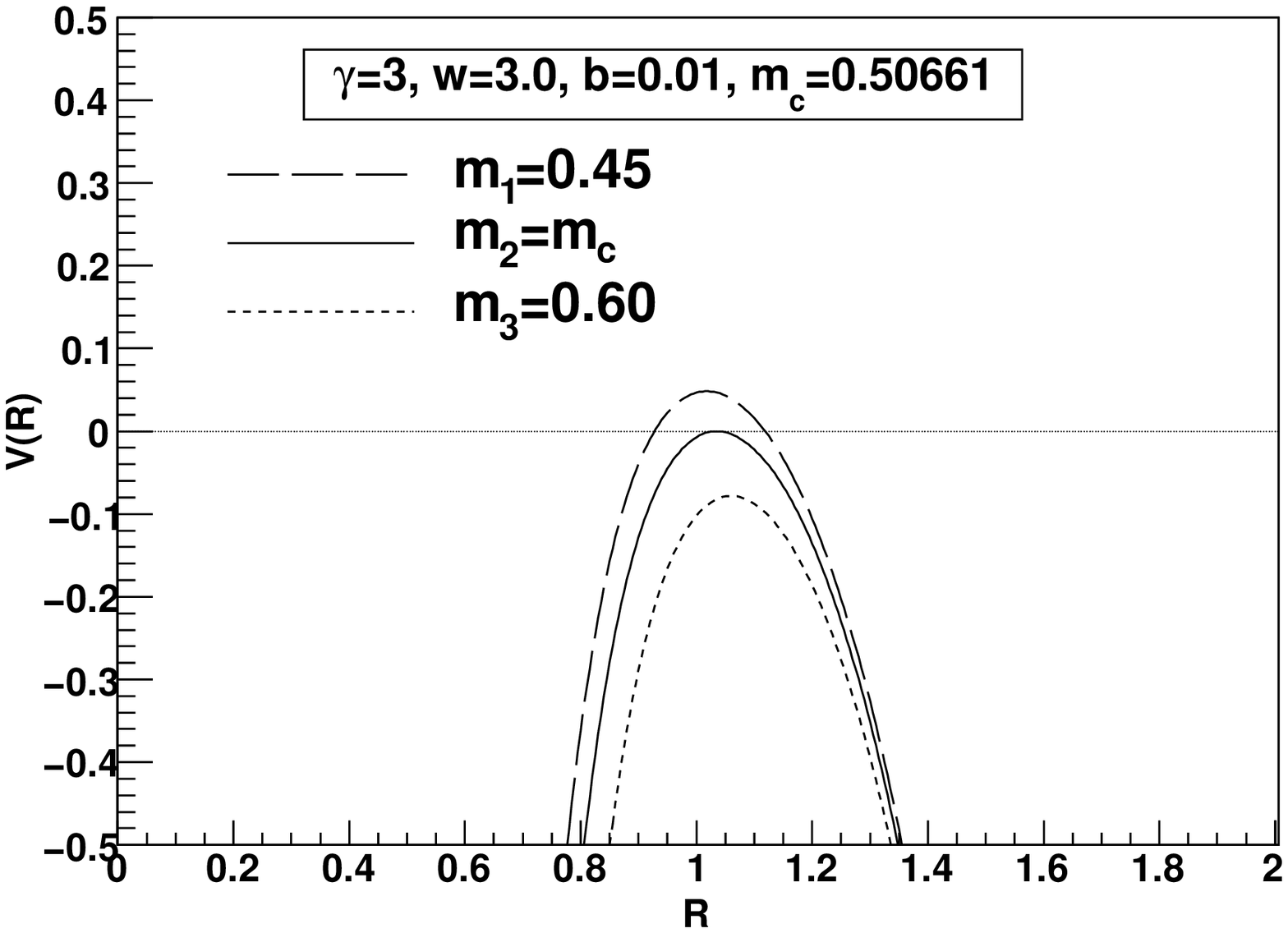,width=3.3truein,height=3.0truein}\hskip
.25in \psfig{figure=ECw3v0b0v01.eps,width=3.3truein,height=3.0truein}
\hskip .5in} \caption{The potential $V(R)$ and the energy conditions EC1$\equiv \rho+p_r+2p_t$, 
EC2$\equiv \rho+p_r$ and EC3$\equiv \rho+p_t$, for $\gamma=3$,
$\omega=3$, $b=0.01$ and $m_c=0.50661$. {\bf Case I}}
\label{fig29}
\end{figure}

\begin{figure}
\vspace{.2in}
\centerline{\psfig{figure=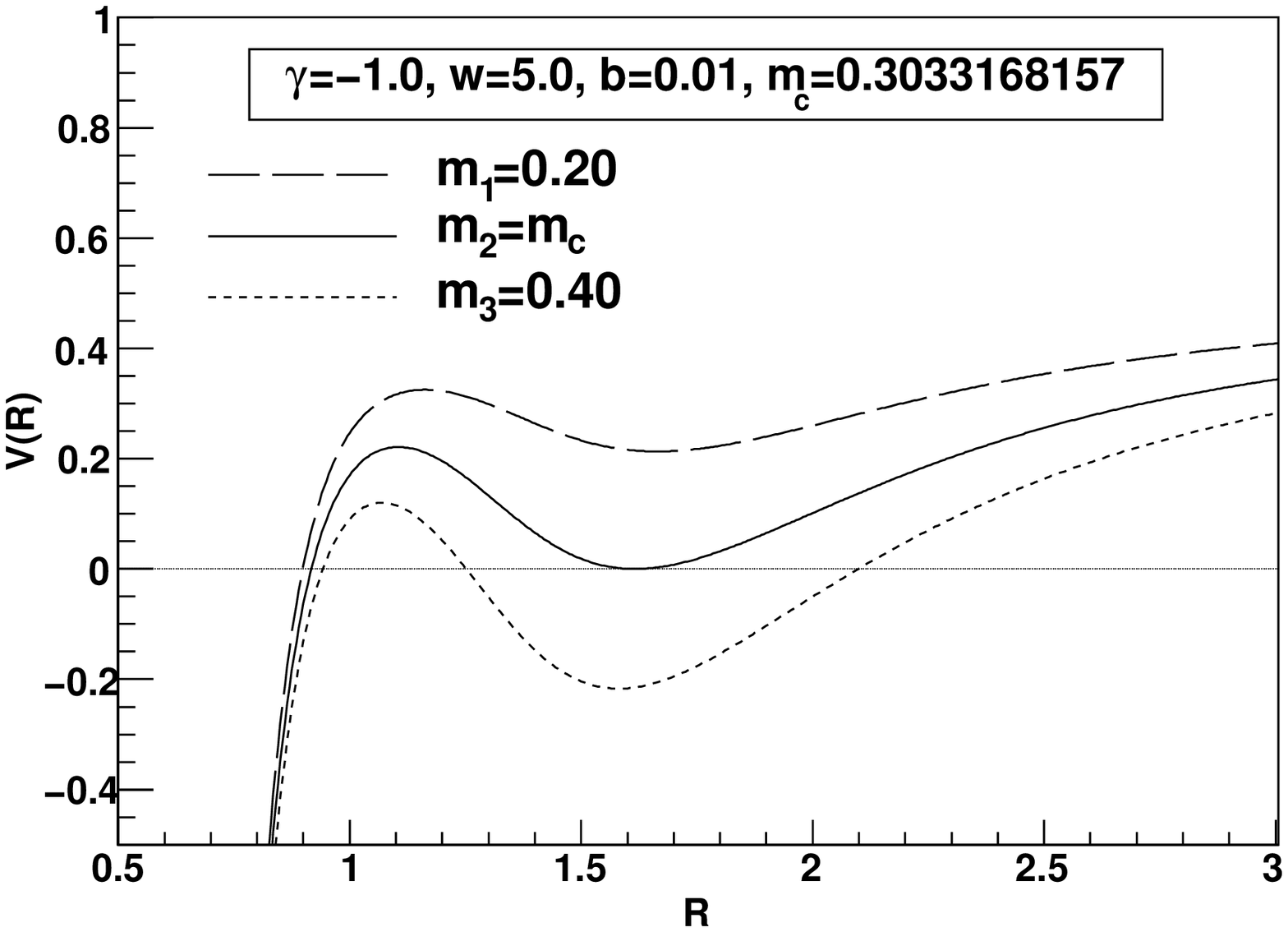,width=3.3truein,height=3.0truein}\hskip
.25in \psfig{figure=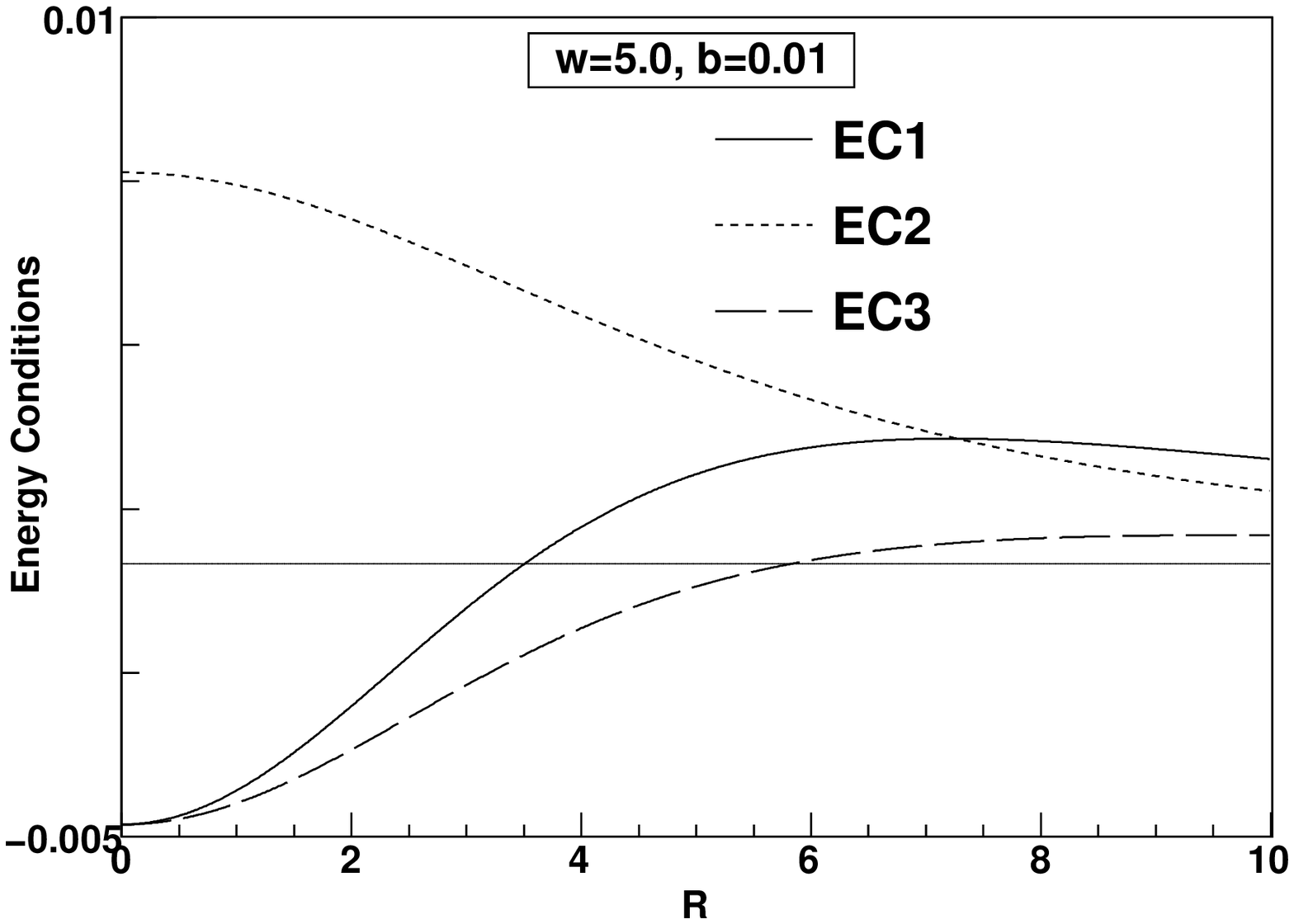,width=3.3truein,height=3.0truein}
\hskip .5in} \caption{The potential $V(R)$ and the energy conditions EC1$\equiv \rho+p_r+2p_t$, 
EC2$\equiv \rho+p_r$ and EC3$\equiv \rho+p_t$, for $\gamma=-1$,
$\omega=5$, $b=0.01$ and $m_c=0.3033168157$. {\bf Case G}}
\label{fig30}
\end{figure}

\begin{figure}
\vspace{.2in}
\centerline{\psfig{figure=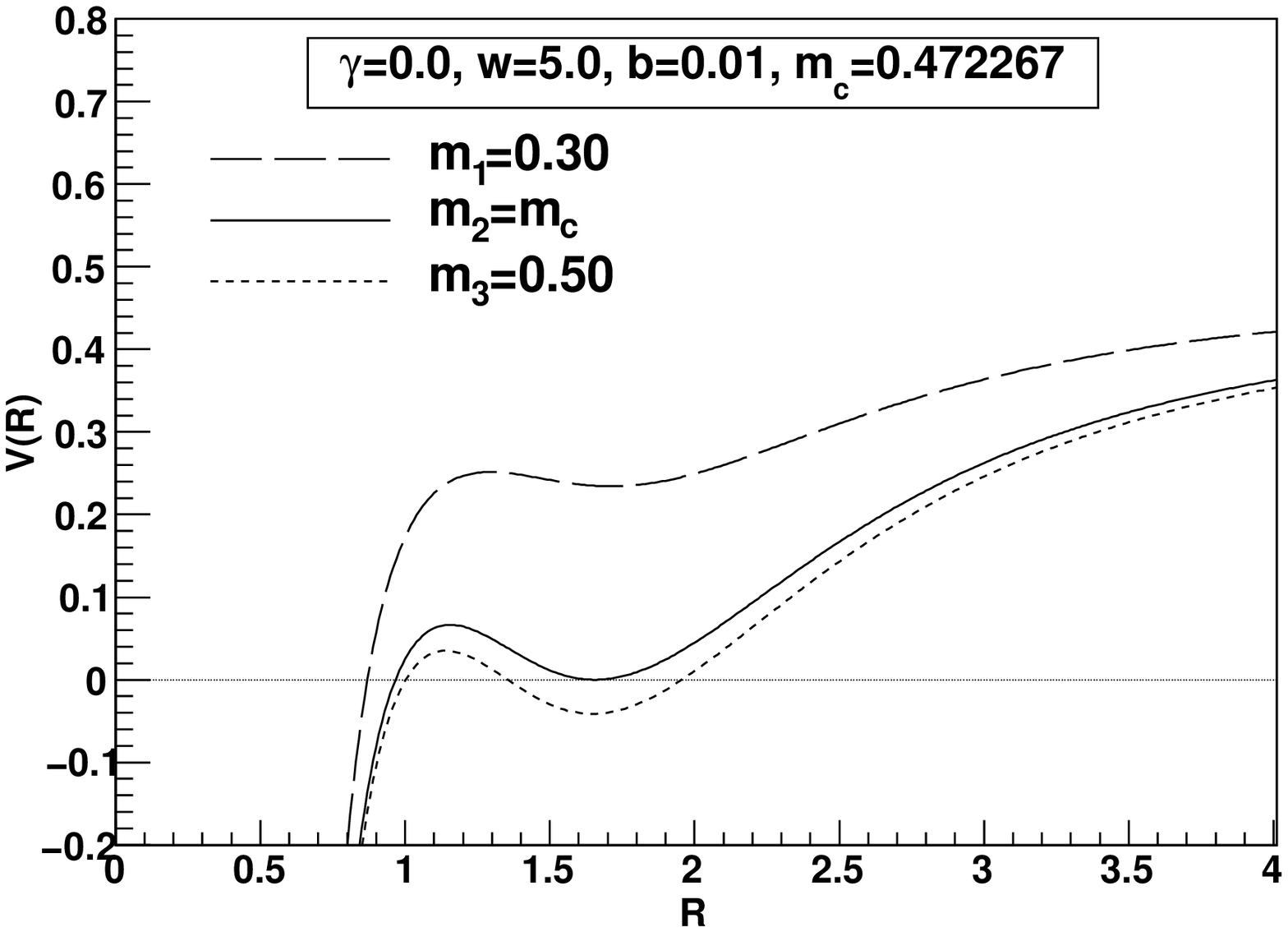,width=3.3truein,height=3.0truein}\hskip
.25in \psfig{figure=ECw5v0b0v01.eps,width=3.3truein,height=3.0truein}
\hskip .5in} \caption{The potential $V(R)$ and the energy conditions EC1$\equiv \rho+p_r+2p_t$, 
EC2$\equiv \rho+p_r$ and EC3$\equiv \rho+p_t$, for $\gamma=0$,
$\omega=5$, $b=0.01$ and $m_c=0.472267$. {\bf Case G}}
\label{fig31}
\end{figure}

\begin{figure}
\vspace{.2in}
\centerline{\psfig{figure=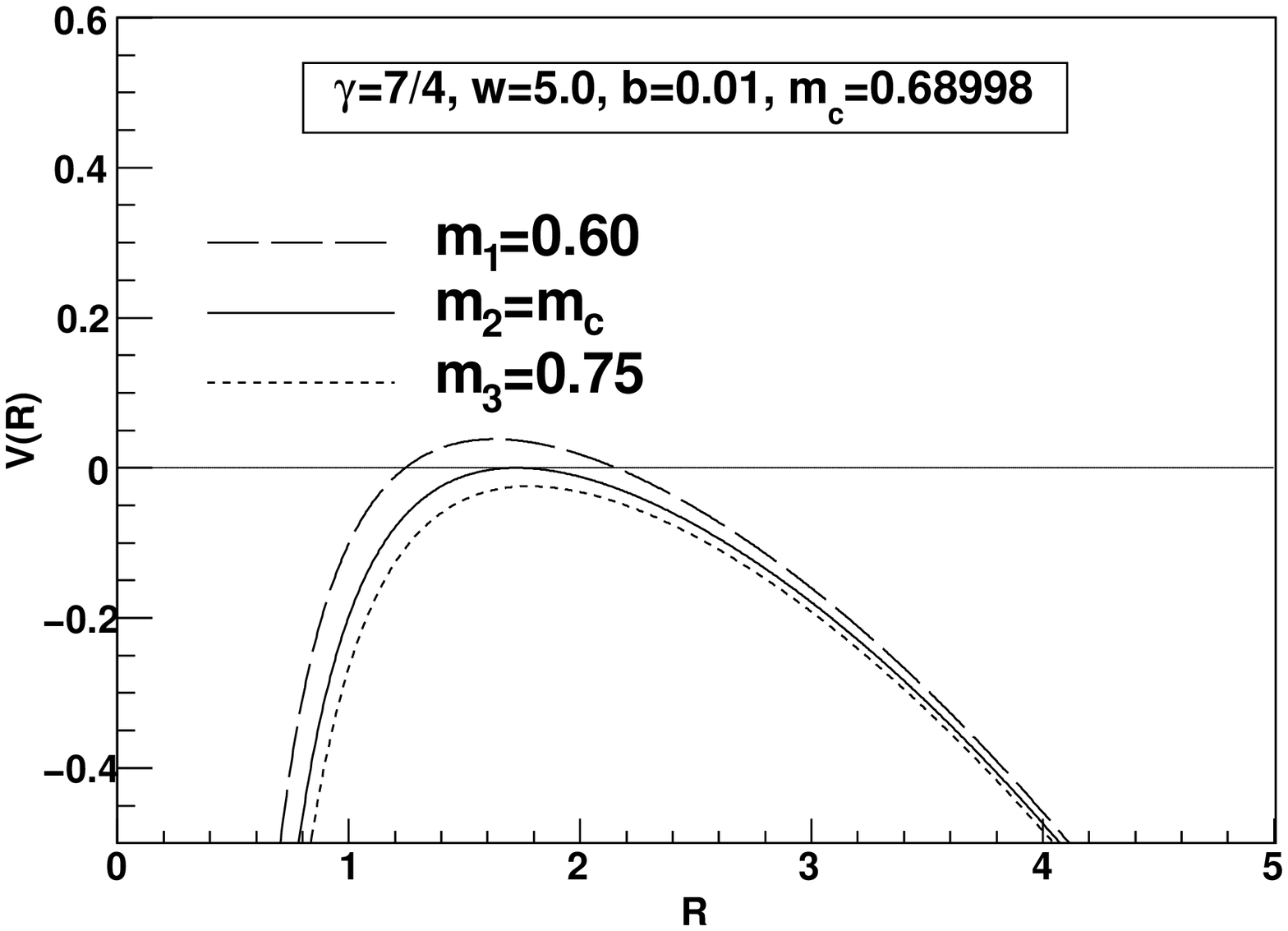,width=3.3truein,height=3.0truein}\hskip
.25in \psfig{figure=ECw5v0b0v01.eps,width=3.3truein,height=3.0truein}
\hskip .5in} \caption{The potential $V(R)$ and the energy conditions EC1$\equiv \rho+p_r+2p_t$, 
EC2$\equiv \rho+p_r$ and EC3$\equiv \rho+p_t$, for $\gamma=7/4$,
$\omega=5$, $b=0.01$ and $m_c=0.68998$. {\bf Case H}}
\label{fig32}
\end{figure}

\begin{figure}
\vspace{.2in}
\centerline{\psfig{figure=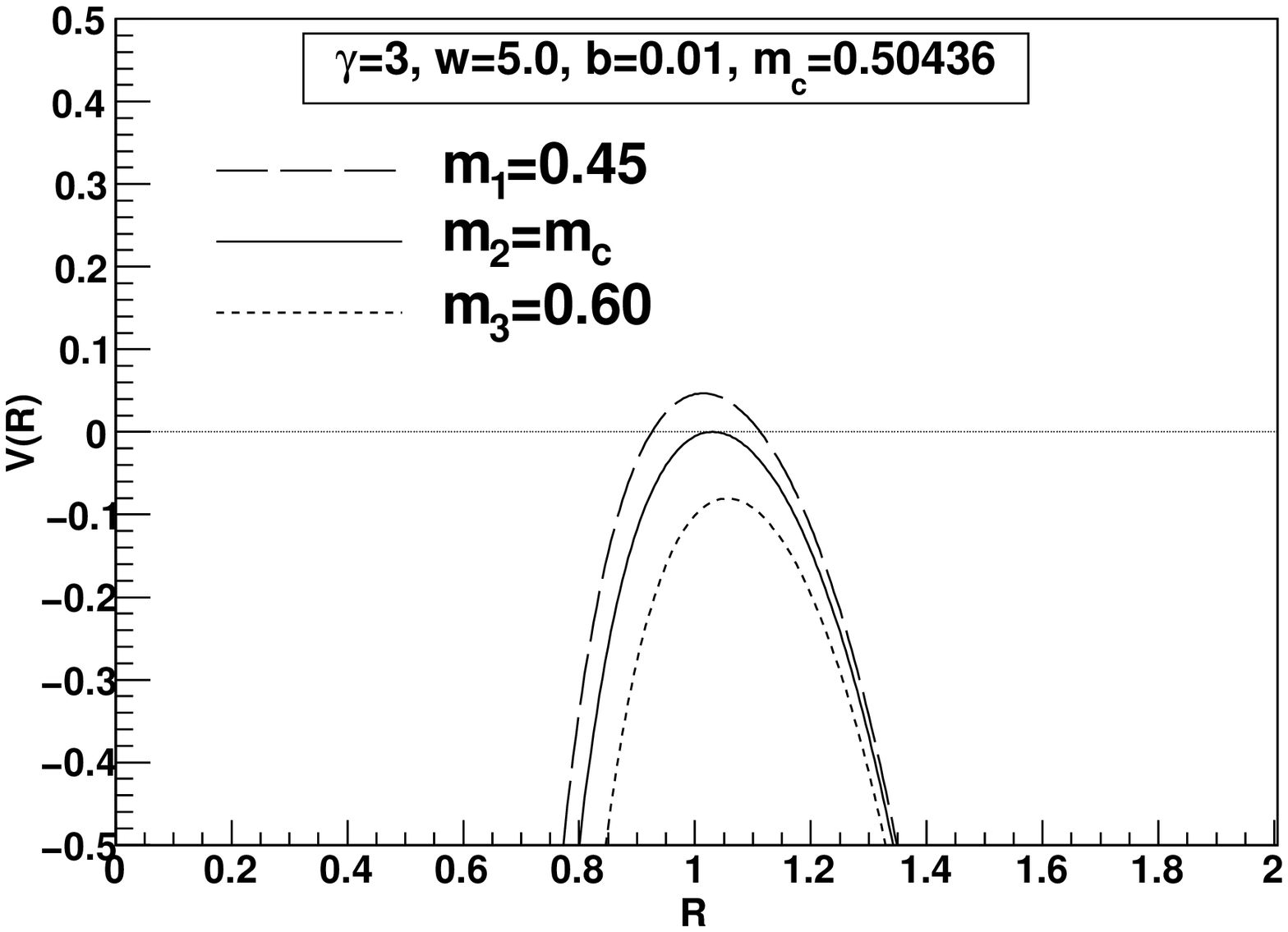,width=3.3truein,height=3.0truein}\hskip
.25in \psfig{figure=ECw5v0b0v01.eps,width=3.3truein,height=3.0truein}
\hskip .5in} \caption{The potential $V(R)$ and the energy conditions EC1$\equiv \rho+p_r+2p_t$, 
EC2$\equiv \rho+p_r$ and EC3$\equiv \rho+p_t$, for $\gamma=3$,
$\omega=5$, $b=0.01$ and $m_c=0.50436$. {\bf Case I}}
\label{fig33}
\end{figure}

\begin{figure}
\vspace{.2in}
\centerline{\psfig{figure=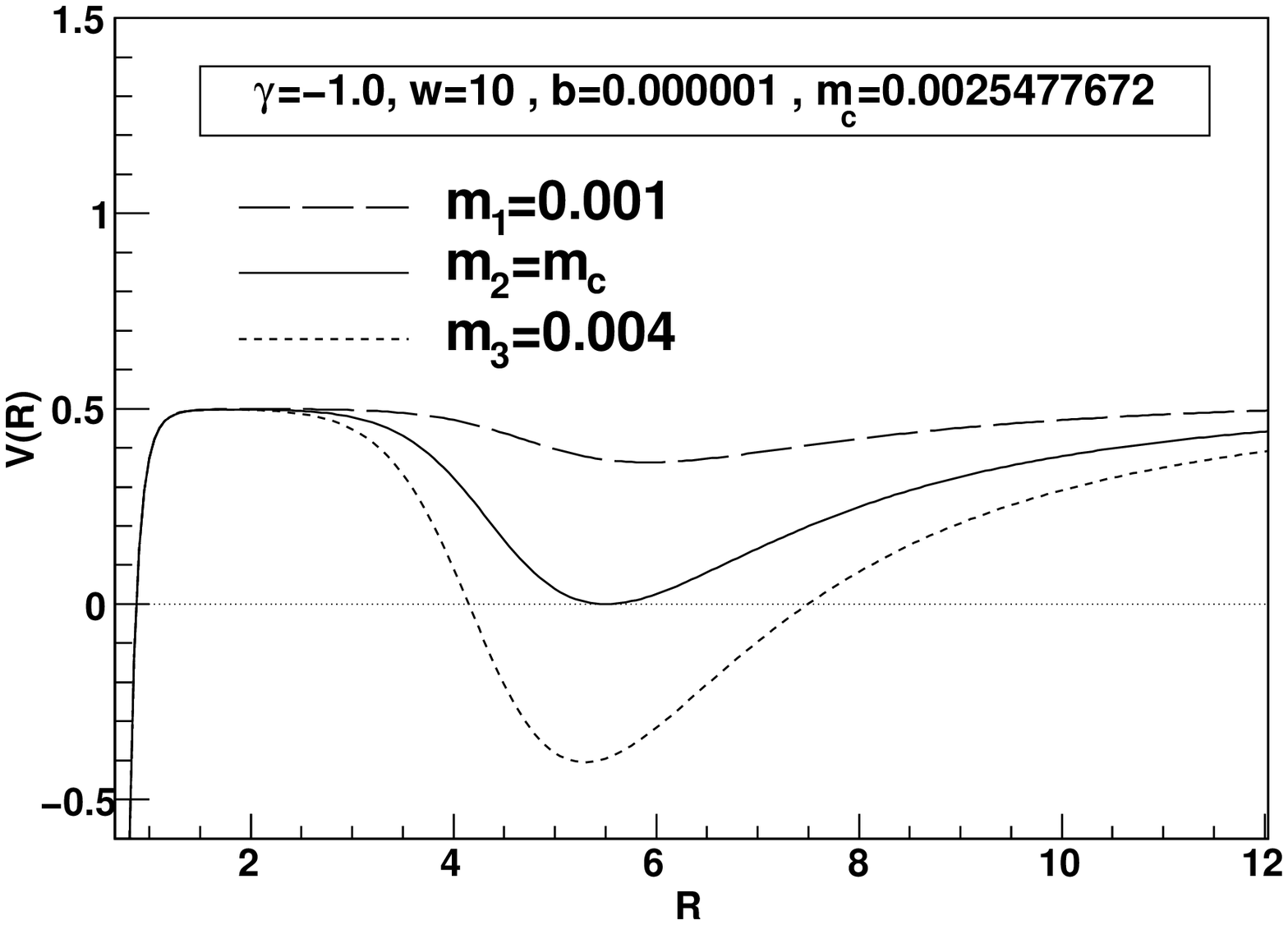,width=3.3truein,height=3.0truein}\hskip
.25in \psfig{figure=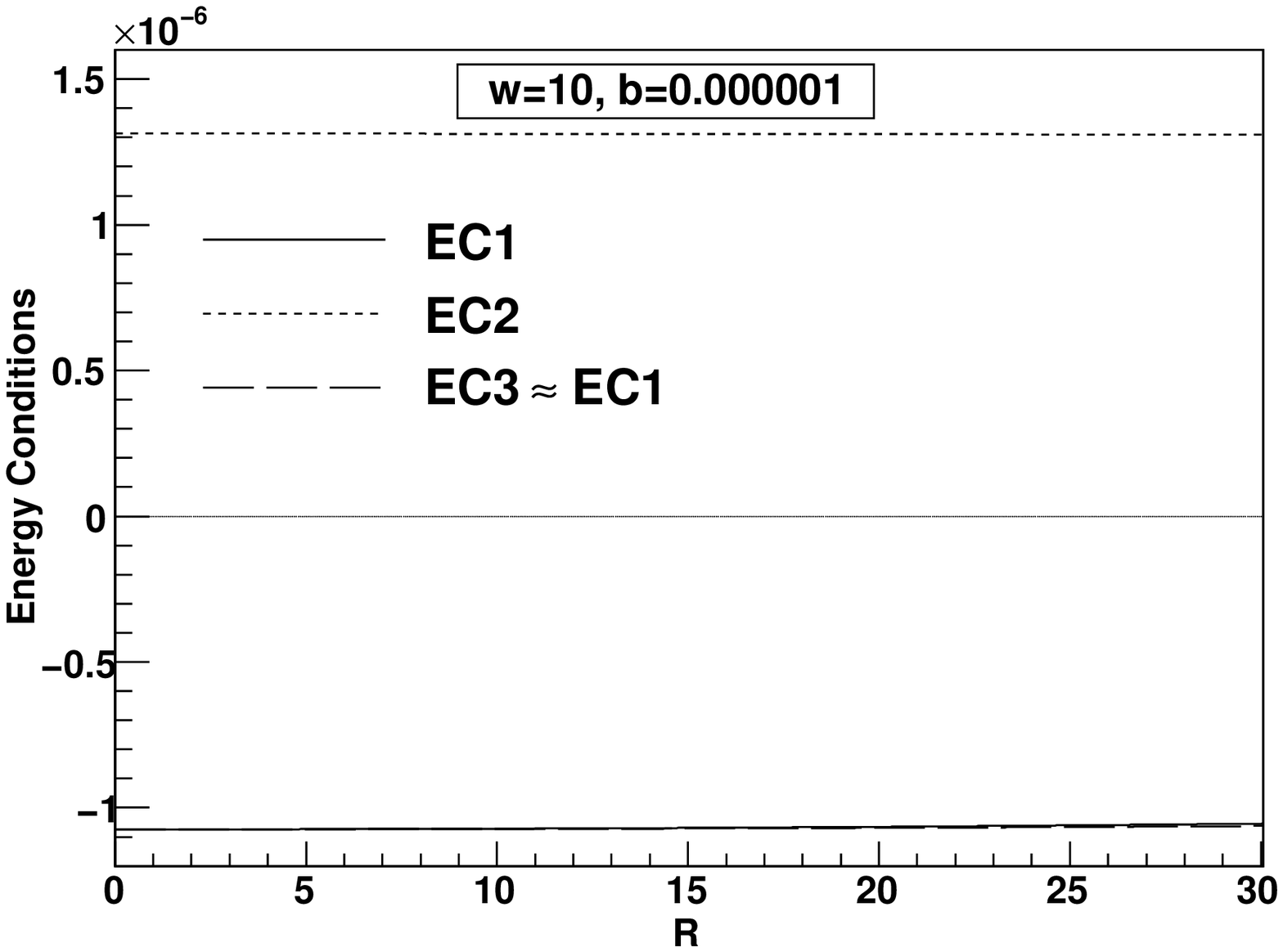,width=3.3truein,height=3.0truein}
\hskip .5in} \caption{The potential $V(R)$ and the energy conditions EC1$\equiv \rho+p_r+2p_t$, 
EC2$\equiv \rho+p_r$ and EC3$\equiv \rho+p_t$, for $\gamma=-1$,
$\omega=10$, $b=0.000001$ and $m_c=0.0025477672$. {\bf Case G}}
\label{fig34}
\end{figure}

\begin{figure}
\vspace{.2in}
\centerline{\psfig{figure=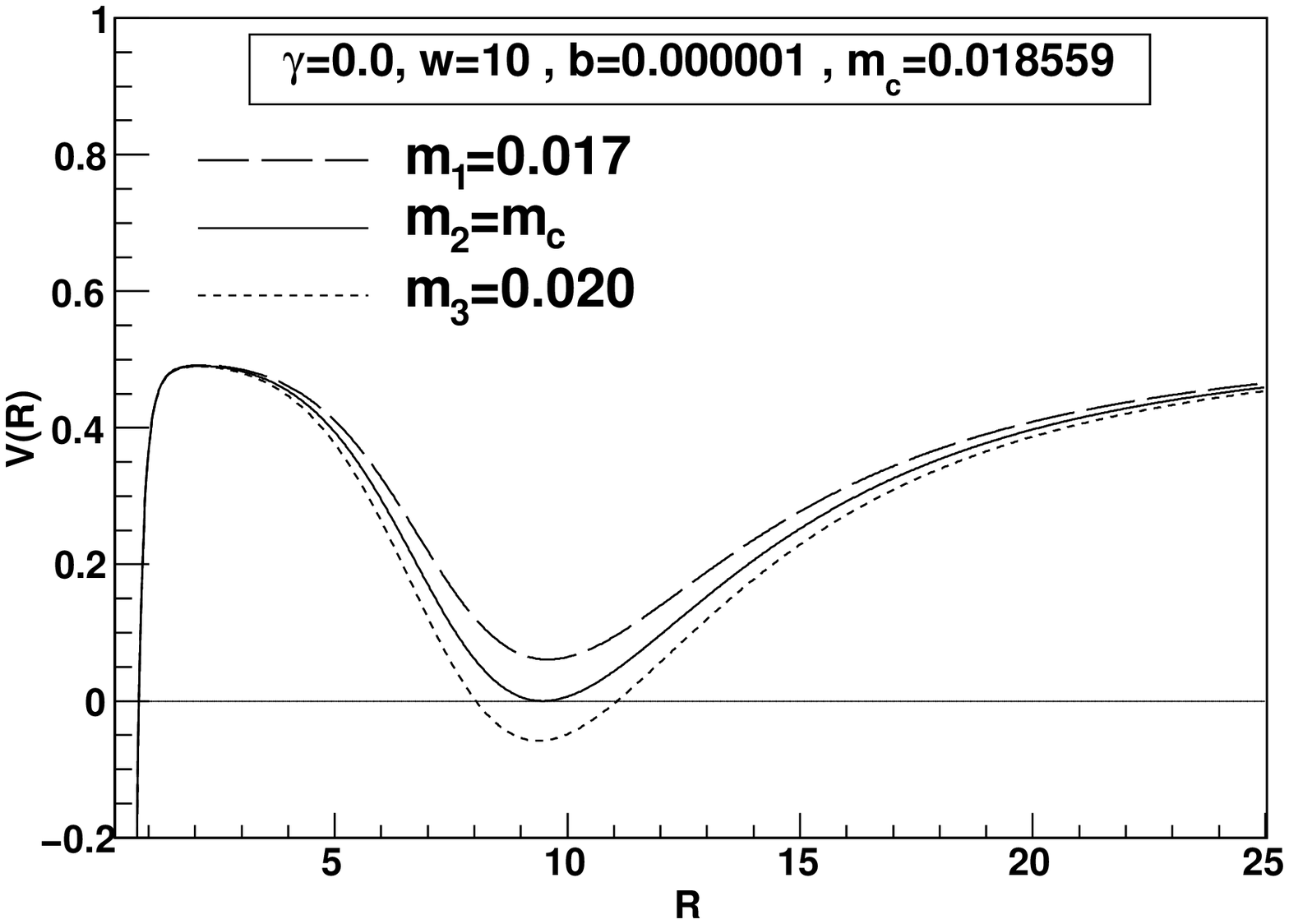,width=3.3truein,height=3.0truein}\hskip
.25in \psfig{figure=ECw10b0v000001novo.eps,width=3.3truein,height=3.0truein}
\hskip .5in} \caption{The potential $V(R)$ and the energy conditions EC1$\equiv \rho+p_r+2p_t$, 
EC2$\equiv \rho+p_r$ and EC3$\equiv \rho+p_t$, for $\gamma=0$,
$\omega=10$, $b=0.000001$ and $m_c=0.018559$. {\bf Case G}}
\label{fig35}
\end{figure}

\begin{figure}
\vspace{.2in}
\centerline{\psfig{figure=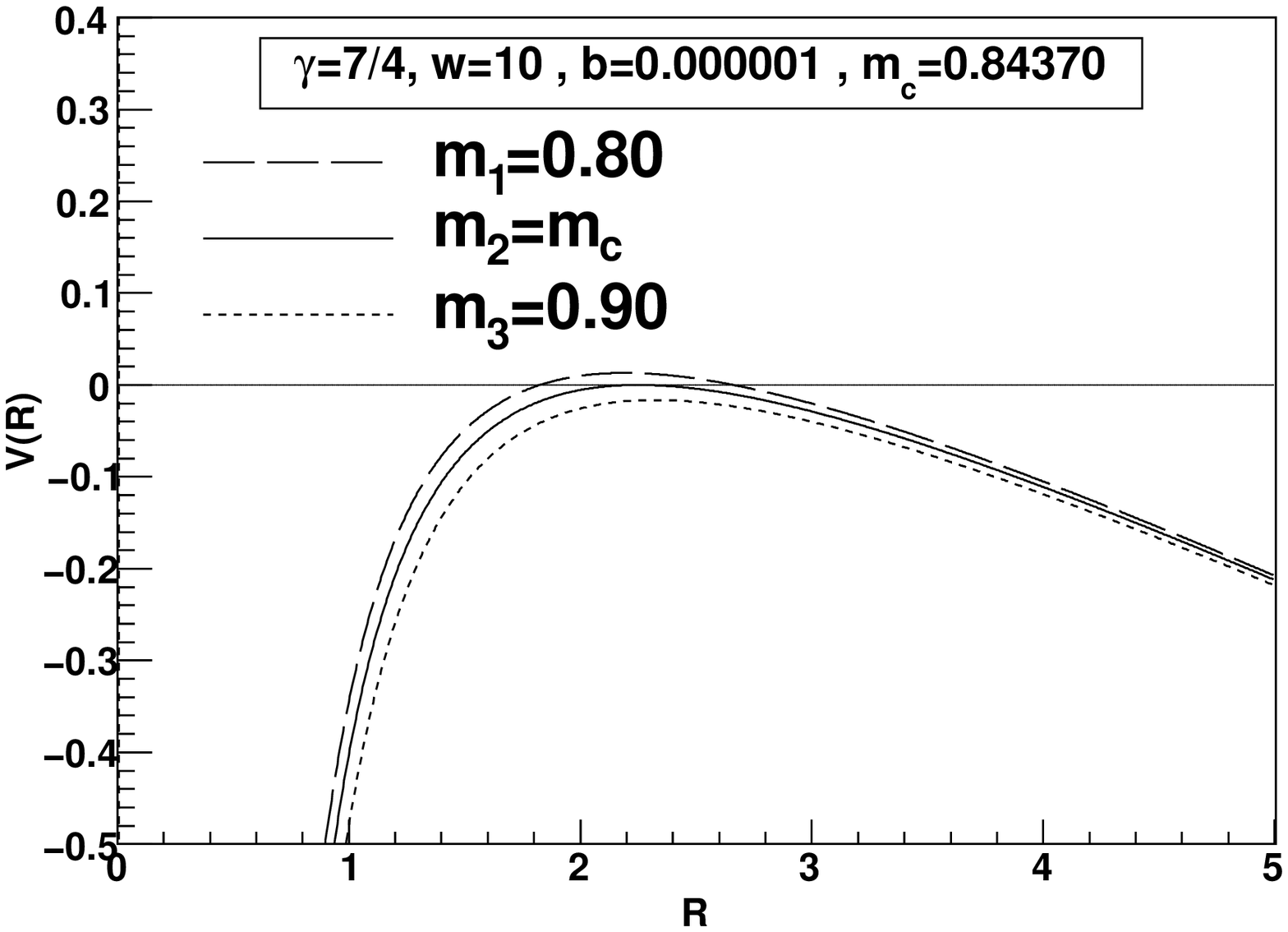,width=3.3truein,height=3.0truein}\hskip
.25in \psfig{figure=ECw10b0v000001novo.eps,width=3.3truein,height=3.0truein}
\hskip .5in} \caption{The potential $V(R)$ and the energy conditions EC1$\equiv \rho+p_r+2p_t$, 
EC2$\equiv \rho+p_r$ and EC3$\equiv \rho+p_t$, for $\gamma=7/4$,
$\omega=10$, $b=0.000001$ and $m_c=0.84370$. {\bf Case H}}
\label{fig36}
\end{figure}

\begin{figure}
\vspace{.2in}
\centerline{\psfig{figure=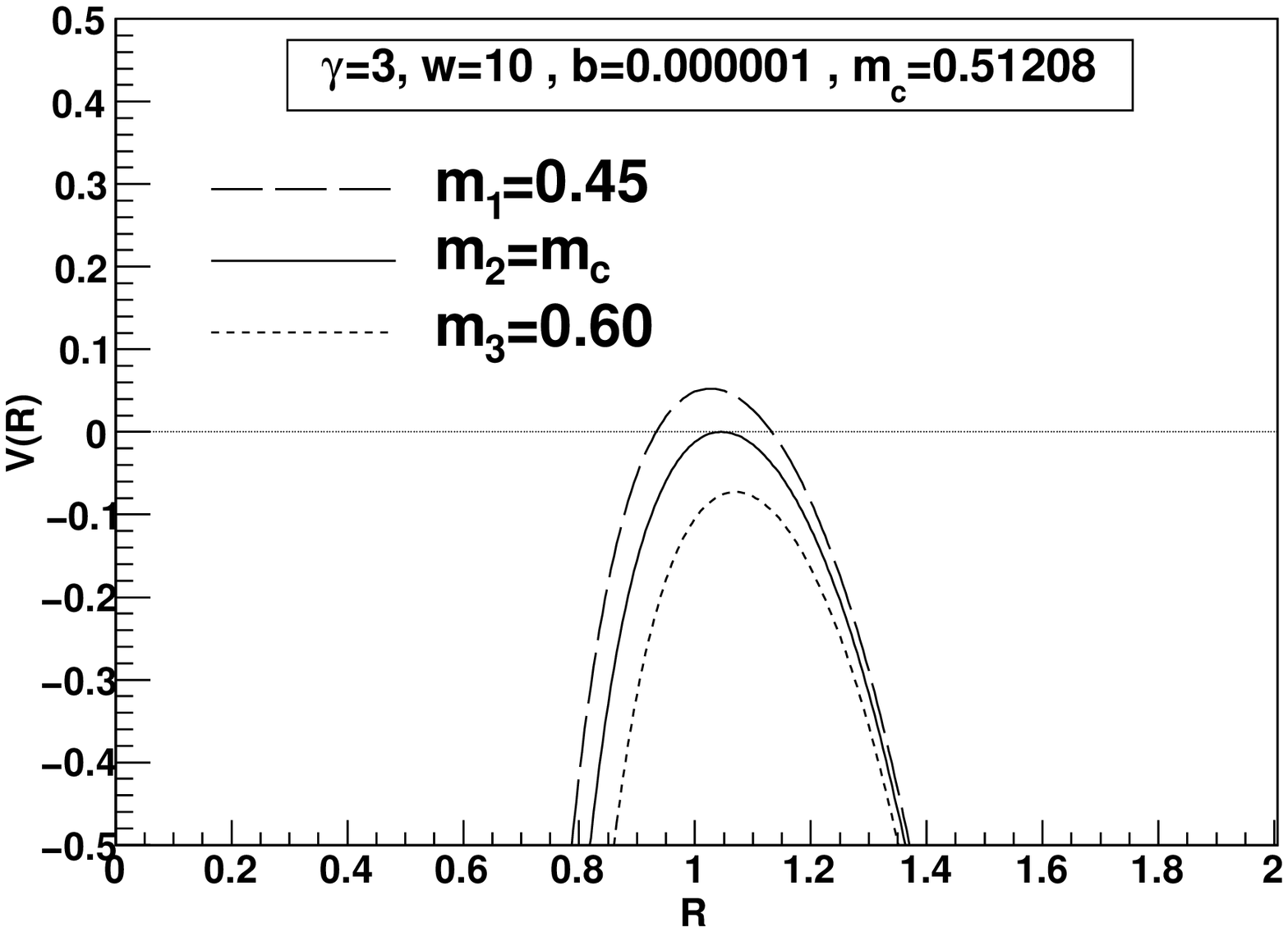,width=3.3truein,height=3.0truein}\hskip
.25in \psfig{figure=ECw10b0v000001novo.eps,width=3.3truein,height=3.0truein}
\hskip .5in} \caption{The potential $V(R)$ and the energy conditions EC1$\equiv \rho+p_r+2p_t$, 
EC2$\equiv \rho+p_r$ and EC3$\equiv \rho+p_t$, for $\gamma=3$,
$\omega=10$, $b=0.000001$ and $m_c=0.51208$. {\bf Case I}}
\label{fig37}
\end{figure}

\begin{figure}
\vspace{.2in}
\centerline{\psfig{figure=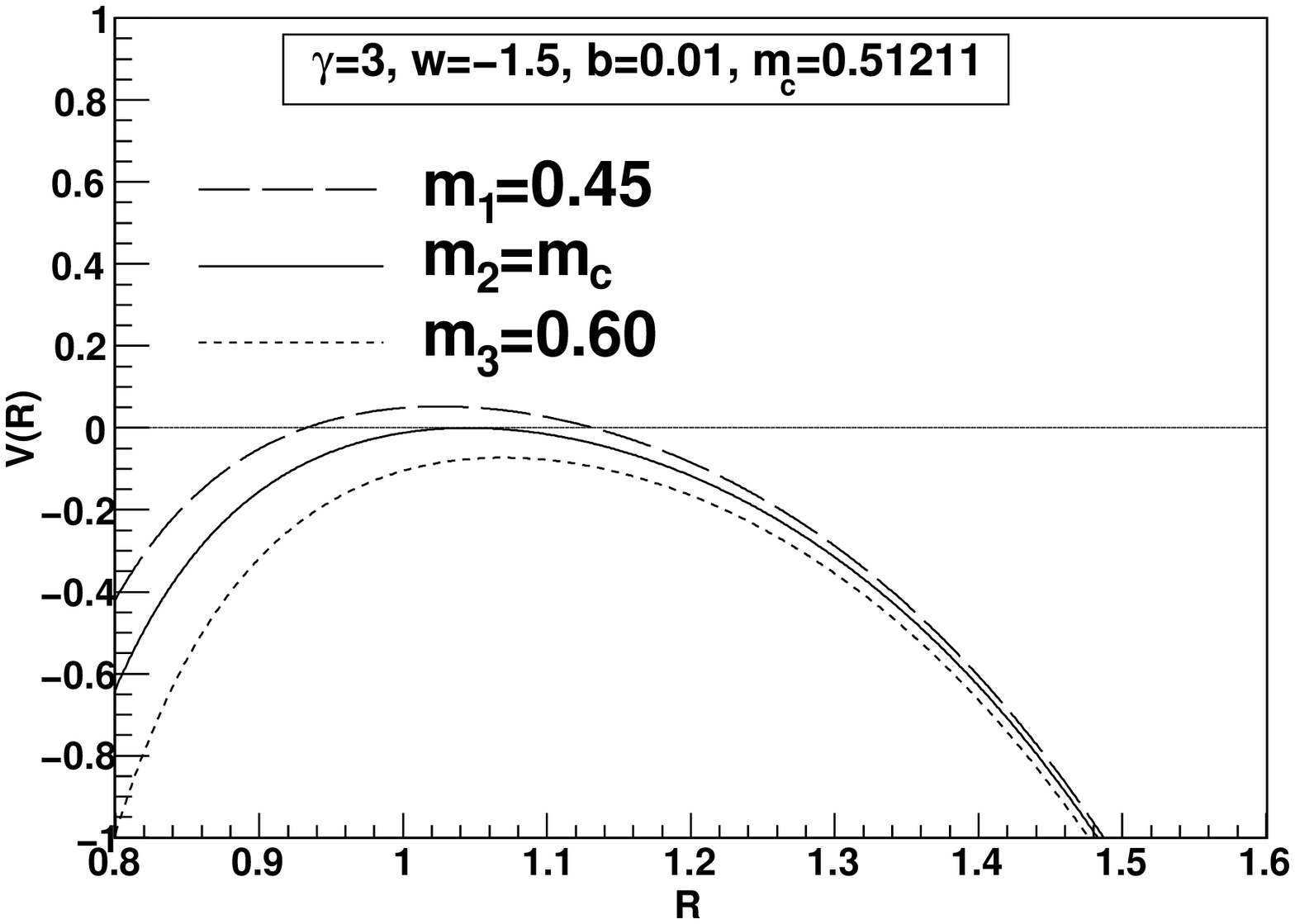,width=3.3truein,height=3.0truein}\hskip
.25in \psfig{figure=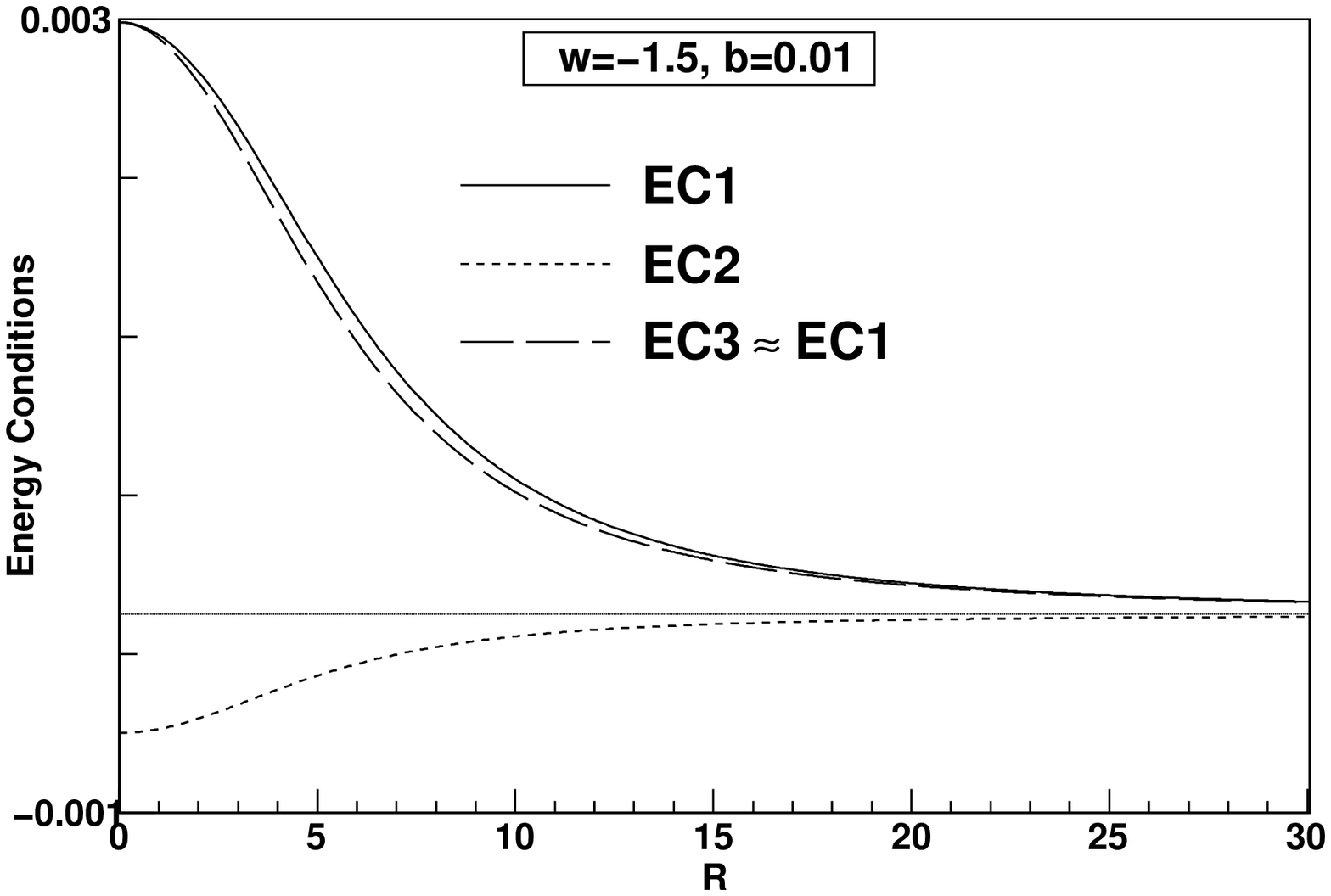,width=3.3truein,height=3.0truein}
\hskip .5in} \caption{The potential $V(R)$ and the energy conditions EC1$\equiv \rho+p_r+2p_t$, 
EC2$\equiv \rho+p_r$ and EC3$\equiv \rho+p_t$, for $\gamma=3$,
$\omega=-1.5$, $b=0.01$ and $m_c=0.51211$. {\bf Case K}}
\label{fig38}
\end{figure}

\begin{figure}
\vspace{.2in}
\centerline{\psfig{figure=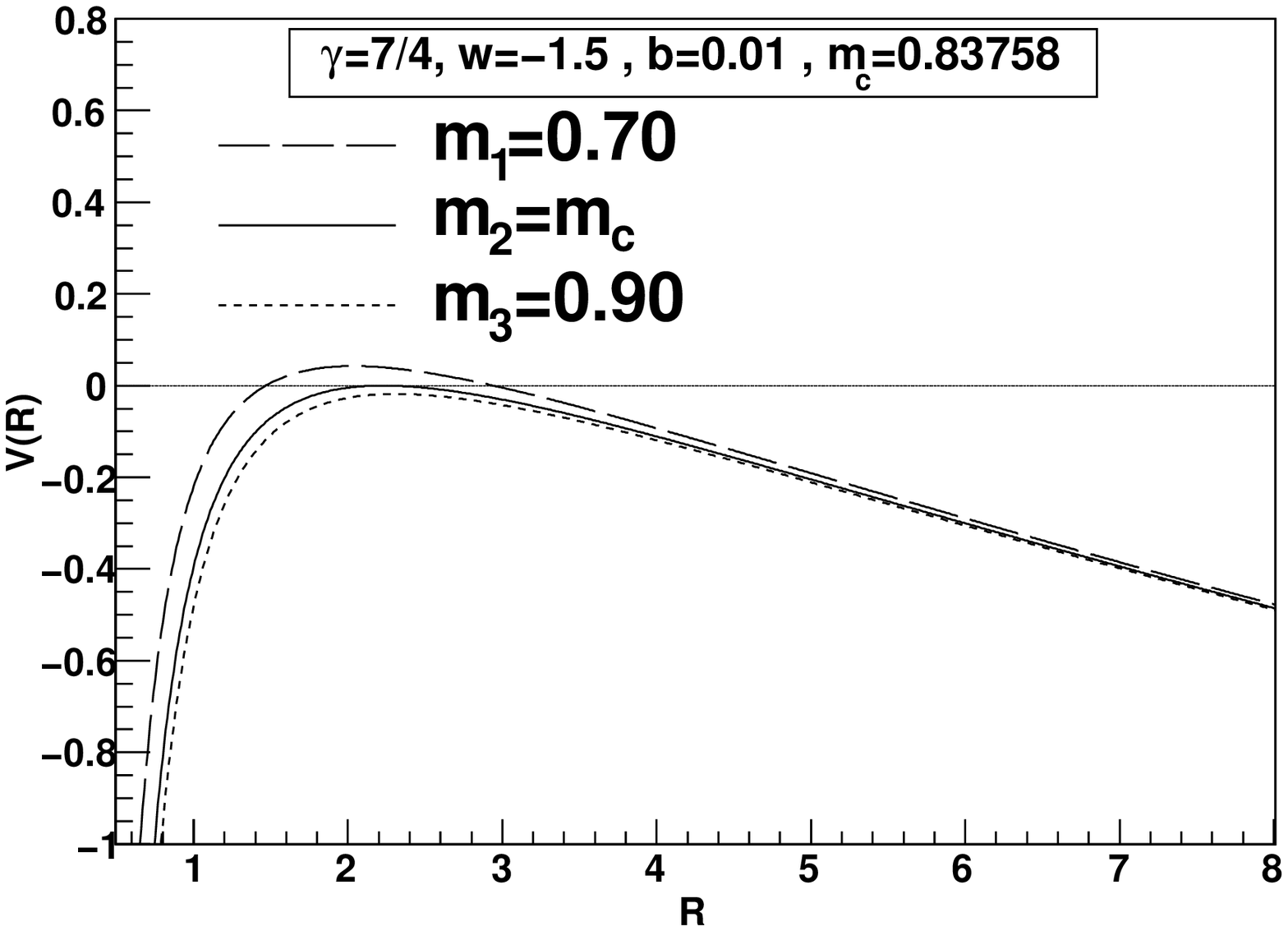,width=3.3truein,height=3.0truein}\hskip
.25in \psfig{figure=ECwneg1v5b0v01.eps,width=3.3truein,height=3.0truein}
\hskip .5in} \caption{The potential $V(R)$ and the energy conditions EC1$\equiv \rho+p_r+2p_t$, 
EC2$\equiv \rho+p_r$ and EC3$\equiv \rho+p_t$, for $\gamma=7/4$,
$\omega=-1.5$, $b=0.01$ and $m_c=0.83759$. {\bf Case L}}
\label{fig39}
\end{figure}

\begin{figure}
\vspace{.2in}
\centerline{\psfig{figure=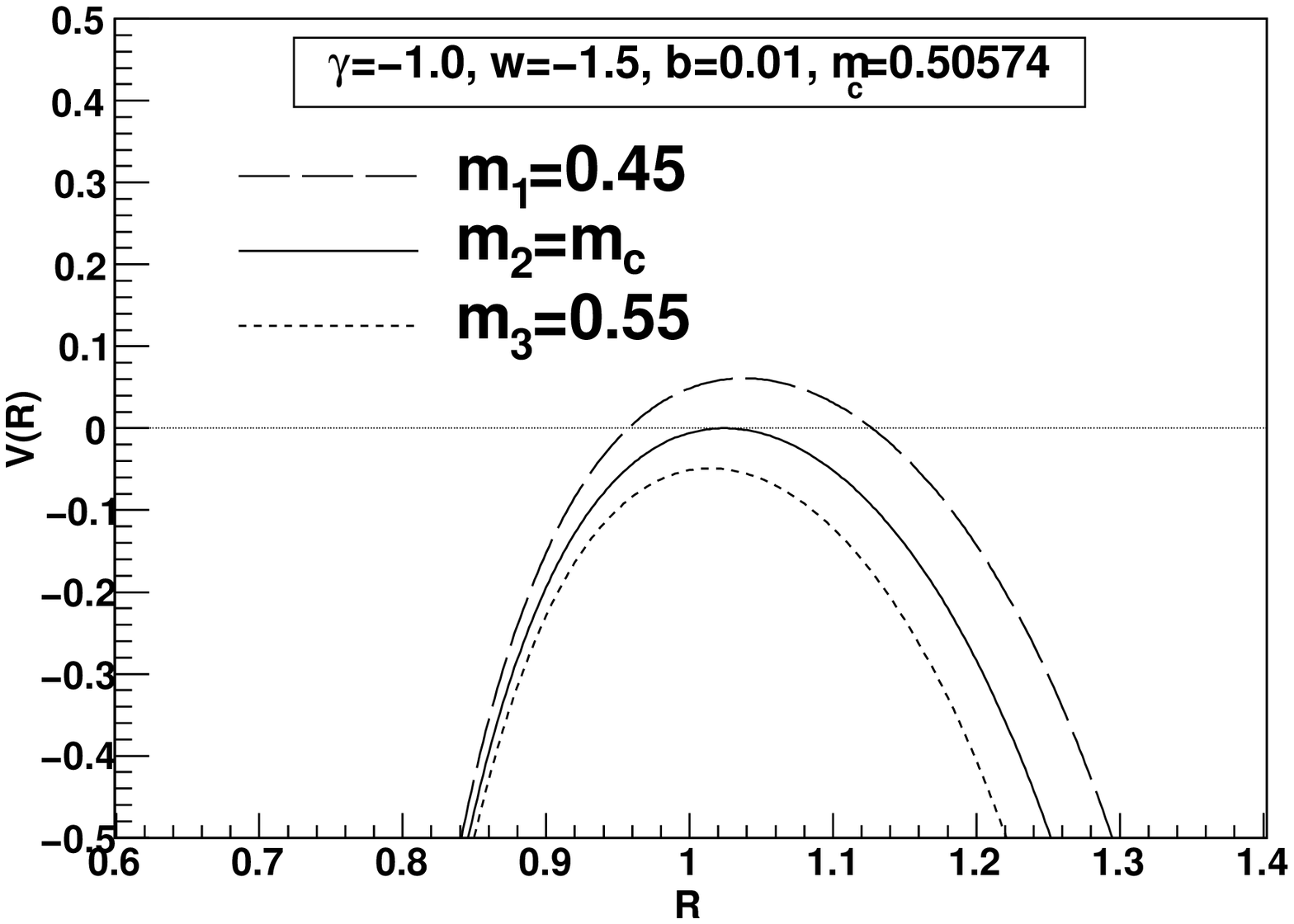,width=3.3truein,height=3.0truein}\hskip
.25in \psfig{figure=ECwneg1v5b0v01.eps,width=3.3truein,height=3.0truein}
\hskip .5in} \caption{The potential $V(R)$ and the energy conditions EC1$\equiv \rho+p_r+2p_t$, 
EC2$\equiv \rho+p_r$ and EC3$\equiv \rho+p_t$, for $\gamma=-1$,
$\omega=-1.5$, $b=0.01$ and $m_c=0.50574$. {\bf Case J}}
\label{fig40}
\end{figure}

\begin{figure}
\vspace{.2in}
\centerline{\psfig{figure=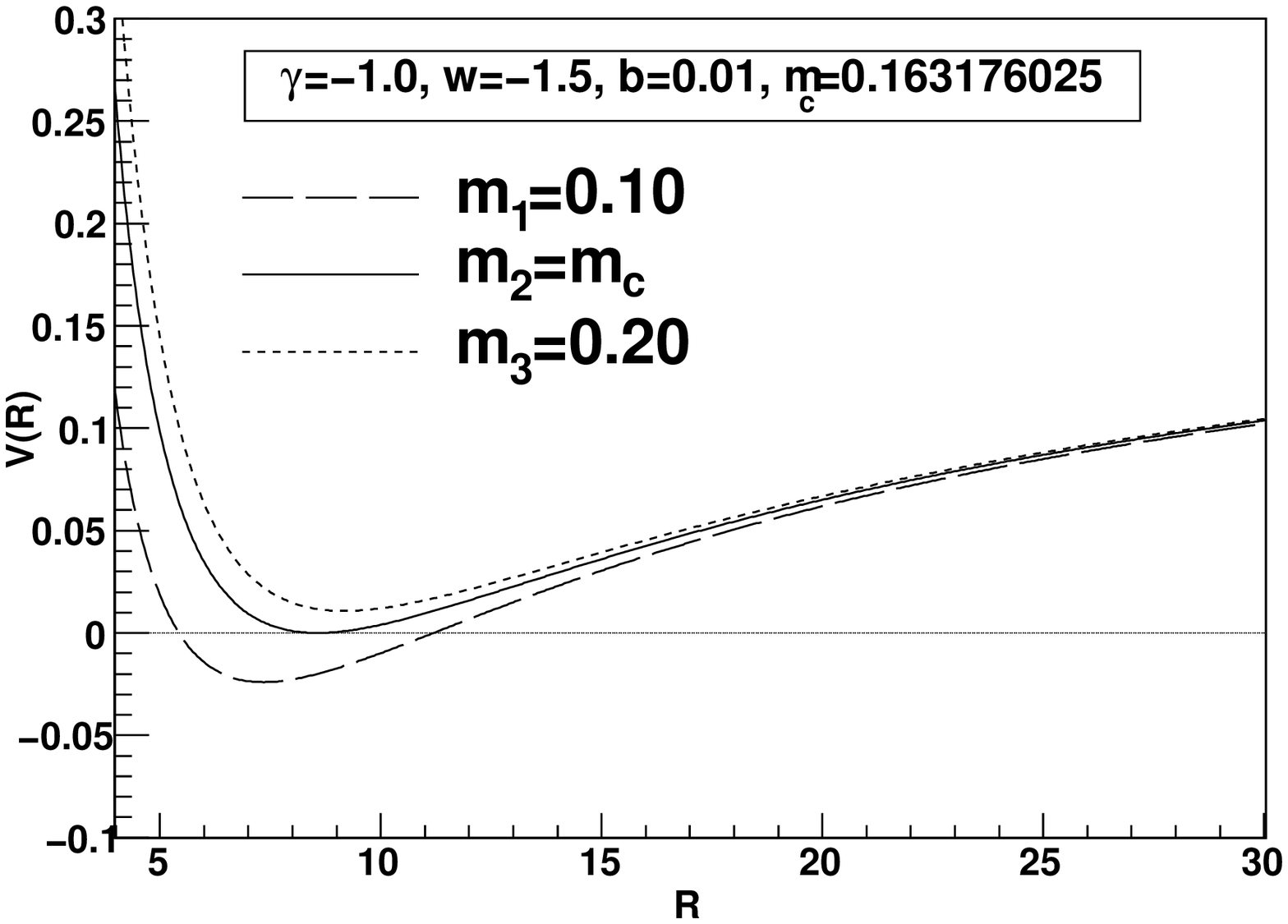,width=3.3truein,height=3.0truein}\hskip
.25in \psfig{figure=ECwneg1v5b0v01.eps,width=3.3truein,height=3.0truein}
\hskip .5in} \caption{The potential $V(R)$ and the energy conditions EC1$\equiv \rho+p_r+2p_t$, 
EC2$\equiv \rho+p_r$ and EC3$\equiv \rho+p_t$, for $\gamma=-1$,
$\omega=-1.5$, $b=0.01$ and $m_c=0.163176025$. {\bf Case J}}
\label{fig41}
\end{figure}

\begin{figure}
\vspace{.2in}
\centerline{\psfig{figure=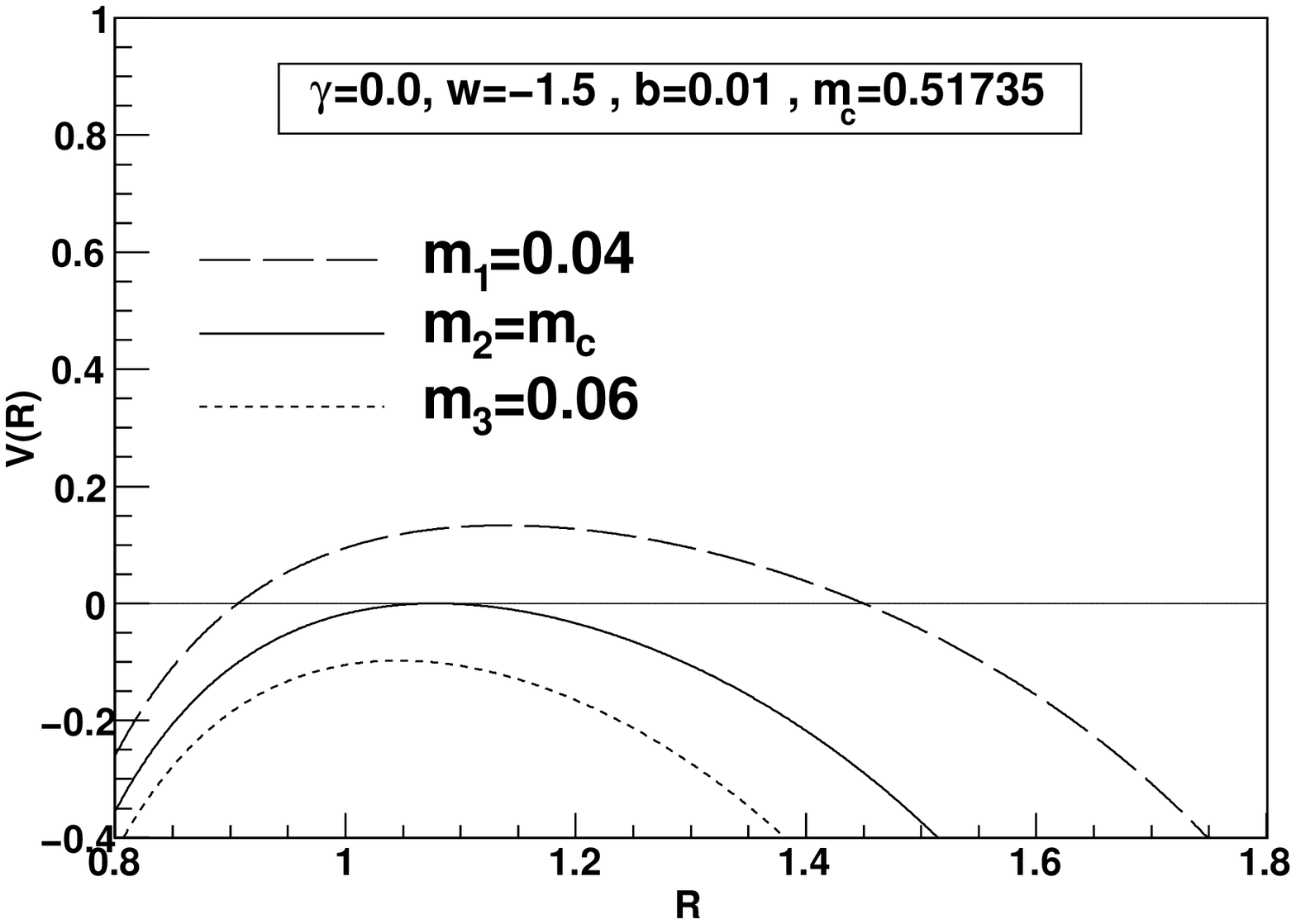,width=3.3truein,height=3.0truein}\hskip
.25in \psfig{figure=ECwneg1v5b0v01.eps,width=3.3truein,height=3.0truein}
\hskip .5in} \caption{The potential $V(R)$ and the energy conditions EC1$\equiv \rho+p_r+2p_t$, 
EC2$\equiv \rho+p_r$ and EC3$\equiv \rho+p_t$, for $\gamma=0$,
$\omega=-1.5$, $b=0.01$ and $m_c=0.51735$. {\bf Case J}}
\label{fig42}
\end{figure}

\begin{figure}
\vspace{.2in}
\centerline{\psfig{figure=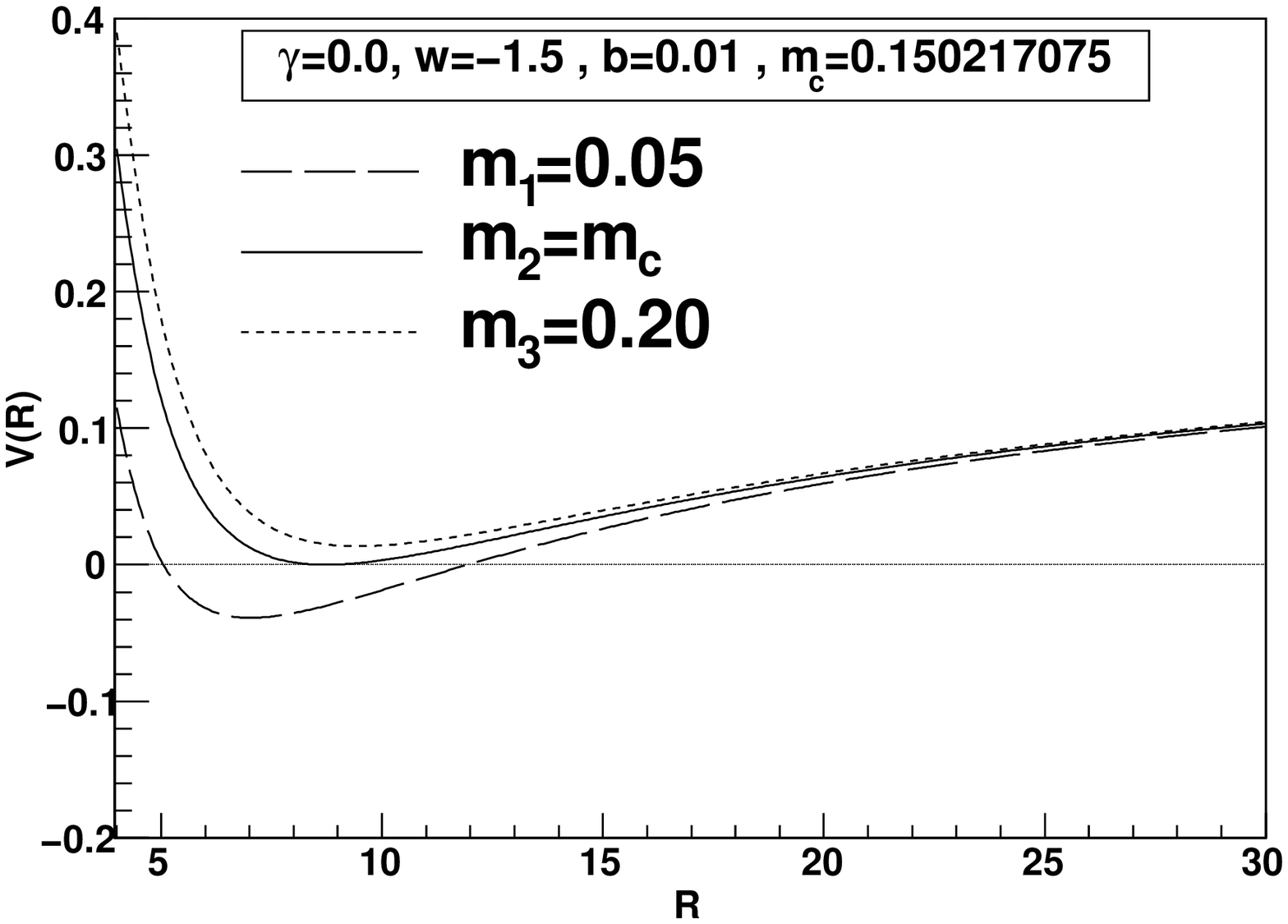,width=3.3truein,height=3.0truein}\hskip
.25in \psfig{figure=ECwneg1v5b0v01.eps,width=3.3truein,height=3.0truein}
\hskip .5in} \caption{The potential $V(R)$ and the energy conditions EC1$\equiv \rho+p_r+2p_t$, 
EC2$\equiv \rho+p_r$ and EC3$\equiv \rho+p_t$, for $\gamma=0$,
$\omega=-1.5$, $b=0.01$ and $m_c=0.150217075$. {\bf Case J}}
\label{fig43}
\end{figure}

\begin{acknowledgments}
The financial assistance from 
FAPERJ/UERJ (MFAdaS) are gratefully acknowledged. The
author (RC) acknowledges the financial support from FAPERJ (no.
E-26/171.754/2000, E-26/171.533/2002 and E-26/170.951/2006). 
The authors (RC and MFAdaS) also acknowledge the financial support from 
Conselho Nacional de Desenvolvimento Cient\'{\i}fico e Tecnol\'ogico - 
CNPq - Brazil.  The author (MFAdaS) also acknowledges the financial support
from Financiadora de Estudos e Projetos - FINEP - Brazil (Ref. 2399/03).
\end{acknowledgments}

\end{document}